\documentclass[a4paper,12pt]{article}

\usepackage[utf8]{inputenc}
\usepackage[obeyspaces]{url}
\usepackage{physics}
\usepackage{mathtools}
\usepackage{bm}
\usepackage{booktabs,tabularx}
\usepackage{graphicx}
\usepackage{xspace}
\usepackage[usenames]{xcolor}
\usepackage{listings}
\usepackage[absolute]{textpos}
\usepackage[many]{tcolorbox}
\usepackage{xparse}
\usepackage[font=small,labelfont=bf,format=plain]{caption}
\usepackage{bbm}
\usepackage{subcaption}
\usepackage{dsfont}
\usepackage[normalem]{ulem}
\usepackage{dirtree}
\usepackage[text,italic]{esdiff}
\usepackage{multirow}
\usepackage{makecell}
\usepackage{tabularx}
\usepackage{footnote}
\usepackage{amsmath,amssymb}
\usepackage{fullpage}
\usepackage{lmodern}
\definecolor{darkblue}{rgb}{0.0, 0.0, 0.55}
\usepackage[colorlinks=true,allcolors=darkblue,linktocpage=true]{hyperref}
\usepackage{cleveref}
\usepackage[numbers,sort&compress]{natbib}

\bibstyle{elsarticle-num}

\makeatletter
\g@addto@macro\bfseries{\boldmath}
\makeatother

\newcommand{\refcite}[1]{Ref.~\cite{#1}}
\newcommand{\Refcite}[1]{Ref.~\cite{#1}}

\newcommand{\gev}{\ensuremath{\,\text{GeV}}\xspace}
\newcommand{\msbar}{\ensuremath{\overline{\text{MS}}}\xspace}
\newcommand{\tree}{\text{tree}}
\newcommand{\hight}{\text{HT}}
\newcommand{\ztwo}{\ensuremath{\mathbb{Z}_2}\xspace}
\newcommand{\os}{\text{OS-like}}
\newcommand{\cw}{\text{CW}}

\newcommand{\gammaew}{\ensuremath{\gamma_{\rm EW}}\xspace}
\newcommand{\Rxi}{\ensuremath{R_\xi}\xspace}

\newcommand{\half}{\frac{1}{2}}
\newcommand{\recip}[1]{\frac{1}{#1}}

\newcommand{\atvevZT}[1]{\left. #1 \right\vert_{\vec{v}}}
\newcommand{\field}{\vec{\phi}}
\newcommand{\vp}{(\field)}
\newcommand{\vv}{(\vec{v})}

\newcommand{\mitwopv}{m_i^2\vp}
\newcommand{\mitwovv}{m_i^2\vv}
\newcommand{\mifourpv}{m_i^4\vp}
\newcommand{\mifourvv}{m_i^4\vv}
\renewcommand{\vec}[1]{\bm{#1}}
\newcommand{\mt}{\bar{m}}

\newcommand{\xibw}{\xi}

\newcommand{\pt}{\texttt{Phase\-Tracer}\@\xspace}

\newcommand{\fs}{\texttt{Flexible\-SUSY}\@\xspace}
\newcommand{\sarah}{\texttt{SARAH}\@\xspace}
\newcommand{\softsusy}{\texttt{SOFT\-SUSY}\@\xspace}

\crefname{figure}{Fig.{}}{Figs.{}}
\crefname{section}{Sec.{}}{Secs.{}}
\crefname{table}{Tab.{}}{Tabs.{}}

\title{How arbitrary are perturbative calculations of the electroweak phase transition?}
\author{Peter Athron$^{a,b}$\footnote{peter.athron@njnu.edu.cn},
Csaba Bal\'azs$^b$\footnote{csaba.balazs@monash.edu},
Andrew Fowlie$^a$\footnote{andrew.j.fowlie@njnu.edu.cn},
Lachlan Morris$^b$\footnote{lachlan.morris@monash.edu},
\\
Graham White$^c$\footnote{grahamwhite@g.ecc.u-tokyo.ac.jp},
Yang Zhang$^{d,e}$\footnote{zhangyangphy@zzu.edu.cn}}
\date{}

\begin{document}

\maketitle
\thispagestyle{empty}
\begin{center}
  \it
  $^a$Department of Physics and Institute of Theoretical Physics, Nanjing Normal University, Nanjing, Jiangsu 210023, China\\[0.5em]
  $^b$School of Physics and Astronomy, Monash University, \\ Melbourne 3800 Victoria, Australia\\[0.5em]
  $^c$Kavli IPMU (WPI), UTIAS, The University of Tokyo, \\ Kashiwa, Chiba 277-8583, Japan\\[0.5em]
  $^d$School of Physics, Zhengzhou University, Zhengzhou 450000, China\\
  $^e$CAS Key Laboratory of Theoretical Physics, Institute of Theoretical Physics, \\Chinese Academy of Sciences, Beijing 100190, China\\
\end{center}
\begin{abstract}
We investigate the extent to which perturbative calculations of the electroweak phase transition are arbitrary and uncertain, owing to their gauge, renormalisation scale and scheme dependence, as well as treatments of the Goldstone catastrophe and daisy diagrams. Using the complete parameter space of the Standard Model extended by a real scalar singlet with a $\mathbb{Z}_2$ symmetry as a test, we explore the properties of the electroweak phase transition in general $R_\xi$ and covariant gauges, OS and $\overline{\text{MS}}$ renormalisation schemes, and for common treatments of the Goldstone catastrophe and daisy diagrams. Reassuringly, we find that different renormalisation schemes and different treatments of the Goldstone catastrophe and daisy diagrams typically lead to only modest changes in predictions for the critical temperature and strength of the phase transition. On the other hand, the gauge and renormalisation scale dependence may be significant, and often impact the existence of the phase transition altogether. 
\end{abstract}

\clearpage
\setcounter{tocdepth}{2}
\tableofcontents

\section{Introduction}
A central motivation for the ambitious program of the next generation experiments that probe physics beyond the Standard Model (SM) is to understand aspects of electroweak symmetry breaking in the cosmological context~\cite{Caprini:2019egz,Ramsey-Musolf:2019lsf,Barrow:2022gsu}.  If the electroweak phase transition (EWPT) was strongly first order, it would satisfy Sakharov's third condition for baryogenesis~\cite{Sakharov:1967dj} and the collision of sound shells could create a stochastic gravitational wave background that is expected to peak at a frequency range visible to LISA~\cite{Caprini:2019egz}.  In the SM, however, the EWPT is predicted to be a crossover~\cite{Kajantie:1995kf,Kajantie:1996mn,Kajantie:1996qd,Csikor:1998eu,DOnofrio:2015gop} and new states near the electroweak scale would be needed to change it to a strong first-order transition~\cite{Pietroni:1992in,Cline:1996mga,Ham:2004nv,Funakubo:2005pu,Barger:2008jx,Chung:2010cd,Espinosa:2011ax,Chowdhury:2011ga,Gil:2012ya,Carena:2012np,No:2013wsa,Dorsch:2013wja,Curtin:2014jma,Huang:2014ifa,Profumo:2014opa,Kozaczuk:2014kva,Jiang:2015cwa,Curtin:2016urg, Vaskonen:2016yiu,Dorsch:2016nrg,Huang:2016cjm,Chala:2016ykx,Basler:2016obg,Beniwal:2017eik,Bernon:2017jgv,Kurup:2017dzf,Andersen:2017ika,Chiang:2017nmu,Dorsch:2017nza,Beniwal:2018hyi,Bruggisser:2018mrt,Athron:2019teq,Kainulainen:2019kyp,Bian:2019kmg,Li:2019tfd,Chiang:2019oms,Xie:2020bkl,Bell:2020gug}.  Either a 27 TeV upgrade to the LHC~\cite{Papaefstathiou:2020iag} or the ambitious 100 TeV proposal~\cite{Kotwal:2016tex} could potentially confirm the SM prediction~\cite{Kajantie:1995kf,Kajantie:1996mn,Kajantie:1996qd,Csikor:1998eu,DOnofrio:2015gop}. 

Phenomenological studies usually assess the impact of new states on the EWPT through the perturbative effective potential.
As has long been recognised, the perturbative effective potential depends on a choice of gauge and gauge fixing parameter $\xi$~\cite{Patel:2011th}, a choice of renormalisation scheme~\cite{Martin:2001vx, McKeon:2015zxa} and a choice of renormalisation scale~\cite{Coleman:1973jx}.  Besides these choices, there are different ways to remedy infrared (IR) divergences caused by Goldstone boson loops and different treatments of daisy diagrams~\cite{Arnold:1992fb}.  These subjective factors lead to theoretical uncertainties and a degree of arbitrariness associated with the perturbative EWPT predictions. For a recent update on non-perturbative approaches, see \refcite{Gould:2022ran,Schicho:2022wty}. An effective field theory approach is also being developed that promises a gauge-independent evaluation of the effective potential free of some of the problems of the traditional approaches \cite{Schicho:2022wty, Ekstedt:2022ceo, Gould:2022ran, Ekstedt:2022zro, Ekstedt:2022bff}.   

In this work we are predominantly motivated by the possibility of a first-order EWPT.  In that case, the ground state makes a discontinuous jump while electroweak symmetry breaks.  As is customary, we characterise a transition by a critical temperature, below which the transition is possible, and a transition strength. More detailed descriptions could include, for example, the latent heat, which characterises the energy available to phenomena such as gravitational waves, and the nucleation temperature. A critical temperature is defined as any temperature at which the effective potential contains two degenerate minima, i.e.,
\begin{equation}
    V(\Phi, T_c) = V(\Theta, T_c) ,
\end{equation}
where $\Phi$ and $\Theta$ are distinct locations in the field space where the potential has a minimum at the temperature $T_c$. 
For non-manifestly gauge covariant calculations the strength of the transition is typically quantified by 
\begin{equation}
    \gamma_{\text{EW}} = \frac{\sqrt{\sum_i (\Phi_i - \Theta_i)^2}}{T_c} ,
\end{equation}
where the sum only includes components of the fields that break EW symmetry.
Within electroweak baryogenesis, the condition that $\gamma_{\text{EW}} \gtrsim 1$ approximately corresponds to avoiding washing-out baryon asymmetries by sphaleron processes~\cite{Patel:2011th}. The rate of sphaleron processes is governed by the spaleron action, which in the context of baryogenesis could be a more appropriate characterization of a phase transition than $\gamma_{\text{EW}}$. The sphaleron action is gauge independent if computed carefully~\cite{Garny:2012cg,Fuyuto:2014yia,Fuyuto:2015jha}.
However, computing it requires solving a system of coupled differential equations, and hence it tends to be calculated for benchmarks rather than scans. Such computationally expensive calculations are not currently feasible for this study because we wish to explore the arbitrariness of perturbative calculations across the parameter space of a model. So we leave that for future work. We compute $T_c$ and $\gamma_{\text{EW}}$ numerically from the effective potential using \pt; see the manual~\cite{Athron:2020sbe} for an overview of the numerical methods used.  For the purposes of the work described here we extended\footnote{The code for this is in the development branch of our github repository and will later be released in a future stable version.} \pt to include other approaches such as the so called $\hbar$ expansion approach of \refcite{Patel:2011th}. These features should be available in a future release of \pt and are not to our knowledge implemented in any other public software.  

In this work, for the first time we present a comprehensive study of uncertainties that arise in the simplest widely used techniques in theoretical calculations.\footnote{For example, we do not consider lattice calculations, or the $\hbar$-expansion in dimensional reduction.} Such techniques include gauge-independent methods that lack resummation at leading order~\cite{Patel:2011th} and gauge-dependent methods that include leading-order resummation. Additionally, we examine numerical differences arising from different renormalisation schemes and gauge sensitivity arising from $R_\xi$ and covariant gauges and gauge-dependent tadpole constraints.
Quantifying gauge dependence is difficult, as although there are indications from perturbativity, it is unclear how much to change the gauge parameter. Our investigation of renormalisation scheme and scale dependence, and the two gauge-independent approaches, on the other hand, do not depend on these choices.
Approximate as they may be, such techniques are fast and simple and have been used in phenomenological analyses in many models, including large multi-parameter scans. In this paper we are agnostic about the theoretical properties, assumptions and shortcomings in the methods, and instead focus on their numerical differences and abitrariness, and leave it to the reader to decide the most appropriate method for their analysis.

As our benchmark model, we consider the SM augmented by a real scalar singlet (the SSM) and a ${\mathbb Z}_2$ discrete symmetry imposed on the potential. This model is ideal in its simplicity for highlighting key features of the parameter space as it includes both tree-level and thermal barriers depending on whether the discrete symmetry is broken at non-zero temperatures (but is enforced at zero temperature). It also includes cases where the gauge boson thermal loops are sufficiently important that gauge dependence might become crucial, as it does in the SM with a light Higgs boson~\cite{Patel:2011th}.  We build on recent progress in examining the impact of theoretical uncertainties on the EWPT. The gauge dependence of tunneling rates and the implications for gravitational wave signals in gauge-dependent techniques were examined in \refcite{Garny:2012cg,Chiang:2017zbz,Arunasalam:2021zrs} and the scale dependence in multiple models was investigated in \refcite{Chiang:2018gsn,Croon:2020cgk,Gould:2021oba}. Here, we consider the combined impacts of choices of scale, gauge and gauge fixing parameter, treatment of higher-order daisies, and the Goldstone catastrophe (GC).

The paper is organised as follows. In \cref{sec:treatments} we define our model and discuss various treatments of the effective potential, including choices of renormalisation scheme and gauge. In \cref{sec:modifications}, we describe modifications to the effective potential to account for daisy diagrams and avoid the GC. In \cref{sec:results}, we show our numerical results on the impact of the choice of renormalisation scheme, scale, resummation methods, and gauge. Finally, the conclusions and discussions are given in \cref{sec:conc}.

\section{The effective potential}\label{sec:treatments}

\begin{figure}[t!]
    \centering
    \includegraphics[height=4cm]{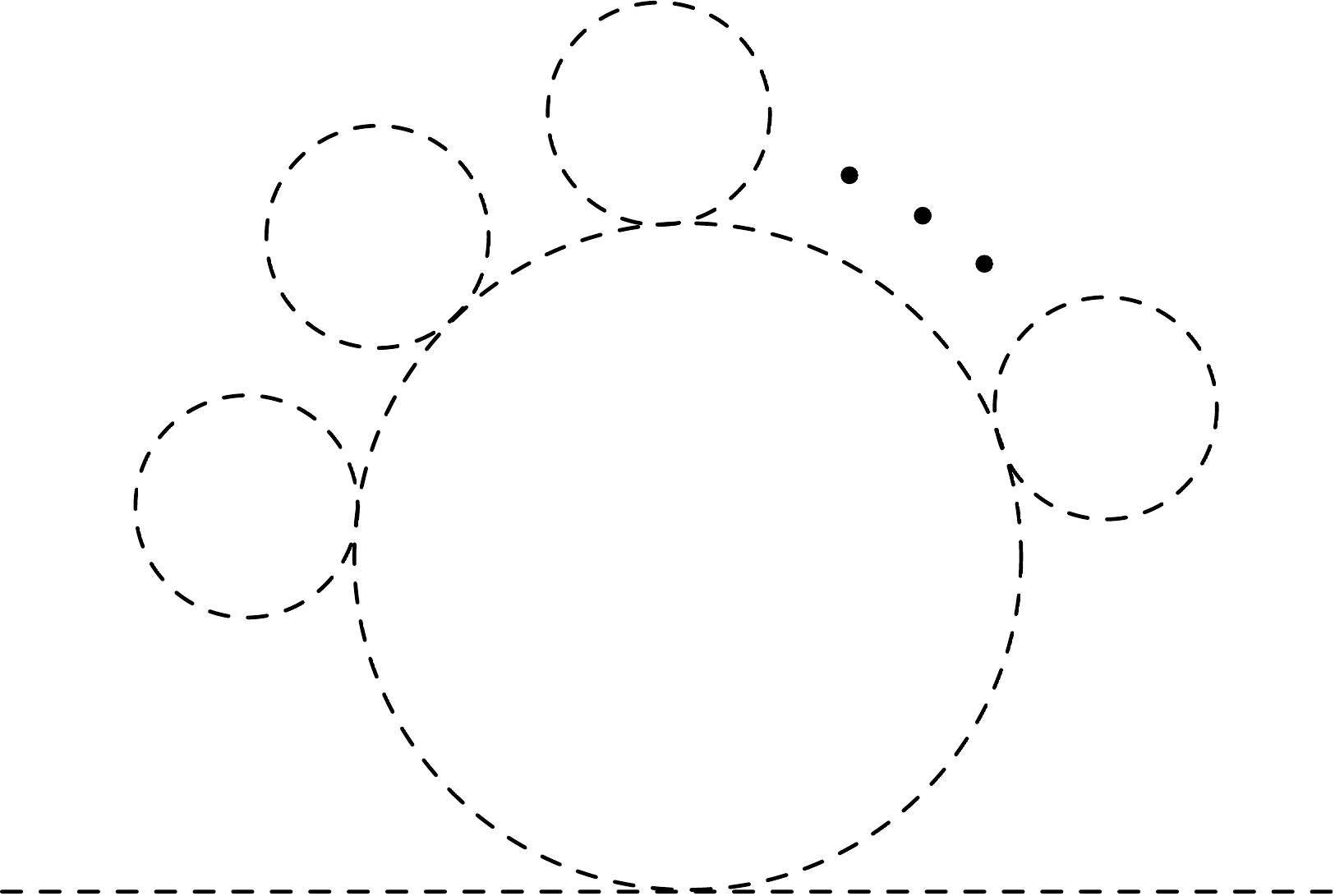}
    \caption{Daisy contributions to the two-point function from the quartic interaction. Each additional loop contributes a factor $\lambda T^2 / m^2$, spoiling perturbativity unless appropriately resummed.}
    \label{fig:daisy}
\end{figure}

Evidently, analysing the dynamics of the phase transition requires an accurate treatment of the effective potential. In the following subsections, we discuss the methods considered in this paper. Unfortunately, all methods suffer from some of the following potential theoretical ambiguities:
\begin{enumerate}
    \item Gauge dependence --- the effective potential is fundamentally gauge dependent as it only contains one-particle irreducible loop diagrams. This doesn't include external leg corrections that would otherwise be included in scattering amplitudes to cancel the gauge dependence with other diagrams and maintain gauge independence~\cite{Jackiw:1974cv}. The choice of gauge can strongly impact the EWPT predictions; for example, in the SM the strength of the EWPT at leading order is proportional to terms involving the gauge parameter,~$\xi^{3/2}$~\cite{Patel:2011th}.\footnote{Though \refcite{Garny:2012cg} notes that such terms come with opposite signs and that the gauge dependence isn't strong for gauges consistent with a perturbative expansion.}
    \item Scale dependence --- the one-loop potential and resulting phenomenological predictions could strongly depend on the scale $Q$ and choice of renormalisation scheme. The dependence originates from regularizing the loop corrections in a particular scheme, as usual.
    \item The IR problem~\cite{PhysRevD.9.3320,Weinberg:1974hy,Kirzhnits:1974as} --- the perturbative expansion may break down near the critical temperature. The loop expansion involves the mode occupation $g \to g n_B \sim g T/m$ for gauge bosons and $\lambda \to \lambda T^2/m^2$ for scalars as shown in the daisy diagram in \cref{fig:daisy}.\footnote{In the SM it is common to use $\lambda \sim g^2$ which reduces the theory to power counting in a single coupling. This though is model dependent, see for instance \refcite{Ekstedt:2021kyx}.} This expansion diverges when $T \gg m$. Specifically the Goldstone contributions can get large deep in the broken phase and the gauge boson contributions can get large for small field values when $m \simeq 0$.  There exists two prescriptions that alleviate the IR problem by resumming the most dangerous daisy diagrams to all orders; see \cref{sec:daisy}.
    \item Goldstone catastrophe (GC)~\cite{Elias-Miro:2014pca,Martin:2014bca,PhysRevD.97.056020} --- Goldstone bosons could be massless, e.g.,  in the $\xi = 0$ gauge in the tree-level vacuum, and lead to divergences in the second-derivatives of the potential. Although terms in one-loop corrections to the potential of the form $x^2 \ln x \to 0$ vanish as $x\to 0$, the second derivatives diverge, meaning that we could not compute e.g., the Higgs masses at one-loop. Whilst the photon is also massless, its mass isn't field-dependent, and doesn't lead to these problems. See \cref{sec:goldstone} for a discussion of the treatments of this problem considered in this work.
    \item Tadpole constraints --- fixing parameters to ensure agreement at one-loop with experimentally measured Higgs mass and vacuum expectation value (VEV) can reintroduce gauge dependence if not done carefully, even in an otherwise gauge-independent method (see \cref{sec:hbar}). To do so one should only impose the tadpole conditions at tree-level.  However in this case one must include one-loop tadpoles diagrams in corrections to the masses and other loop calculations.   
\end{enumerate}
The occurrences of these problems in the treatments in this paper are summarised in \cref{tab:compare}. In addition, uncertainties on the SM nuisance parameters, such as the top mass, result in parametric uncertainties. 

\begin{table}[t]
    \centering
    \footnotesize
    \renewcommand{\arraystretch}{1.25}
    \begin{tabular}{lllll}
    \toprule
    Name & Order & $\xi$ dependence & $\mu$ dependence & Concern\\
    \midrule
    PRM & \makecell[tl]{Tree minima;\\ 1-loop potential} & Tadpoles & Explicit & No daisies \\
    High-$T$ & 1-loop leading terms & Tadpoles & Implicit & Accuracy \\
    \msbar & 1-loop potential & Explicit &  Explicit & $\xi$ \& $\mu$ dependence \\
    \msbar + RGE & \makecell[tl]{1-loop potential;\\ 2-loop RGE} & Explicit & Reduced by RGE & $\xi$ dependence \\
    OS & 1-loop potential & Explicit & No & $\xi$ dependence \\
    Covariant gauge & 1-loop potential & Explicit &  Explicit & $\xi$ \& $\mu$ dependence \\
    \bottomrule
    \end{tabular}
    \caption{Comparison of perturbative treatments of the effective potential explored in this work. In the final column, we show our initial concerns about the treatment that we investigate in detail in \cref{sec:results}.}
    \label{tab:compare}
\end{table}

The benchmark model (the SSM) used in our calculations is the SM extended by a real scalar singlet. The tree-level Higgs potential for the Higgs doublet and the real scalar field respects a \ztwo symmetry, $S \rightarrow -S$,
\begin{equation}
V_0(H,S) = -\mu_h^2 H^{\dag}H + \lambda_h(H^{\dag}H)^2 - \frac{\mu_s^2}{2} S^2 + \frac{\lambda_s}{4} S^4 + \frac{\lambda_{hs}}{2} H^{\dag}H S^2.
\end{equation}
Note the convention of the negative sign in front of the $\mu_h^2$ and $\mu_s^2$ terms.
Denoting the fields as 
\begin{equation}
H = \frac{1}{\sqrt 2}\begin{pmatrix}
        G^\pm\\
        \phi_h + i G^0
    \end{pmatrix}
\quad \text{and} \quad
S = \phi_s ,
\end{equation}
where we treat $\phi_h$ and $\phi_s$ as real background fields, the tree level effective potential takes the form 
\begin{equation}
V_0\vp = -\frac{\mu_h^2}{2} \phi_h^2 + \frac{\lambda_{h}}{4}\phi_h^4 - \frac{\mu_s^2}{2} \phi_s^2 + \frac{\lambda_s}{4} \phi_s^4 + \frac{\lambda_{hs}}{4} \phi_h^2 \phi_s^2 ,
\label{Eq:tree}
\end{equation}
where $\field \equiv (\phi_h, \phi_s)$.

We impose constraints on the free parameters in the potential to fix the masses of the scalar particles and the vacuum. First, we require that at zero temperature an extremum occurs at $\phi_h = v_h \simeq 246\gev$ and $\phi_s = v_s = 0$,
\begin{equation}
    \atvevZT{\pdv{V}{\phi_h}} = 0 \quad \text{and} \quad \atvevZT{\pdv{V}{\phi_s}} = 0 , \label{Eq:tadpoleEWSB}
\end{equation}
where $\vec{v} \equiv (v_h, v_s)$ and the potential is to be evaluated at $T = 0$.  We could impose this constraint on the tree-level or one-loop potential.
The second condition in \cref{Eq:tadpoleEWSB} is satisfied trivially with a vanishing singlet VEV, while, at tree-level, the first condition for the Higgs requires, 
\begin{equation}
\mu_h^2 = \lambda_h v_h^2.\label{eq:tadpoleEWSB-tree}
\end{equation}
We only fix the zero-temperature vacuum in our model --- at finite temperature the vacuum needn't satisfy \cref{Eq:tadpoleEWSB}. Typically, at high-temperature the vacuum lies at the origin, $\field = (0, 0)$. As the Universe cools, there is a smooth transition in the singlet direction, $(0, 0) \to (0, v_s(T))$. From that state, there is the possibility of a FOPT that breaks EW symmetry, $(0, v_s(T)) \to (v_h(T), 0)$. Subsequently, the ground state smoothly evolves to the desired $(v_h, 0)$ state at zero temperature. This was demonstrated to be the only possible pathway for a FOPT in a high-temperature approximation~\cite{Ghorbani:2020xqv}. For further discussion see \cref{app:phasestructure}.

Second, we require that the vacuum reproduces the experimentally measured Higgs mass, $m_h = 125.25\gev$~\cite{ParticleDataGroup:2020ssz}, and the desired scalar singlet mass, which we treat as an input in this work.  If we neglect the momentum in the self energy (see remarks in \cref{sec:RGEs} detailing the impact of this approximation) this corresponds to,   \begin{equation}
    \atvevZT{\pdv[2]{V}{\phi_h}} = m_h^2 \quad \text{and} \quad \atvevZT{\pdv[2]{V}{\phi_s}} = m_s^2. \label{Eq:tadpoleMass}
\end{equation}
To ensure the potential has the correct form for this we eliminate the $\lambda_h$ quartic by fixing $m_h$ and exchange $\mu_s^2$ for the singlet mass $m_s$.
This leaves only $m_s$, $\lambda_s$ and $\lambda_{hs}$ as free parameters in our model.

The desired squared masses are positive, which ensures the zero-temperature VEVs correspond to minima of the potential. After fixing the Lagrangian parameters, we check that the global minimum of the potential at zero temperature is
\begin{equation}\label{eq:vacuum}
\langle \phi_h \rangle = v_h \simeq 246\gev \quad \text{and} \quad \langle \phi_s \rangle =  v_s = 0,
\end{equation}
as desired. We may impose \cref{Eq:tadpoleMass} on the tree-level or one-loop potential. The tree-level scalar masses are 
\begin{align}\label{eq:mh}
\overline{m}_h^2& = -\mu_h^2 + 3 \lambda_h v_h^2 = 2 \lambda_h v_h^2,\\
\overline{m}_s^2&= -\mu_s^2 + \frac{\lambda_{hs}}{2} v_h^2.\label{eq:ms}
\end{align}
where we use $\overline{m}_{h,s}$ to emphasise that these are scheme-dependent tree-level masses.
When solving this at higher order with the one-loop Higgs pole mass and one-loop tadpoles, we numerically solve for $\lambda_h$ including the one-loop tadpoles iteratively. We detail treatments of the one-loop corrections to this potential in the following subsections. 

\subsection{\msbar + $\Rxi$: The \msbar scheme in the standard $\Rxi$ gauge} \label{sec:msbar}

This is a standard treatment of the effective potential. The one-loop corrections to the potential in the standard $\Rxi$ \cite{Fujikawa:1972fe} gauges\footnote{Although we refer to this as the standard $\Rxi$ gauges, it is important to note that we do not use fixed VEVs in the gauge fixing Lagrangian, as is standard when calculating most observables.  However since the field in the effective potential should be allowed to vary, the use of a fixed VEV would spoil cancellations of the mixing between the scalar (goldstones) and the gauge sectors, so instead the field is allowed to fluctuate away from the minimum.  Although we still refer to this as the standard $\Rxi$ gauge here, it is sometimes referred to as the background field $\Rxi$ gauge in the literature, see e.g. \cite{Martin:2018emo}, to distinguish it from the case with a fixed VEV.} and \msbar scheme are given by \cite{Coleman:1973jx, Patel:2011th}
\begin{equation}\label{eq:VCW}
\begin{split}
V_{\cw}\vp = \frac{1}{64\pi^2} \Bigg[
& \sum_{i \in h} n_i m_i^4(\field,\xi) \left(\log(\frac{m_i^2(\field,\xi)}{Q^2})-\frac{3}{2}\right)\\
+ & \sum_{i \in V} n_i m_i^4\vp \left(\log(\frac{m_i^2\vp}{Q^2})-\frac{5}{6}\right)\\
- & \sum_{i \in \text{FP}} n_i m_i^4(\field,\xi) \left(\log(\frac{m_i^2(\field,\xi)}{Q^2})-\frac{3}{2}\right)\\
- & \sum_{i \in f} n_i m_i^4\vp \left(\log(\frac{m_i^2\vp}{Q^2})-\frac{3}{2}\right)
\Bigg],
\end{split}
\end{equation}
where $Q$ is the renormalisation scale and $\overline{m}_i$ are field-dependent \msbar masses (for their specific forms in our model see \cref{App:MassSpectrum}), and we have suppressed the scale dependence of the masses. The sum over $h$ represents a sum over the scalar fields, which at the electroweak breaking minimum can be separated into physical Higgs and Goldstone bosons. The sum over $V$ includes transverse and longitudinal massive gauge bosons.  The third term is a gauge dependent sum over Faddeev-Popov (FP) ghosts that contribute to the effective potential in $\xi \neq 0$ gauges. There is one FP ghost corresponding to each vector boson with mass
\begin{equation}\label{eq:fp_ghost_mass}
    m_\text{FP}^2 = \xi m_{V}^2.
\end{equation}
Lastly, the sum over $f$ includes relevant fermion fields.

The one-loop thermal corrections in the $R_\xi$ gauge are
\begin{equation}\label{eq:VT}
\begin{split}
V_{T}(\field,T) = \frac{T^4}{2\pi^2} \Bigg[
& \sum_{i \in h} n_i J_{B} \left(\frac{m_i^2(\field,\xi)}{T^2}\right) 
+ \sum_{i \in V} n_i J_{B} \left(\frac{m_i^2\vp}{T^2}\right)\\
- & \sum_{i \in \text{FP}} n_i J_B\left(\frac{ m^2_i(\field,\xi)}{T^2}\right) 
+ \sum_{i \in f} n_i J_F \left(\frac{m_i^2\vp}{T^2}\right) 
\Bigg].
\end{split}
\end{equation}
Note that in the tree-level minimum, the Goldstone boson masses are proportional to the corresponding gauge boson masses and $\xi$ and thus equal the FP ghost masses in \cref{eq:fp_ghost_mass},
\begin{equation}
m_G^2 = \xi m_{V}^2 = m_\text{FP}^2,
\end{equation}
and that the degrees of freedom match, $n_G = n_\text{FP}$. As a result, the $\xi$-dependence from Goldstone and FP ghost contributions cancel in \cref{eq:VCW,eq:VT}, as expected from Nielsen's identity.

We construct the total effective potential by adding the one-loop terms to the tree-level potential,
\begin{align}
V(\field, T) = V_0\vp + V_{\cw}\vp + V_{T}(\field,T).
\end{align}
Additionally, there are further possible modifications to the potential to treat the GC and resum daisy diagrams; see \cref{sec:modifications}.

\subsection{\msbar + \texttt{cov}: The \msbar scheme in the covariant gauge} \label{sec:covariant}

The previous \msbar treatment considered the standard $R_\xi$ gauge. The standard $R_\xi$ gauge used for effective potentials includes scalar fields that we vary in the gauge fixing Lagrangian.  However this means that a different gauge is used for calculating the potential at each field value, which may be a cause for some concern~\cite{Arnold:1992fb,Laine:1994bf,Andreassen:2013hpa}.\footnote{It has also been pointed out recently that this condition is not Renormalisation Group invariant \cite{Martin:2018emo}.}    To remedy this, the effective potential of the SSM using the covariant gauge (sometimes called the Fermi gauges in the literature) \cite{Laine:1994bf, Andreassen:2013hpa, Andreassen:2014gha,Andreassen:2014eha} in the \msbar scheme has been used in recent work~\cite{Papaefstathiou:2020iag}. The gauge fixing Lagrangian in this gauge is,
\begin{equation}
{\cal L}_{\text{g.f.}} = - \frac{1}{2\xi_W} (\partial^\mu W^a_\mu)^2 -  \frac{1}{2\xi_B} (\partial^\mu B_\mu)^2.
\end{equation}
This avoids the issue of a fixed VEV at the expense of mixing between the Goldstones and the gauge sector. We here use the same potential as derived in \refcite{Papaefstathiou:2020iag}, but with a \ztwo discrete symmetry enforced,
\begin{align}
\begin{split}
V_{\cw}\vp ={}& \sum_{i \in h} n_i \frac{m_i^4(\field,\xi_W,\xi _B)}{64 \pi^2} \left[ \log(\frac{m_i^2(\field,\xi_W,\xi_B)}{Q^2}) - \frac{3}{2} \right] \\ 
              & + \sum_{i \in V} n_i \frac{m_i^4\vp}{64 \pi^2} \left[\log(\frac{m_i^2\vp}{Q^2}) - \frac{5}{6} \right]\\
              & - \sum_{i \in f} n_i \frac{m_i^4\vp}{64 \pi^2} \left[\log\left(\frac{m_i^2\vp}{Q^2} \right) - \frac{3}{2} \right]\, ,\\
\end{split}\\[2ex]
& V_T(\field,T) = \frac{T^4}{2\pi^2} \left[\sum_{i \in h,V} n_i J_B \left(\frac{m_i^2(\field,\xi_W,\xi_B)}{T^2} \right) + \sum_{i \in f} n_i J_F \left(\frac{m_i^2\vp}{T^2} \right) \right] \,.
\end{align}
In the above, the field-dependent fermion, Higgs and gauge boson masses take their usual form.  There are two gauge-fixing parameters, $\xi_W$ and $\xi_B$, that appear only implicitly through the masses of the mixed Goldstone-ghost scalars,\footnote{They separate into Goldstone and ghost modes when $\xi=0$.}
\begin{align}
    m_{1,\pm }^2 &= \frac{1}{2}\left(\chi \pm \Re\sqrt{\chi^2 - \Upsilon_W } \right) , \label{Eq:cov-m1pm}\\
    m_{2,\pm }^2 &= \frac{1}{2}\left(\chi \pm \Re\sqrt{\chi^2 - \Upsilon_Z } \right) , \label{Eq:cov-m2pm}
\end{align}
 where 
\begin{equation}
\chi =\recip{\phi_h} \frac{\partial V_0}{\partial \phi_h} =  -\mu^2_h + \lambda_h \phi_h^2 + \frac{1}{2} \lambda_{hs} \phi_s^2 ,
\end{equation}
and the gauge dependence appears only in the terms
\begin{equation}
 \Upsilon_W = \chi g_2^2 \xi_W \phi_h^2 
\quad\text{and}\quad
  \Upsilon_Z = 
  \chi \left(g_2^2  \xi_W + g_1^2 \xi_B \right) \phi_h^2.
\end{equation}
Because the $\Upsilon$ terms  are proportional to the derivative of the tree-level potential, the $\xi$-dependence vanishes in the tree-level minima, as expected from Nielsen's identity. If $\Upsilon > 0$, by increasing $\xi$ sufficiently it may be possible to make the square roots in \cref{Eq:cov-m1pm,Eq:cov-m2pm} imaginary. As we take the real part, in this regime the potential would become independent of $\xi$.

The multiplicities for the Goldstone-like modes are $(n_1,n_2)=(2,1)$ respectively to account for six terms in the effective potential, that is double the number of Goldstone modes. This is because the Goldstone and ghost contributions mix in the covariant gauge, so the total number of terms in the gauge-dependent effective potential is the same as in the $R_\xi$ gauge. The positive and negative roots in \cref{Eq:cov-m1pm,Eq:cov-m2pm} may both produce zero eigenvalues, potentially resulting in GC. The zeros occur in the tree-level vacuum, where $\Upsilon = \chi = 0$ for all $\xi$. For the negative roots, however,  when $\xi = 0$ the zero eigenvalues are field-independent and so cannot contribute to the GC.

\subsection{\hight: The high-temperature approximation}\label{sec:hight}

The high-temperature (HT) approximation of the total effective potential is 
\begin{equation}
V_{\hight}(\field, T) = V_0\vp + \frac{1}{2}c_h T^2\phi_h^2 + \frac{1}{2}c_s T^2\phi_s^2,
\label{Eq:HT}
\end{equation}
where the Debye coefficients $c_h$ and $c_s$ are given in \cref{eq:debye_coefficient_h,eq:debye_coefficient_s}. At zero temperature, this reduces to the tree-level potential, $V_{\hight}(\field, T=0) = V_{0}\vp$.

\Cref{Eq:HT} is obtained by considering only the leading terms of a high-temperature approximation of the finite-temperature correction in \cref{eq:VT}. Namely, for $y \equiv m / T  \ll 1$, the thermal functions are approximated by the leading terms in the expansions~\cite{PhysRevD.9.3320}
\begin{align}\label{eq:JB}
J_B(y^2) &\simeq -\frac{\pi^4}{45} + \frac{\pi^2}{12}y^2 , \\ 
J_F(y^2) &\simeq \frac{7\pi^4}{360} -\frac{\pi^2}{24}y^2 .
\end{align}
The zero-temperature Coleman-Weinberg corrections are neglected. Importantly, the gauge parameter $\xi$ and renormalisation scale $Q$ do not appear in this high-temperature expansion, leaving an explicitly gauge- and scale-independent high-temperature approximation to the potential. However, the fact that $Q$ does not explicitly appear in the potential may in fact worsen the scale dependence, as explicit and implicit scale dependence are expected to partially cancel. The tadpole constraints are applied at tree-level, which avoids a potential source of gauge dependence.

In this scheme, the potential contains only quadratic and quartic terms, and the critical temperature and transition strength can be solved analytically,
\begin{align}
T_c^2 &= \frac{\lambda_h c_s \mu_s^2 -\lambda_s c_h \mu_h^2 -\sqrt{\lambda_h \lambda_s} \left|c_s\mu_h^2-c_h\mu_s^2 \right|}{\lambda_s c_h^2 -\lambda_h c_s^2} , \label{Eq:TC_HT}\\
\gamma &= \frac{1}{T_c} \sqrt{\frac{\mu_h^2+\frac{1}{2}c_h T_c^2}{\lambda_h}+\frac{\mu_s^2+\frac{1}{2}c_s T_c^2}{\lambda_s}} .
\end{align}

\subsection{PRM: The $\hbar$ expansion}\label{sec:hbar}

This is a technique for computing the critical temperature and transition strength in a gauge-independent way~\cite{Patel:2011th}. The method exploits the Nielsen identity~\cite{Nielsen:1975fs}, which, when truncating the effective potential to one-loop, means that the one-loop effective potential is gauge independent at tree-level turning points. To maintain gauge independence, we ultimately use the tree-level, one-loop and high-temperature expansion of the effective potential in \cref{sec:hight}. As usual, the renormalisation scale enters explicitly in the Coleman-Weinberg potential and implicitly in the Lagrangian parameters.

We first find the minima of the tree-level potential, $\phi_\tree$. It is possible that the structure of minima in the tree-level potential does not match that of the one-loop potential. Therefore if there is only one minimum at tree-level, we further consider a saddle point. If there are no saddles, we consider a maximum. To ensure gauge independence of the critical temperature, the two points must be turning points.\footnote{In our modified version of \pt, this choice is model-dependent and not automated.} To ensure that no gauge dependence creeps into the tree-level potential, we fix Lagrangian parameters only using tree-level tadpoles and tree-level constraints on masses. To ensure the validity of Nielsen's identity, we cannot modify the field-dependent masses in the Coleman-Weinberg potential, even by terms that are formally order $\mathcal{O}(\hbar^2)$ or resum daisy terms by either the Parwani or Arnold-Espinosa methods. 

We then turn to the one-loop potential with no daisy resummation. We find the temperature at which the two points considered are degenerate in the one-loop potential. This is the critical temperature, $T_c$. Lastly, at that critical temperature, we find the minima of the gauge-invariant high-temperature expansion of the potential, $\phi_\hight$. We use this in our estimate of the transition strength. For the strength of a transition involving one field that breaks electroweak symmetry,
\begin{equation}
    \gamma_{\text{EW}} = \frac{|\phi_\hight|}{T_c}.
\end{equation}
In multi-field cases beyond our model, we would use the distance from $\phi_\hight$ to the $h$-axis.

Importantly, because we use tree-level constraints and tadpoles to maintain gauge independence, we predict $v \simeq 246.22\gev$ and $m_h \simeq 125.25\gev$ only at tree-level. The most common approach for calculating the masses and observables in BSM theories to fix $v=246.22\gev$ and use the loop corrected effective potential to fix parameters in the potential such that the minimum lies at $v=246.22\gev$.  In this way, one-loop tadpole diagrams vanish and do not need to be included in calculating observables.  This is the approach implemented in most public tools, and therefore one needs to proceed with care when using the PRM method. The one-loop predictions for them could be quite different and disagree with measurements, and, depending on how they are computed, be gauge dependent. As we shall discuss, there is a risk that we are merely trading sensitivity to the choice of gauge for imprecision in agreement with experimental measurements; neither are desirable. Alternatives approaches, such as the Fleischer–Jegerlehner treatment of tadpoles \cite{Fleischer:1980ub} may provide an approach to ensure that $\lambda$ is extracted from the Higgs mass at the order $\hbar$ without re-introducing gauge dependence into to effective potential (see also \refcite{Braathen:2021fyq} for a recent discussion about the differences in these approaches). Later we study the impact of extracting $\lambda_h$ at one loop by reintroducing gauge dependence via one-loop EWSB conditions and label this PRM + \texttt{1LHiggs\_1LTad}. 

Lastly, we note that whilst the daisy treatments are an all-order resummation, and thus do not naturally fit into an $\hbar$-expansion scheme, \refcite{Patel:2011th} proposed a gauge-invariant modification to \cref{eq:daisy} that could be incorporated into the PRM scheme. This method requires the effective three-dimensional theory and, as far as we know, is only rigorously proven to capture the $\mathcal{O}(\hbar^2 T^3)$ terms correctly. The conflict between IR divergences, which appear to require an all-order resummation, and the $\hbar$ expansion is further discussed in \refcite{Ekstedt:2018ftj}.

\subsection{OS-like: The on-shell-like scheme} \label{sec:os}

The on-shell-like (OS-like) scheme is often used in studies of first-order phase transitions (see e.g.\ \refcite{Anderson:1991zb, Megevand:2007sv,Ashoorioon:2009nf, Megevand:2014dua, Chiang:2018gsn, Beniwal:2018hyi, Alanne:2019bsm, Abdussalam:2020ssl,Xie:2020wzn,Azatov:2022tii,Ellis:2022lft}). This scheme has the practical advantage that the tree-level VEVs and mass eigenstates are preserved at the one-loop level. That is, the masses extracted from the one-loop potential are the same as those extracted from the tree-level potential, and both potentials have a minimum at the same location in field space. Therefore in this work we compare results from the \msbar scheme and this OS-like scheme to indicate the differences in results that may be expected between the two most commonly used schemes.

Uncertainties from zero-temperature higher-order corrections are often estimated by comparing results in different renormalisation schemes.  While both schemes can provide controlled approximations allowing predictions that can be tested in experiments, the predictions will not in general be the same, with the difference coming from theoretical uncertainties arising from missing higher-order corrections in the respective schemes. Therefore, comparing the calculation in different schemes provides sensitivity to missing higher-order corrections and can be used as an estimate of this theoretical uncertainty.  

\Refcite{Anderson:1991zb} demonstrated that for field-dependent masses of the form
\begin{equation}\label{eq:os_assumed}
    m^2(\phi) = m^2 + g \phi^2,
\end{equation}
we may pick counter-terms that satisfy the OS conditions and bring the zero-temperature one-loop corrections in the Landau gauge into the simple form
\begin{equation}
    V_1 \vp = \recip{64 \pi^2} \sum_i \tilde{n}_i \left[\mifourpv \left(\log(\frac{\mitwopv}{\mitwovv}) - \frac{3}{2} \right) + 2 \mitwovv \mitwopv - \half \mifourvv \right] ,
    \label{Eq:os}
\end{equation}
where $i$ ranges over the entire field-dependent mass spectrum (including scalars, gauge bosons and fermions), and we have defined $\tilde{n}_i \equiv (-1)^{2 s_i} n_i$ to absorb the sign for particle species of spin $s_i$. The last term is field-independent and so is often ignored. The finite-temperature corrections are identical to those in \cref{sec:msbar}.

We make a common simplification: we ignore the limitation \cref{eq:os_assumed} and use \cref{Eq:os} in our SSM model with the field-dependent masses in \cref{App:MassSpectrum}. The OS conditions remain satisfied; however, \cref{Eq:os,eq:VCW} differ by terms that do not appear in the tree-level Lagrangian and cannot be interpreted as counter-terms. However, we predominantly consider transitions between ground states in which mixing between the Higgs and singlet vanishes, that is, between $(v_h(T), 0)$ and $(0, v_s(T))$ ground states. This potential is correct along these axes as the field-dependent masses are in the form \cref{eq:os_assumed}. See e.g.\ \refcite{Basler:2016obg} for further discussion about implementing an OS-like scheme in scalar extensions of the SM.

\section{Modifications to the effective potential}\label{sec:modifications}
The treatments in \cref{sec:treatments} may be modified to alleviate the Goldstone catastrophe and resum daisy diagrams.

\subsection{Goldstone catastrophe}\label{sec:goldstone}
The Goldstone catastrophe (GC) occurs when a massless goldstone boson entering loop functions leads to an infrared divergence.  The GC should appear directly in the \msbar $\xi=0$ potential at three-loop order and higher and in the electroweak symmetry breaking conditions at two-loop order and higher \cite{Martin:2014bca}.   Furthermore it is very common in BSM physics investigations for the effective potential to be calculated using the effective potential approximation or OS-like schemes, and as discussed in \cref{sec:treatments} this causes the second derivative of the one-loop potential to diverge. For example, in the \msbar case this happens when the tree-level masses that appear in loop functions of the Coleman-Weinberg potential are fixed using a tree-level EWSB condition\footnote{Note the  EWSB constraints in \cref{Eq:tadpoleEWSB} are still applied at one-loop to fix the parameters appearing explicitly in the tree-level potential, here we just describe the standard treatment of tree-level \msbar masses that enter the loop functions. }, such that for $\xi=0$ we have massless Goldstone bosons. This particular issue could be avoided simply by relating the parameters to the Higgs pole masses with self energies evaluated with the momentum set equal to the Higgs  mass \cite{Cline:1996mga} (see also \cref{sec:RGEs} for comments on the impact of including the momentum), but here we consider instead two more direct alternatives that avoid the massless Goldstone and could be used to avoid the GC appearing directly in the potential and electroweak symmetry breaking conditions:   
\begin{enumerate}
    \item  \texttt{GC\_SelfEnergy\_Sol}: Adding the tree-level Goldstone boson mass with a one-loop self-energy correction to all Goldstone masses~\cite{Martin:2014bca,Elias-Miro:2014pca} \begin{equation}
       M_{G^{0,\pm}}^2(\field) = m_{G^{0,\pm}}^2(\field) + \Sigma_{G}(\field, p=0),
    \end{equation}
    where the self energy contribution for the Goldstone can be given by,
    \begin{equation}
        \Sigma_G(\field, p=0) = \recip{\phi_h} \pdv{V_{\cw}^\prime}{\phi_h} ,
    \end{equation}
     where $V_{\cw}^\prime$ does not include Goldstone boson contributions. This correction depends on the renormalisation scale and gauge.\footnote{We also apply this to the \os\ scheme and this introduces a renomalisation scale which we simply fix to ensure $ \atvevZT{d^2 V_{\rm CW} / d \phi_h^2} = (1/v_h) \atvevZT{d V_{\rm CW}/d \phi_h}$.}
    \item \texttt{GC\_Tadpole\_Sol}: Applying EWSB constraints \cref{Eq:tadpoleEWSB} at one-loop when calculating the masses that enter the one-loop corrections to the potential.    
\end{enumerate}
In both cases this effectively adds a correction to the potential which is formally beyond the one-loop precision we aim for, while shifting the Goldstone masses away from the tree-level value.  The former is formally re-summing a sub-class of corrections to all-orders, providing a principled solution (please see the extensive discussion in Refs.\ \cite{Martin:2014bca,Elias-Miro:2014pca} for details), while the latter is simply a convenient way to avoid the massless tree-level Goldstone that is easy to apply and done in the literature \cite{Chiang:2018gsn}, while only introducing spurious two-loop pieces to the effective potential.

\subsection{Daisy diagrams}\label{sec:daisy}

There are two popular methods for resumming daisy diagrams:
\begin{enumerate}
\item Parwani~\cite{Parwani:1991gq}: This resums the finite-temperature daisy diagrams. To use the Parwani method, the mass eigenvalues appearing throughout the one-loop potential are replaced by thermal mass eigenvalues (see \cref{App:MassSpectrum}) and one forgoes the inclusion of any explicit daisy term. With this method it is important to only substitute in the thermal masses with lowest order from the high temperature expansion to avoid a introducing spurious linear terms \cite{Dine:1992vs,Dine:1992wr}. 

\item Arnold-Espinosa~\cite{Arnold:1992rz}: This resums only the zeroth Matsubara modes in the daisy diagrams. We add the daisy term
\begin{equation}\label{eq:daisy}
V_\text{daisy}(\field, T) = -\frac{T}{12\pi} \sum_i \left[ 
    \left(\mt_i^2(\field, T)\right)^{3/2} -  \left(m_i^2(\field)\right)^{3/2}
\right]
\end{equation}
where $\mt_i^2(\field, T)$ are thermal masses including Debye corrections (see \cref{App:MassSpectrum}), and we have suppressed their gauge and scale dependence.
\end{enumerate}
Both methods introduce terms to the effective potential that are cubic in the coupling $g$ for gauge bosons and to the $3/2$ power in scalar couplings. The thermal mass is, by contrast, quadratic in $g$ and linear in scalar couplings. Neither treatment changes the zero-temperature potential. 
See \refcite{Schicho:2021gca, Niemi:2021qvp} for recent progresses about the two-loop order computations.

\subsection{Renormalisation group equations}
\label{sec:RGEs}
We take into account both the implicit and explicit scale dependence of the effective potential by using renormalisation group equations (RGEs) to run the \msbar parameters 
and compute the full \msbar renormalised effective potential.
To do this we implement a \fs~\texttt{2.6.1}~\cite{Athron:2014yba,Athron:2017fvs}\footnote{\fs uses \sarah~\texttt{4.14.3}~\cite{Staub:2009bi,Staub:2010jh,Staub:2012pb,Staub:2013tta} and parts of the \softsusy~\cite{Allanach:2001kg,Allanach:2013kza} code.} 
version of the effective potential in \pt. The \fs spectrum generator has two-loop RGEs, one-loop threshold corrections for the extraction of the gauge and Yukawa couplings and it also extracts the Higgs potential parameters using one-loop electroweak symmetry breaking conditions and one-loop self energies. The latter include the $p^2$ contributions, which are neglected when masses are calculated using the one-loop effective potential approximation. We have checked that it will lead to 1.3\% difference on extracted $\lambda_h$ and 1.0\% difference on output $T_c$ for benchmark point defined later in \cref{eq:benchmark}.  Due to the small size of this correction we do not consider it in our uncertainties.  Since \fs uses the $\xi=1$ gauge all the calculations using RGEs are fixed to that gauge and there are no Goldstone IR divergences for this choice. The inputs of the \fs model file are $m_h$, $m_s$, $\lambda_s$ and $\lambda_{hs}$, while $\lambda_h$, $\mu_h$ and $\mu_s$ are determined by EWSB conditions.

\section{Results}\label{sec:results}

We now investigate different sources of uncertainty in the analysis of a PT, including gauge dependence, renormalisation scale dependence, treatment of the Goldstone catastrophe (GC), resummation of daisy diagrams, and renormalisation schemes.  We do so by applying the treatments discussed in \cref{sec:treatments,sec:modifications} to the SSM.  After considering them separately, we compare their typical sizes across the available parameter space.  The numerical uncertainties associated with our computational methods are discussed in \cref{app:numerical}.

\subsection{Gauge dependence}\label{sec:results:gauge}

\begin{figure}[t!]
\centering
\includegraphics[width=0.99\textwidth]{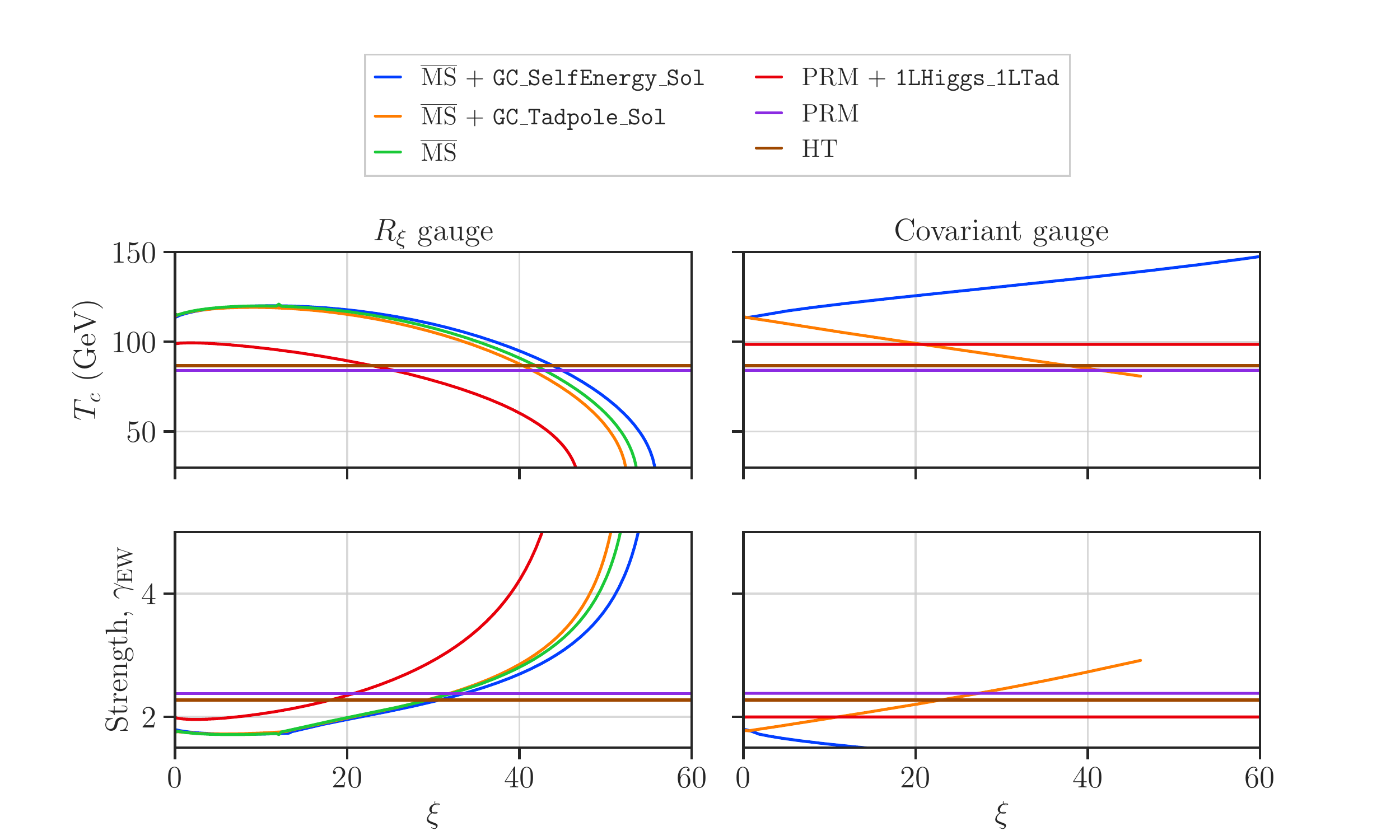}
\caption{The gauge dependence from different methods for the critical temperature (top) and transition strength (bottom) in $\Rxi$ (left) and covariant gauge (right) for the benchmark point in \cref{eq:benchmark}.}
\label{fig:1d_xi}
\end{figure}

\subsubsection{Benchmark point}

First, in \cref{fig:1d_xi} we show the $\xi$-dependence of the critical temperature $T_c$ and transition strength \gammaew in the $\Rxi$ and covariant gauges for different methods, using a benchmark point of
\begin{equation}\label{eq:benchmark}
\lambda_s=0.1, \quad \lambda_{hs}=0.3 \quad\text{and}\quad m_s=65\gev .
\end{equation}
This benchmark lies near the center of the region of parameter space in which a FOPT could occur.
We vary the gauge parameter between $\xi = 0$ and $60$. Note that the upper limit may lie beyond constraints from perturbativity~\cite{Laine:1994bf,Garny:2012cg}, as increasing the gauge parameter aggravates IR divergences that spoil perturbation theory.  The legend follows the notation introduced in \cref{sec:treatments,sec:modifications}.  In the three \msbar methods, we used the Arnold-Espinosa technique to resum daisy terms, and set $Q = m_t$.  The gauge dependence of the OS-like scheme is not presented as \cref{Eq:os} is only valid at $\xi=0$.

As a reference, in \cref{fig:1d_xi} we show the PRM and HT methods which, of course, are independent of the gauge parameter $\xi$.  The left panels of \cref{fig:1d_xi} show results for the $\Rxi$ gauge.  
In the \msbar scheme, with $\xi$ increasing, the critical temperature $T_c$ increases slightly for small values of $\xi$, and then decreases significantly for large values of $\xi$.
With $\xi$ increasing even further, there is no FOPT for this benchmark point anymore in the gauge-dependent schemes. The choice of treatment of the GC has limited impact.  

\begin{figure}[t!]
\centering
\includegraphics[width=0.99\textwidth]{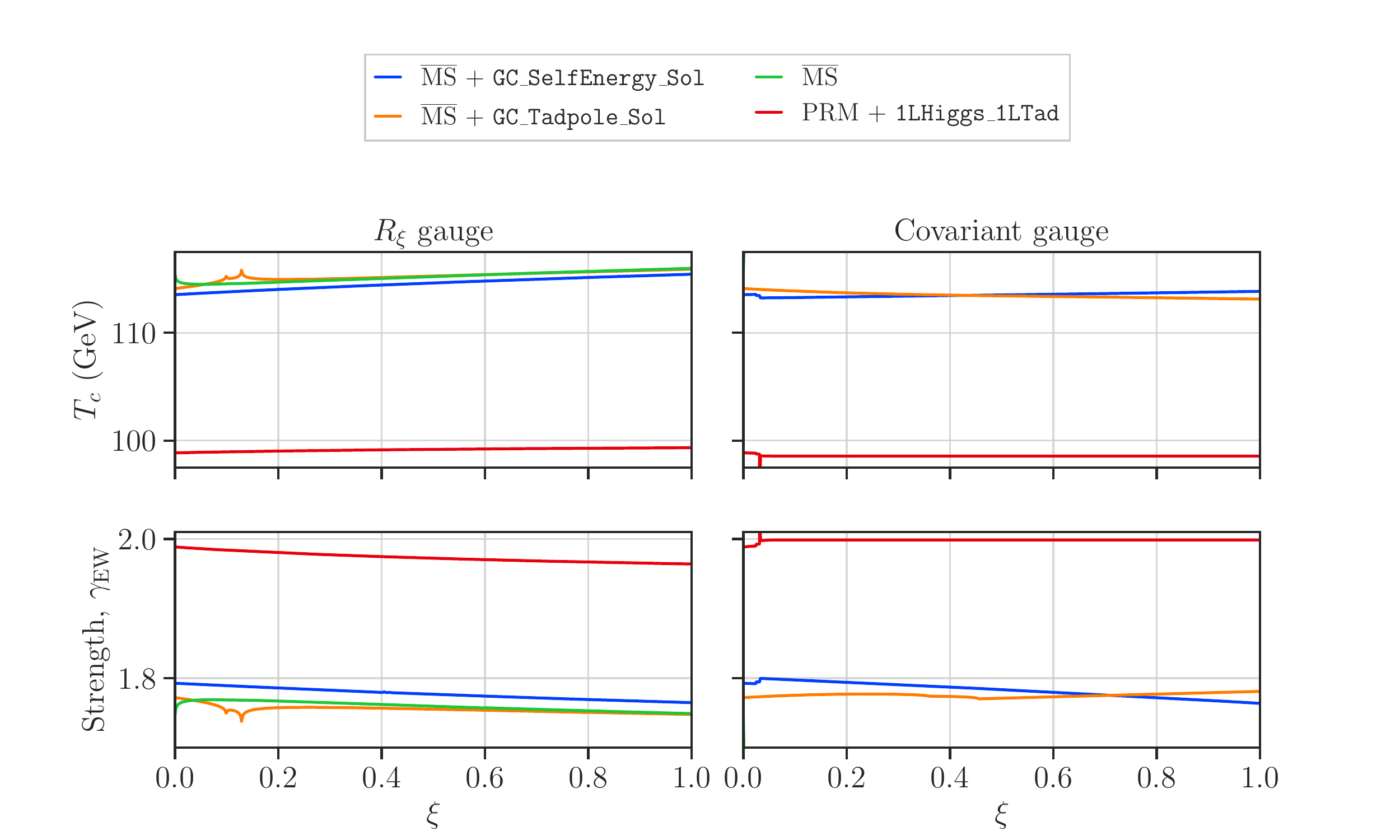}
\caption{As per \cref{fig:1d_xi}, but for $\xi \le 1$. Lines for PRM and HT are not shown here, as they are independent of $\xi$ and exactly the same as in \cref{fig:1d_xi}.}
\label{fig:1d_xi_zoom}
\end{figure}

In  \cref{fig:1d_xi} we investigated $\xi \le 60$. However, there may be constraints on perturbativity that impose $\xi \lesssim 1$~\cite{Laine:1994bf,Garny:2012cg}. We see in \cref{fig:1d_xi_zoom} that varying the gauge between $\xi = 0$ and $1$ results in only slight changes in the properties of the FOPT. We furthermore see that some methods do not depend smoothly on $\xi$. For example, the $T_c$ of \msbar blows up at $\xi=0$.  This originates from the GC: due to the catastrophe, the second-derivatives of the one-loop potential are singular, and so the Lagrangian parameters required in the tree-level potential to maintain finite scalar masses diverge.  Although using the modified $\mu$ from one-loop tadpole cures the GC at $\xi=0$, the Goldstone may be massless at tree-level in the one-loop vacuum for other values of $\xi$.  For this benchmark point, when $\xi=0$, $M_{G^{0}}^2 \simeq M_{G^{\pm}}^2 \simeq -840 \gev^2$;
when $\xi=0.0995$, $M_{G^{0}}^2\simeq 0 \gev^2$ and $M_{G^{\pm}}^2\simeq -200 \gev^2$;
when $\xi=0.1286$, $M_{G^{0}}^2 \simeq 250 \gev^2$ and $M_{G^{\pm}}^2 \simeq 0 \gev^2$. 
This is why we see two singularities (zeros) in the $T_c$ (\gammaew) of  the orange curve for \msbar + \texttt{ GC\_Tadpole\_Sol} at $\xi \simeq 0.1$ and $\xi \simeq 0.13$.  The catastrophe could occur at other $\xi$ for other choices of benchmark point.  In contrast, adding the one-loop self-energy to the Goldstone masses avoids a catastrophe at any $\xi$ as seen in the blue curve for \msbar + \texttt{GC\_SelfEnergy\_Sol}. 

In PRM + \texttt{1LHiggs\_1LTad}, we use one-loop tadpole conditions to set the Lagrangian parameter $\mu_h$ and extract $\lambda_h$ and $\mu_s$ at the one-loop level from the measured Higgs pole mass and the input singlet pole mass, respectively.  The one-loop tadpoles are required if one wants to extract $\lambda_h$ from the Higgs mass at the one-loop level using the standard approach we applied in the \msbar calculation, where the tadpoles vanish in the mass corrections and do not need to be explicitly included\footnote{Though they must be accounted for if you use the tree-level EWSB condition to rewrite the tree-level Higgs mass as $\overline{m}_h=\sqrt{2\lambda}v$. }. A one-loop extraction of $\lambda_h$ is required for a consistent order $\hbar$ calculation of the effective potential consistent with the measured Higgs pole mass.  As a result, the $T_c$ and \gammaew calculations in this method are closer in spirit to those in the \msbar method and numerically are closer when far away from the $\xi$ values where the \msbar and pure PRM results match. This, however, introduces gauge dependence, which contradicts the purpose of the PRM method. This gauge dependence originates from the gauge dependence of the \msbar scalar masses (which are extracted at the one-loop level) used in the one-loop tadpole conditions. Just as in the \msbar scheme, however, $T_c$ significantly changes with increasing $\xi$ for this case at this benchmark point.

We show results in the covariant gauge in the right panels of \cref{fig:1d_xi,fig:1d_xi_zoom}, where $\xibw \equiv \xi_B = \xi_W$.  As anticipated, results from the covariant and $\Rxi$ gauges are identical at $\xi = 0$.  The results for the \msbar calculation without any modification to solve the GC cannot be seen as the GC exists for all $\xibw$ in the covariant gauge. 

Different to the $\Rxi$ gauge, in the covariant gauge the $T_c$ for \msbar + \texttt{GC\_SelfEnergy\_Sol} and + \texttt{GC\_Tadpole\_Sol} respectively increase and decrease almost linearly with $\xi$ increasing. Meanwhile, there are some discontinuities in the lines of the covariant gauge when $\xibw$ is small, while the lines then become continuous when $\xibw$ is large.  This is because we take the real part of the square root in \cref{Eq:cov-m1pm,Eq:cov-m2pm} and for sufficiently large $\xibw$, the $\xibw$-dependent parts are imaginary. This behavior can also be seen in the results of PRM + \texttt{1LHiggs\_1LTad} in the covariant gauge, i.e.~there are two steps in the results at small $\xibw$ and the results are independent of $\xibw$ for large $\xibw$.  This gauge independence, however, does not necessarily mean that this method is a better choice, because it is simply due to the fact that we ignored the imaginary gauge-dependent parts of the squared masses. There are no results when $\xi>46.5$ for \msbar + \texttt{GC\_Tadpole\_Sol} because our
iterative solver for $\lambda_h$ using one-loop EWSB did not converge. 
The gauge dependence of the \msbar + \texttt{GC\_SelfEnergy\_Sol} in the covariant gauge is also fairly strong, though it shows a different tendency.

The gauge dependence in the calculations of $T_c$ and \gammaew originates from two sources: indirectly from the Lagrangian parameters chosen by solving one-loop tadpole conditions and explicitly in the one-loop potential. The indirect dependence causes $T_c$ to decrease with increasing $\xi$, as shown by the PRM + \texttt{1LHiggs\_1LTad} results in $\Rxi$ gauge, which only suffers from this source. The explicit dependence, on the other hand, increases $T_c$, as shown by the \msbar + \texttt{GC\_SelfEnergy\_Sol} results in the covariant gauge with large $\xibw$. In this case, the $\xibw$ dependence vanishes in the zero-temperature vacuum, as the square roots in \cref{Eq:cov-m1pm,Eq:cov-m2pm} are imaginary, such that the tadpoles are gauge independent. However, the potential remains gauge dependent at other field values.  When both contributions appear at the same time, such as in the \msbar scheme with the $\Rxi$ gauge and the \msbar + \texttt{GC\_Tadpole\_Sol} in the covariant gauge,  the indirect dependence through tadpoles overpowers the explicit dependence. 

\subsubsection{Two-dimensional scans}

\begin{figure}[t!]
\centering
\includegraphics[width=0.99\textwidth]{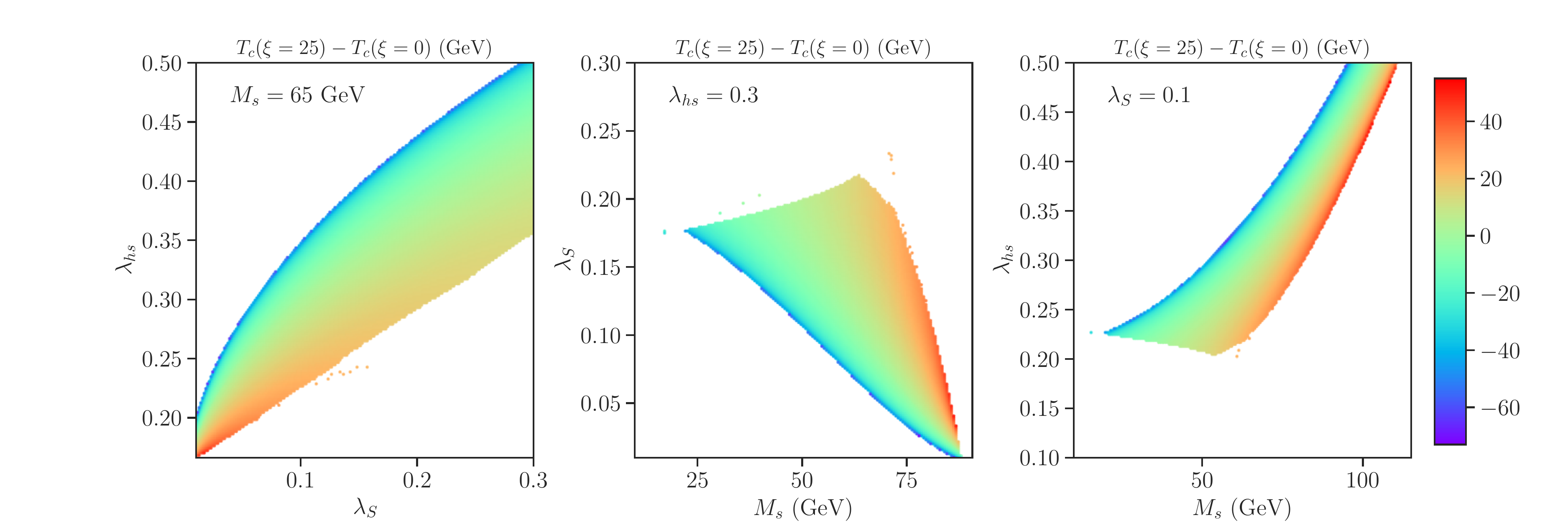}
\includegraphics[width=0.99\textwidth]{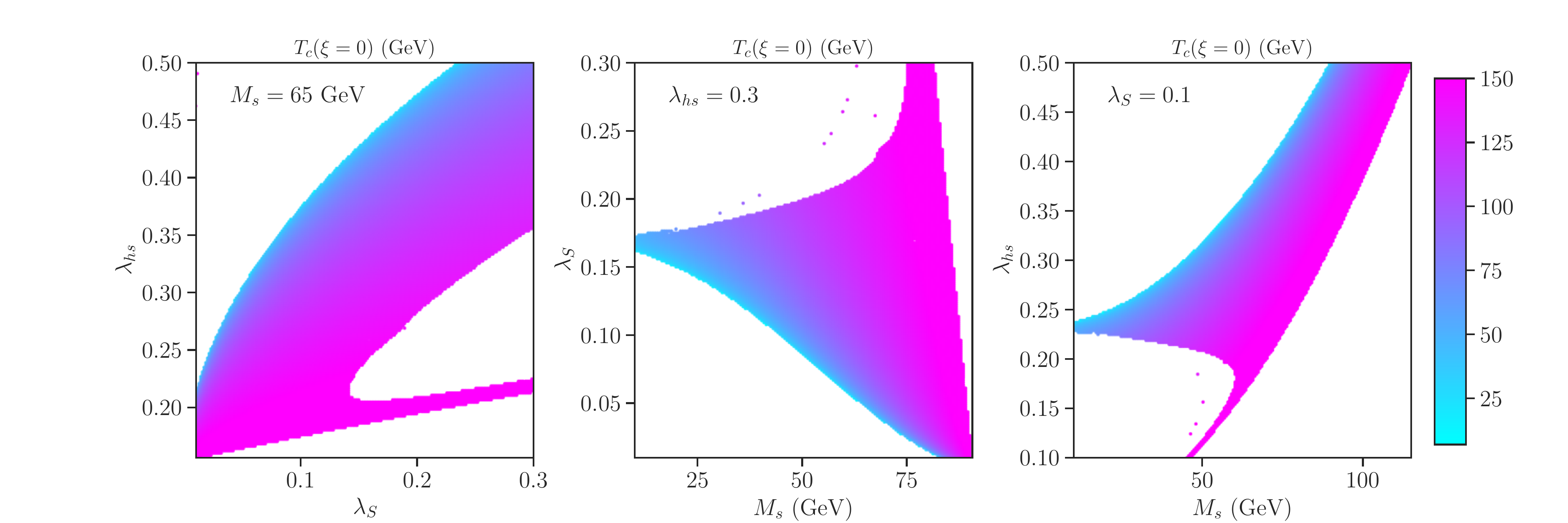}
\caption{Gauge-dependence of the \msbar + \texttt{GC\_SelfEnergy\_Sol} scheme for the critical temperature $T_c$ in two-dimensional scans. We use the difference in \cref{eq:delta_T_c} between $\xi=0$ and $25$
to show the gauge-dependence in the upper panels. We show the corresponding $T_c(\xi=0)$ displayed in the lower panels. In the blank regions there are no points with a FOPT at $\xi=0$ and $\xi=25$.
}
\label{fig:xi_msbar_2d}
\end{figure}

The above discussion is based on a single benchmark point.  The magnitude of the gauge dependence, however, may depend strongly on the Lagrangian parameters.  In \cref{fig:xi_msbar_2d}, we show the uncertainty from the choice of gauge parameter in the \msbar + \texttt{GC\_SelfEnergy\_Sol} prediction for the critical temperature $T_c$ with three two-dimensional scans of the Lagrangian parameters.  We quantify this uncertainty through
\begin{equation}\label{eq:delta_T_c}
    \Delta_\xi T_c \equiv \left|T_c(\xi=0)-T_c(\xi=25)\right|, 
\end{equation}
where $T_c(\xi=0,25)$ is the critical temperature with $\xi=0,25$ in the \msbar + \texttt{GC\_SelfEnergy\_\allowbreak{}Sol} scheme.  As discussed in the introduction, this quantification is somewhat arbitrary. We attempted to pick an upper limit, $\xi=25$, that strikes a balance between understating the uncertainty and constraints from perturbativity. If we were to increase the upper limit further, eventually there would be no points with FOPT simultaneously at $\xi = 0$ and the upper limit.
We scan over two of the three free parameters, while fixing $m_s$, $\lambda_{hs}$ and $\lambda_s$ to their values in \cref{eq:benchmark} in the left, middle and right panels, respectively.  To help understand the gauge dependence further, we show $T_c(\xi=0)$ in the lower panels. 

We found no valid points in the blank regions of \cref{fig:xi_msbar_2d}.  The causes for this and the phase structures of points in those regions are discussed in \cref{app:phasestructure}.  In the colored regions, the critical temperature changes with the varying input parameters.  The trends can be derived from the expression of $T_c$ in the high-temperature approximation, \cref{Eq:TC_HT}. Alternatively, consider the two minima at temperature $T$, $\field_s(T) \equiv (v_h^{\rm high}=0, v_s^{\rm high}(T))$ and $\field_h(T) \equiv (v_h^{\rm low}(T), v_s^{\rm low}=0)$. We may Taylor expand the difference in potential around $T=0$~\cite{Huang:2014ifa,Harman:2015gif}
\begin{equation}
    \Delta V(T=T_c) \approx \Delta V(T=0) + T_c \left.\frac{\partial \Delta V}{\partial T}\right|_{T=0} ,
\end{equation} 
where $\Delta V(T) \equiv V(\field_s(T), T) -  V(\field_h(T),T)$. Since this difference vanishes at the critical temperature and $\Delta V(T=0)<0$, we find
\begin{equation}
    T_c \approx - \frac{\Delta V(T=0)}{\left.\frac{\partial \Delta V}{\partial T}\right|_{T=0}} .
\end{equation}
Accordingly, a smaller gap at zero temperature may imply a lower $T_c$.  In the SSM at tree level we have 
\begin{equation}
    \Delta V(T=0) = -\frac{1}{4}\frac{\mu_s^4}{\lambda_s} + \frac{1}{4}\frac{\mu_h^4}{\lambda_h} 
    \approx -\frac{1}{4}\frac{(\frac{1}{2}\lambda_{hs}v_h^2-m_s^2)^2}{\lambda_s} + \frac{1}{8}m_h^2 v_h^2, \label{eq:dvtree}
\end{equation}
where $\frac{1}{2}\lambda_{hs}v_h^2>m_s^2$. We can approximately infer that with increasing $\lambda_s$ and $m_s$, and decreasing $\lambda_{hs}$, the gap $\Delta V(T=0)$ and thus $T_c$ must increase.  This agrees with the lower panels of \cref{fig:xi_msbar_2d}.

In the upper panels of \cref{fig:xi_msbar_2d}, we show the uncertainty in $T_c$ caused by changing the gauge parameter $\xi$ in the regions in which a FOPT could occur for both $\xi=0$ and $\xi=25$.  The change $\Delta_\xi T_c$ varies from $-73\gev$ to $+55\gev$, and increases smoothly with $\lambda_{hs}$ decreasing, $m_s$ increasing and $\lambda_{s}$ increasing. 
In Appendix \ref{appendix:BMs} we also provide \cref{tab:gauge} which shows precise input parameters and the output critical temperatures of the benchmark point in \cref{eq:benchmark} and points around it for $\xi=0$ and $\xi=25$, to give quantitative results that also show the trends discussed here.

\subsection{Renormalisation scale dependence}\label{sec:scale}

The renormalisation scale dependence of the PT properties originates from the scale-dependence of the one-loop corrections to the effective potential.  If RGEs are used, this may be partly balanced by renormalisation-scale dependence of the Lagrangian parameters.  In this work, we run the Lagrangian parameters to different scales using \fs, which uses the \msbar scheme with $\xi=1$ by default. We do not consider the OS-like scheme in this section, since there is no free renormalisation scale to vary for that method. 

\begin{figure}[t!]
\centering
\includegraphics[width=0.99\textwidth]{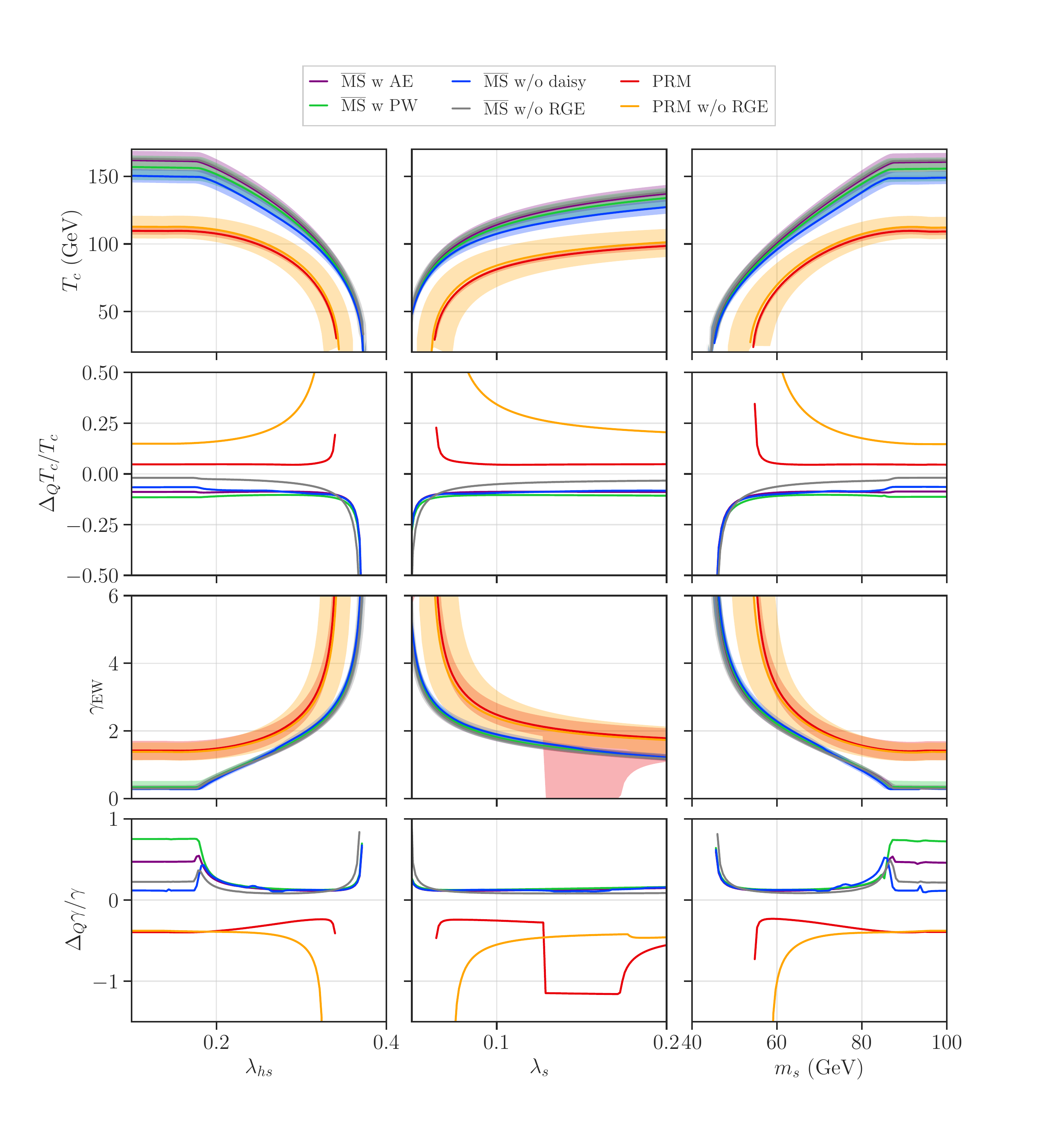}
\caption{The renormalisation scale dependence in the \msbar scheme and PRM method found by varying the renormalisation scale from $Q = \frac12 m_t$ to $2 m_t$.  In each of the columns, we vary one input parameter of the benchmark point in \cref{eq:benchmark}. The lower edges, center and upper edges of the bands in the first and third rows show the $T_c$ and \gammaew at $Q = \frac12 m_t$, $m_t$, and $2 m_t$ separately.  The second and last rows show the fractional change in $T_C$ and $\gammaew$, respectively. The labels `w AE', `w PW' and `w/o daisy' indicate daisy resummation with the Arnold-Espinosa method, the Parwani method, and no resummation, respectively. }
\label{fig:1d_scale}
\end{figure}

\subsubsection{Benchmark point}

In \cref{fig:1d_scale}, we show the renormalisation scale dependence in the \msbar scheme and PRM methods, by varying the renormalisation scale from $Q = \frac12 m_t$ to $Q = 2 m_t$. We choose this interval because it is expected to capture uncertainty from the most important missing high-order corrections from the top quark at $T = 0$.  Here we use the one-loop resummed Goldstone mass to treat the GC.  In each of the panels, we vary one input parameter around the benchmark point in \cref{eq:benchmark}. The top and bottom two rows of panels in \cref{fig:1d_scale} show the uncertainties on $T_c$ and \gammaew, respectively.  The lower edge, center, and upper edge of the bands in the first and third rows indicate results at $Q = \frac12 m_t$, $m_t$, and $2 m_t$, respectively.  To highlight the uncertainties, we separately show the fractional changes, $(T_c(Q=\frac{1}{2}m_t)-T_c(Q=2m_t)) / T_c(Q=m_t)$ and $(\gamma(Q=\frac{1}{2}m_t)-\gamma(Q=2m_t)) / \gamma(Q = m_t)$, directly by colored lines in the second and fourth rows of panels. 

We see almost no parameter dependence in the results at $\lambda_{hs} \lesssim 0.2$ and $m_s \gtrsim 90\gev$.  In these regions the transition is between $(\phi_h=0,\phi_s=0)$ and $(\phi_h\neq0, \phi_s=0)$ and thus is independent of parameters related to the singlet.  Unlike \refcite{Senaha:2018xek}, we show PRM results even if the high-temperature approximation of the effective potential contained a single minimum at the critical temperature. For $\lambda_s \approx 0.15$, that single minima was at the origin for $Q = \frac12 m_t$ leading to $\gammaew = 0$.
The change in $T_c$ for the PRM methods is typically mild and less than about $10\%$, though \gammaew shows more sensitivity, typically less than about $20\%$. 

\begin{figure}[t!]
\centering
\includegraphics[width=0.99\textwidth]{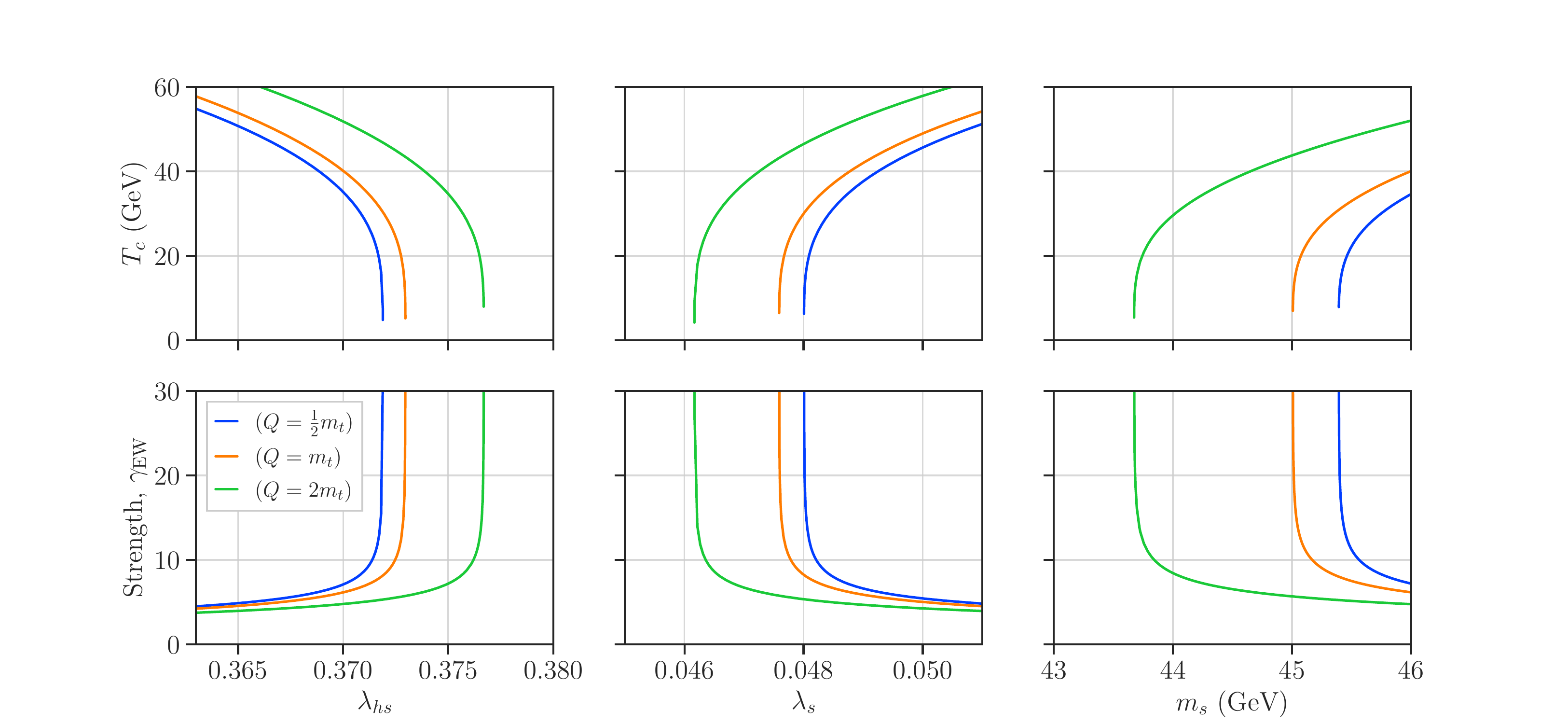}
\caption{Similar to \cref{fig:1d_scale} though for $T_c \lesssim 60\gev$, and only results for the \msbar scheme with AE are shown by colored lines.}
\label{fig:1d_scale_line}
\end{figure}

A common and important feature in all the results, however, is that the uncertainties rise sharply when the critical temperature is sensitive to the Lagrangian parameters.  This is not only true for the uncertainty caused by scale dependence, but also true for the uncertainty caused by gauge dependence discussed in the above subsection, and the uncertainties presented in the following subsection.  As the variation of renormalisation scale or gauge parameter slightly changes the Lagrangian parameters, it is obvious that when the critical temperature is sensitive to the parameters, the critical temperature and transition strength become sensitive to the renormalisation scale or gauge parameter.  This happens in the region of large $\lambda_{hs}$, small $\lambda_s$ and small $m_s$, where the critical temperature is relatively small.  There are no results for $T_c < 30\gev$ in \cref{fig:1d_scale} because we find points in the plots by a grid scan of the parameters, and the grid misses fine-tuned cases in which $T_c<30\gev$.  To find results in that region we performing a more detailed scan in a smaller parameter range where we expect to find smaller $T_c$. We  show the \msbar result for the resulting low temperature range in \cref{fig:1d_scale_line}. To obtain $T_c<5\gev$, the parameters have to be tuned to one part in $10^7$.  The reason can be seen from \cref{eq:dvtree}.  A zero critical temperature means no energy gap between the EW VEV in the $\phi_h$ direction and a local minimum on $\phi_s$ direction, so the parameters should cancel each other.

With RGE running and the Parwani method, the scale uncertainties in the \msbar result for $T_c$ are about 16\gev for $T_c \approx 150\gev$. As $T_c$ decreases, the relative uncertainty changes from approximately constant to sharply increasing in magnitude at $T_c<40\gev$. Using the Arnold-Espinosa method or eliminating daisy resummation results in similar uncertainties with the same tendency.  When $T_c \lesssim 20\gev$, the daisy corrections, which are proportional to the temperature $T^4$ (see \cref{eq:daisy}), are negligible. Consequently, in this regime the scale dependence of different treatments of daisy resummation are approximately the same.

We see that $\Delta_Q T_c$ of the \msbar scheme without RGE running is smaller than the one with RGE running, which is counter-intuitive.  For example at leading order there are cancellations such that\footnote{Note that the left-hand side in \cref{Eq:scale_phi4} includes contributions from wavefunction renormalisation, $- \gamma_\phi \phi \frac{\text{d}V}{\text{d}\phi}$ where $\gamma_\phi$ here is the anomalous dimension, as well as contributions from the $\beta$ functions for the Lagrangian parameters.}~\cite{Gould:2021oba}
\begin{equation}\label{Eq:scale_phi4}
\mu \frac{\text{d}}{\text{d}\mu} \left(V_0 + V_{\rm CW}\right) = 0
\end{equation}
and there can be further cancellations of scale dependence in the temperature-dependent parts of the potential.   In the Parwani method, scale dependence in finite-temperature bosonic contributions cancels scale dependence in temperature-dependent terms introduced in the CW potential by the Parwani method.
The cancellation \cref{Eq:scale_phi4} is, however, spoilt by next-to-leading order terms, such as two-loop running of parameters in the tree-level potential and one-loop running of parameters in the CW potential, and in general requires that we consider the $\beta$-functions for all Lagrangian parameters and the anomalous dimensions of the fields. In our calculation, we apply two-loop RGEs and omit the anomalous dimension, such that \cref{Eq:scale_phi4} no longer holds exactly.  Furthermore, we include finite-temperature corrections from the top quark, and the scale cancellations in the Parwani method only occur for bosonic corrections.
Consequently, results that include RGE running suffer from greater scale uncertainties than results without RGE running, and the Parwani method makes the situation even worse. This is somewhat counter-intuitive, as the derivative of the Parwani improved Coleman Weinberg potential with respect to the RG scale has terms that are quadratic in temperature and negative in sign, so one might expect they naively cancel the scale dependence of the quadratic contribution to the thermal functions. However, the combined $\beta$ function for the thermal mass turns out to be negative as it is dominated by top quark interactions. So the new scale-dependent contribution in Parwani adds to, rather than cancels, the scale dependence of the un-resummed theory.

\begin{figure}[t!]
\centering
\includegraphics[width=0.95\textwidth]{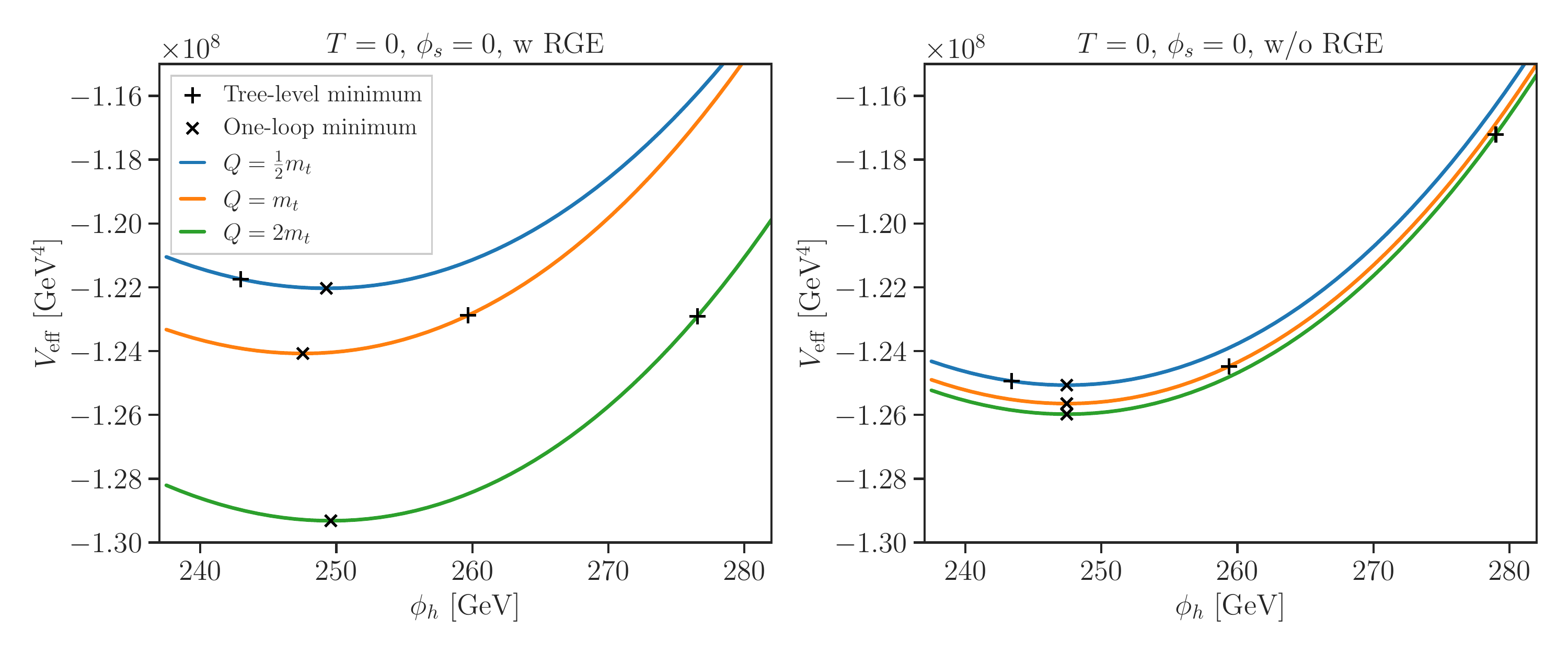}
\caption{The zero-temperature effective potential in the \msbar scheme for the benchmark point in \cref{eq:benchmark} at different renormalisation scales, with (left) and without RGE running (right).}
\label{fig:potential_scale}
\end{figure}

In \cref{fig:potential_scale}, we display the zero-temperature potentials of the \msbar scheme with the Arnold-Espinosa method for the benchmark point of $\lambda_s = 0.1$, $\lambda_{hs} = 0.3$ and $m_s = 65\gev$. We can see clearly that \cref{Eq:scale_phi4} does not hold with RGE running in our calculation, and that the potential suffers from greater scale dependence when we include RGE running. We also mark the tree-level minimum since the PRM method evaluates the one-loop potential at the minima of the tree-level potential. \Cref{fig:potential_scale} shows that changes in  the one-loop potential at the tree-level minima with RGE running are smaller than that without RGE running, as with RGE running the tree-level minima approximately lie  on a horizontal line of constant potential.  Therefore, the $\Delta_Q T_c$ for the PRM method with RGE running is smaller than $\Delta_Q T_c$ without RGE running. Note that for the PRM method, $T_c{(Q=\frac{1}{2}m_t)} > T_c{(Q=2m_t)}$. 

\Cref{fig:1d_scale} shows that $\Delta_Q T_c$ of the PRM method with RGE running is rather small and roughly independent of Lagrangian parameters, typically about 5\% for $T_c$, though it can exceed about 20\% for very small critical temperatures. Without RGE running, $\Delta_Q T_c$ can be larger than 15\%, and increases rapidly with $T_c$ decreasing. This implies that proper RGE running is quite essential in the PRM method to reduce the renormalisation scale dependence. Looking only at $\Delta_Q T_c$, the \msbar scheme has larger scale uncertainties than the PRM method, but considering $\Delta_Q \gamma$, the \msbar scheme has smaller scale uncertainties than the PRM method. The increased uncertainty in $\gamma$ in the PRM method may be connected to the fact that the numerator, $\Delta\phi_h$, and denominator, $T_c$, are obtained from two different potentials in this method.

\begin{figure}[t!]
\centering
\includegraphics[width=0.99\textwidth]{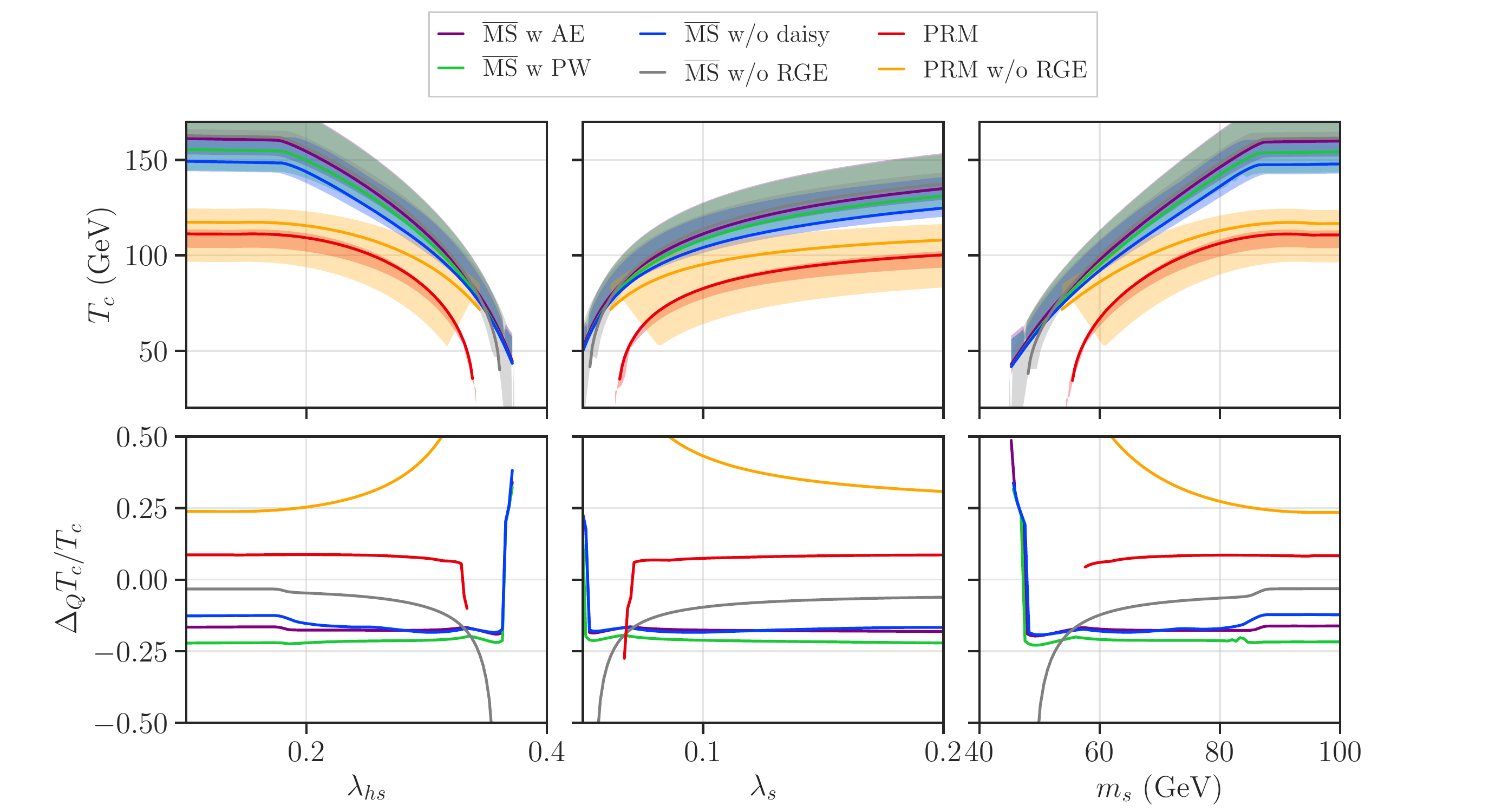}
\caption{Similar to \cref{fig:1d_scale}, but for the renormalisation scale dependence in the \msbar scheme and PRM method found by varying the renormalisation scale from $Q = T/2$ to $2 \pi T$ by solving \cref{eq:iterative}.}
\label{fig:1d_scale_T}
\end{figure}

Finally, we consider a broader variation of the renormalisation scale from $Q = T / 2$ to $2\pi T$. This interval was chosen to capture uncertainty from missing higher-order corrections at finite temperature, specifically involving the soft scale $\sim g T$ and the first Matsubara mode as discussed in \refcite{Croon:2020cgk}. Here, the renormalisation scale changes throughout a calculation, as whenever the potential is evaluated at a temperature $T$, the renormalisation scale is set to a scale $Q \propto T$. We perform calculations at $Q = T$, $Q = T / 2$ and $Q = 2\pi T$ by solving the following equations iteratively for $T$,
\begin{equation}\label{eq:iterative}
    T_c(Q = T) = T, \quad T_c(Q = T / 2) = T, \quad T_c(Q = 2\pi T) = T.
\end{equation}
By doing so, we ensure that at temperature $T = T_c$ the potential was computed at the renormalisation scales $Q = T/2$, $T$ and $2\pi T$, as required. We show the variation in the critical temperature in \cref{fig:1d_scale_T}. We see that the uncertainty bands are broader compared to \cref{fig:1d_scale}, especially for the PRM method for which the relative uncertainty was always greater than about $25\%$. The $\Delta_Q T_c/T_c$ of \msbar scheme flips sign and shoots up at low $T_c$. With scale $Q$ fixed by \cref{eq:iterative}, $T_c$ also decreases with $\lambda_{hs}$ increasing, or $\lambda_s$ and $m_s$ decreasing. The slope of $T_c$ against the input parameters is larger when $Q=2\pi T$ than when $Q=T/2$, especially for low $T_c$. Thus the uncertainty flips sign at certain value of $T_c$. 

\subsubsection{Two-dimensional scans}

\begin{figure}[t!]
\centering
\includegraphics[width=0.99\textwidth]{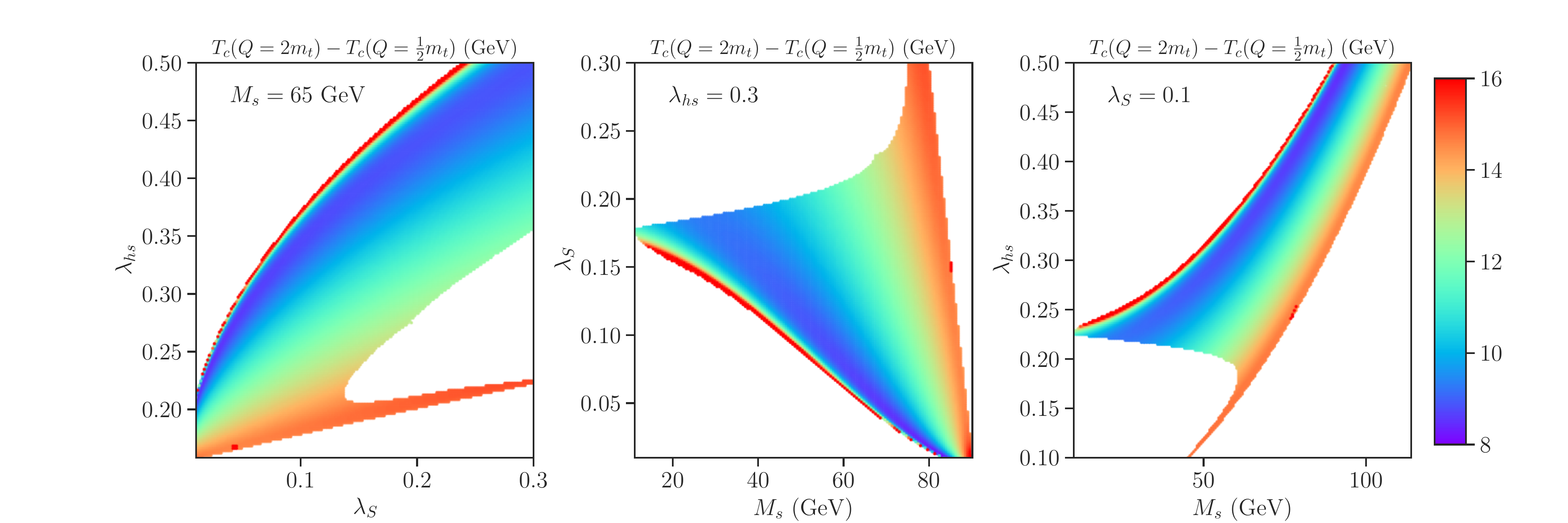}
\caption{The renormalisation scale dependence in the \msbar scheme for the critical temperature $T_c$ found by varying the renormalisation scale from $Q = \frac12 m_t$ to $2 m_t$ in 2-dimensional scans. In the blank regions there are no points with a FOPT at both $Q = \frac12 m_t$ and $2 m_t$. We fix $m_s=62.5\gev$ in the left panel, $\lambda_{hs}=0.3$ in the middle panel, and $\lambda_{s}=0.1$ in the right panel.}
\label{fig:2d_scale_deltaT}
\end{figure}

We now turn to two-dimensional planes of the SSM parameters to check the renormalisation scale dependence throughout the model's parameter space.  Here we change the scale using RGE running, use the \msbar scheme and the $R_\xi$ gauge, and include daisy corrections.  In \cref{fig:2d_scale_deltaT}, we show the difference of critical temperature $\Delta_Q T_c = |T_c{(Q=\frac{1}{2}m_t)} - T_c{(Q=2m_t)}|$  on the $\lambda_s$--$\lambda_{hs}$, $m_s$--$\lambda_{s}$ and $m_s$--$\lambda_{hs}$ planes.  The colored regions are obtained by two-dimensional scans with $m_s=62.5\gev$ fixed in the left panel, $\lambda_{hs}=0.3$ fixed in the middle panel, and $\lambda_{s}=0.1$ fixed in the right panel.  In the blank regions there were no points with a FOPT at both $Q = \frac12 m_t$ and $Q = 2 m_t$.  

The tendency exhibited in the two-dimensional planes in \cref{fig:2d_scale_deltaT} is consistent with that in the one-dimensional results.  The uncertainties at first decrease with $\lambda_{hs}$ increasing, $\lambda_s$ decreasing and $m_s$ decreasing, but subsequently rise sharply.  In most of the parameter space, the uncertainties lie around $8\gev$ -- $16\gev$.  The magnitude of the uncertainties can reach more than 50\gev when $T_c<60\gev$ (see \cref{fig:1d_scale_line}).  The maximum value of $\Delta T_c$ is dependent on the resolution of the scan.  With two-dimensional grid scans of 200 points per dimension in the ranges $0.1<\lambda_{hs}<0.5$, $0.01<\lambda_{hs}<0.3$ and $10\gev<m_{s}<120\gev$, we found a maximum $\Delta T_c$ of 40.2\gev at the parameter point $\lambda_{hs}=0.43$, $\lambda_{s}=0.1$, $m_{s}=76.9\gev$, $T_c{(Q=\frac{1}{2}m_t)}=6.2\gev$ and $T_c{(Q=2m_t)}=46.4\gev$.

\subsection{Daisy resummation dependence}

\begin{figure}[t!]
\centering
\includegraphics[width=0.99\textwidth]{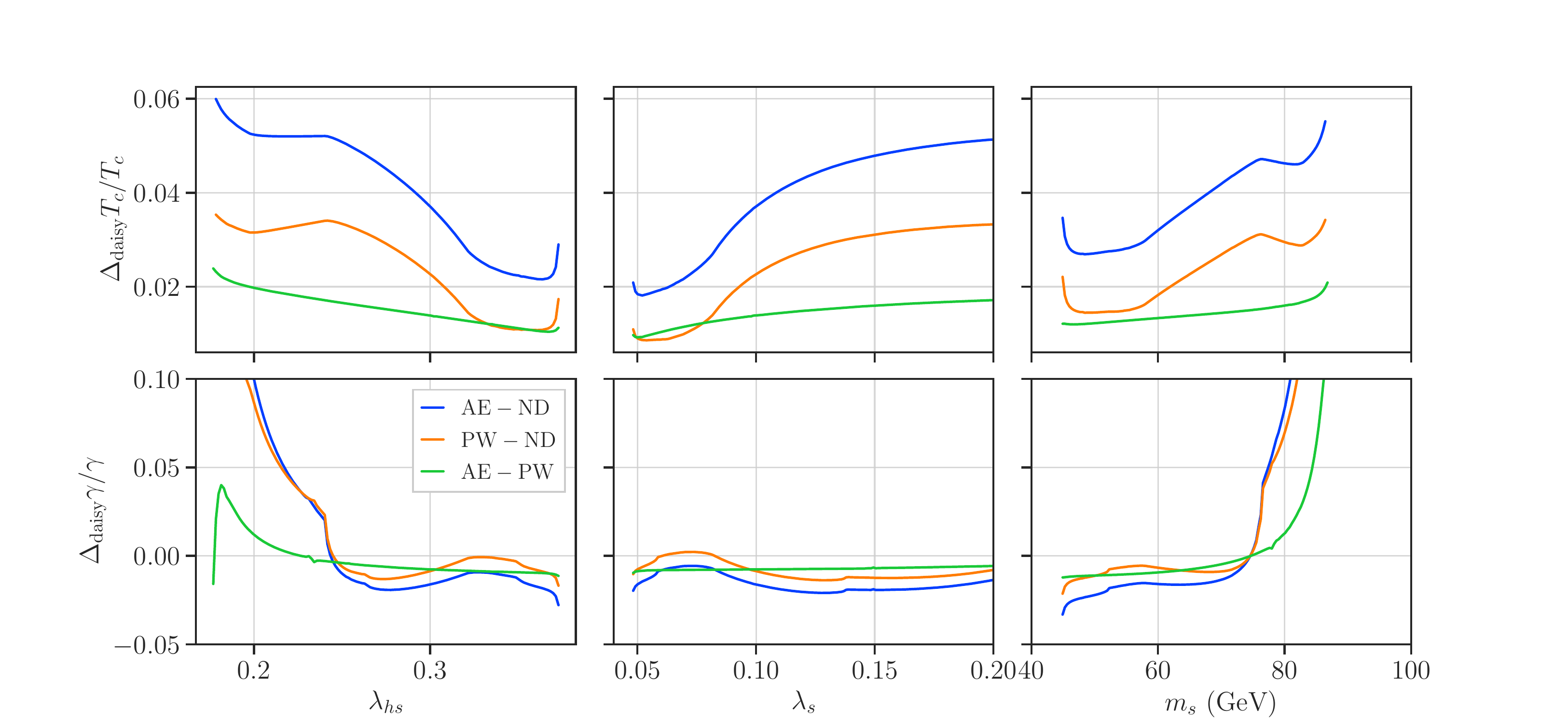}
\caption{Differences in critical temperature and transition strength with different methods for resummation of daisy diagrams, in the \msbar scheme of $\xi=0$ and $Q=m_t$.  `PW' and `AE' stand for adding daisy resummation using the Parwani method and the Arnold-Espinosa method respectively, while `ND' indicates no resummation of daisy diagrams.  In each of the panels we vary one input parameter about the benchmark point in \cref{eq:benchmark}.}
\label{fig:1d_daisy}
\end{figure}

In \cref{fig:1d_daisy}, we present differences in $T_c$ and \gammaew with different treatments of daisy diagrams --- the Parwani method and the Arnold-Espinosa method described in \cref{sec:daisy}, and discarding resummation of daisy diagrams.  We can see that in the SSM the effects of daisy resummation on transition properties are  relatively small (though this effect may be bigger in other models; see e.g.\ \refcite{Cline:1996mga,Basler:2016obg,Biekotter:2021ysx}).

The changes in $T_c$ caused by switching between the Parwani and the Arnold-Espinosa methods are less than 3\% and the changes caused by switching off daisy resummation altogether were less than 6\%.  The changes decrease with $\lambda_{hs}$ increasing or $\lambda_s$ and $m_s$ decreasing, which means $T_c$ decreasing (see top panels of \cref{fig:1d_schemes}). 
Unlike changes caused by varying the gauge parameter or renormalisation scale, varying the treatment for daisy resummation does not alter the Lagrangian parameters.  Therefore, here the changes in $T_c$ do not rise sharply at low $T_c$. On the other hand, the relative changes in \gammaew are typically even smaller, but increase rapidly for large $T_c$, simply because that $\gammaew$ are very small. The absolute changes in \gammaew are less than 0.06 when $T_c > 100\gev$ and $\gammaew<0.4$.
\subsection{Comparison of renormalisation schemes and other methods }

\begin{figure}[t!]
\centering
\includegraphics[width=0.99\textwidth]{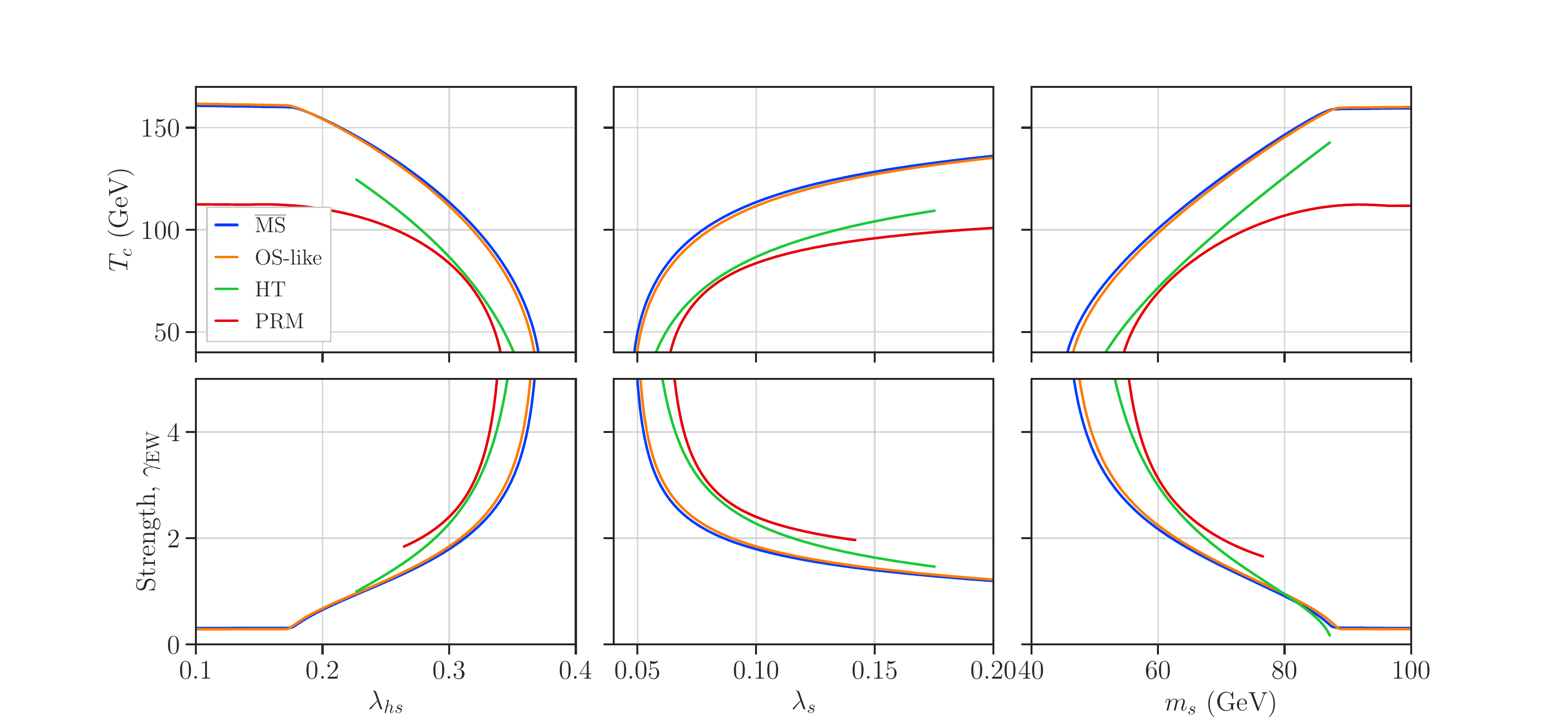}
\caption{Critical temperature and transition strength in the \msbar scheme, OS-like scheme, HT approximation, and PRM method. The other settings ($\xi=0$, $Q=m_t$, using Arnold-Espinosa method and \texttt{GC\_SelfEnergy\_Sol}) are the same if they are optional in that scheme. 
In each of the panels we vary one input parameter about the benchmark point in \cref{eq:benchmark}.  }
\label{fig:1d_schemes}
\end{figure}

We now turn to the renormalisation schemes.  In \cref{fig:1d_schemes}, we compare results of the \msbar and OS-like scheme, together with the high temperature approximation which is independent of renormalisation scheme, and the PRM method which depends on the \msbar scheme.  In both the \msbar scheme and the OS-like scheme, we use $\xi=0$, the Arnold-Espinosa method for daisy resummation, and resum the Goldstone boson mass to solve the GC.  The renormalisation scale in the \msbar scheme is set to $m_t$.  
For the \msbar potential used in the  PRM method we adopt the same settings as in the \msbar scheme, except that we do not resum daisies.  

\begin{figure}[t!]
\centering
\includegraphics[width=0.99\textwidth]{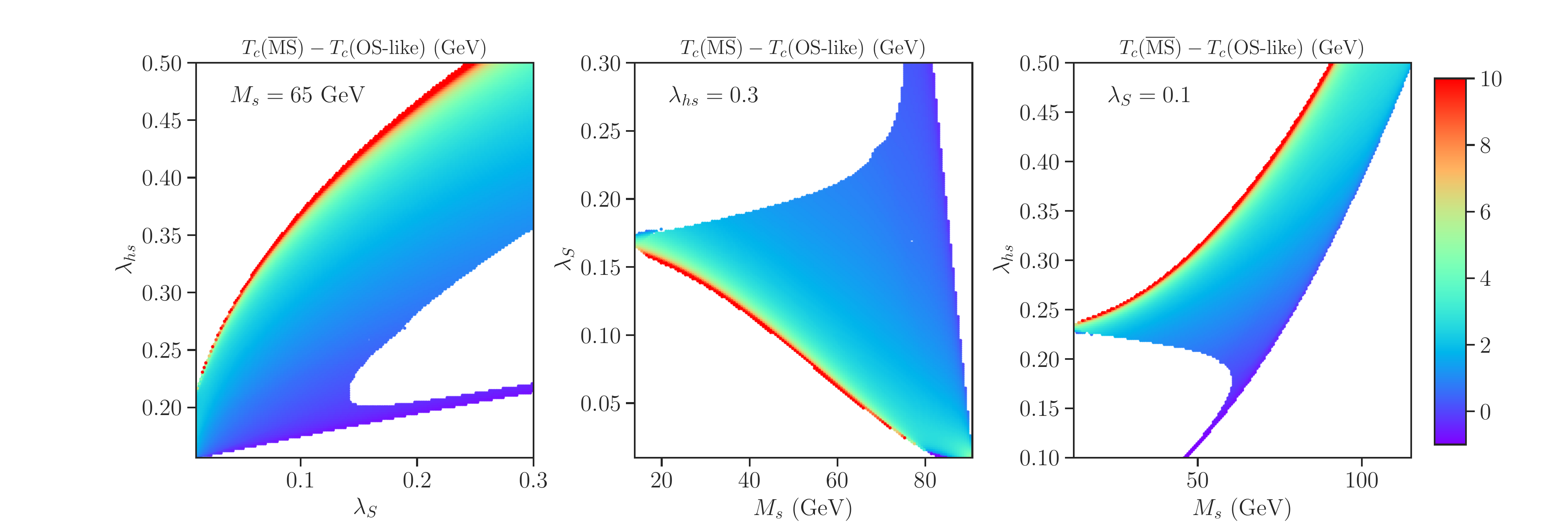}
\caption{The difference in critical temperature $T_c$ between the \msbar and OS-like schemes in 2-dimensional scans, under the same settings.  In the blank regions there are no points with a FOPT in both the \msbar and OS-like schemes.  We fix $m_s=62.5\gev$ in the left panel, $\lambda_{hs}=0.3$ in the middle panel, and $\lambda_{s}=0.1$ in the right panel.}
\label{fig:2d_scheme_deltaT}
\end{figure}

Since the zero-temperature one-loop corrections depend on the choice of renormalisation scheme, the tadpole conditions result in different Lagrangian parameters, $\lambda_h$, $\mu_h$ and $\mu_s$, in each scheme.  Therefore, every part of the effective potential can be different in the \msbar and the OS-like schemes.  However, the differences in $T_c$ and \gammaew are typically mild with the OS and \msbar schemes giving surprisingly similar results for the choice of renormalisation scale we make for the \msbar\ scheme ($Q=m_t$).  It was also found in previous work that the OS-like result lies within the scale variation uncertainty~\cite{Chiang:2018gsn}.  This suggests that the differences to the OS-like scheme is not really accessing any higher-order corrections beyond the \msbar\ scale variation though the very close agreement we see must be coincidental in nature. 

Except for the flat region $\lambda_{hs}<0.17$ in the left panels and $m_s>88\gev$ in the right panels of \cref{fig:1d_schemes}, $T_c$ in the \msbar scheme is larger than $T_c$ in the OS-like scheme, and vice-versa for the transition strengths in the bottom panels.  With critical temperature decreasing, the difference $\Delta T_c = T_c{(\overline{\rm MS})}-T_c(\text{OS-like})$ increases monotonically from about 1\gev to more than 10\gev.  This occurs for the whole parameter space, which can be seen from the results of the two-dimensional scans in \cref{fig:2d_scheme_deltaT}. 

The $T_c$ of the high-temperature approximation and the PRM method are far lower than that of the OS-like scheme and the \msbar scheme.  There is no result for small $\lambda_{hs}$, large $\lambda_s$ and $m_s$, because the expression of $T_c$ of HT, \cref{Eq:TC_HT}, assumes a minimum of the form $(0, v_s^{\rm high})$ exists at $T_c$.  The differences between results from the HT approximation and the OS-like scheme are easy to understand.  They both use tree-level tadpole conditions to solve the Lagrangian parameters, $\lambda_h$, $\mu_h^2$ and $\mu_s^2$, which results in exactly the same tree-level potential.  The one-loop finite-temperature corrections are also supposed to be similar.  Since the zero-temperature one-loop correction in the OS-like scheme is zero at the EW VEV, the only obvious difference is the zero-temperature one-loop correction of the OS-like scheme around the minimum of $\phi_h =0$.  In this minimum, the fermions and vectors are massless, and the only non-vanishing contribution to the zero-temperature one-loop correction comes from scalars and is always positive.  Therefore, to compensate this increase in the potential, the OS-like scheme needs larger thermal corrections, i.e.\ higher temperature than in the HT approximation, to achieve degenerate minima.  As for the PRM method, it also uses tree-level tadpole conditions, and the zero-temperature one-loop corrections in the \msbar scheme reduce the vacuum energy gap, so it leads to lower $T_c$ than the HT approximation.  However, in the HT approximation, both the vacua tend towards the origin as temperature increases, as can be seen in \cref{app:phasestructure}, which also reduces the vacuum energy gap, while the vacua in the PRM method are fixed.  These two impacts cancel each other to some extent, leading to similar results in these two schemes.  In contrast the results can also diverge more at higher critical temperatures.  In that case the HT expansion is, as you would expect, becoming a better approximation of the $\msbar$ calculation and gets closer to the $\msbar$ result. 

\subsection{Comparison in whole parameter space}

\begin{figure}[t!]
\centering
\includegraphics[width=0.99\textwidth]{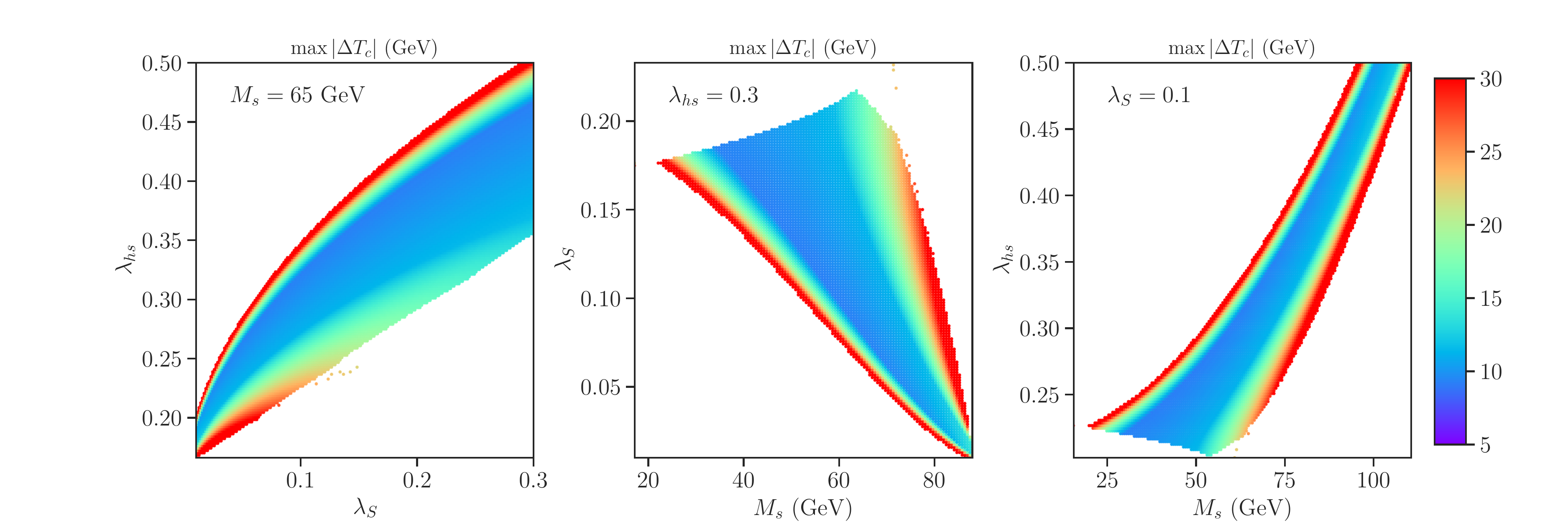}\\
\includegraphics[width=0.99\textwidth]{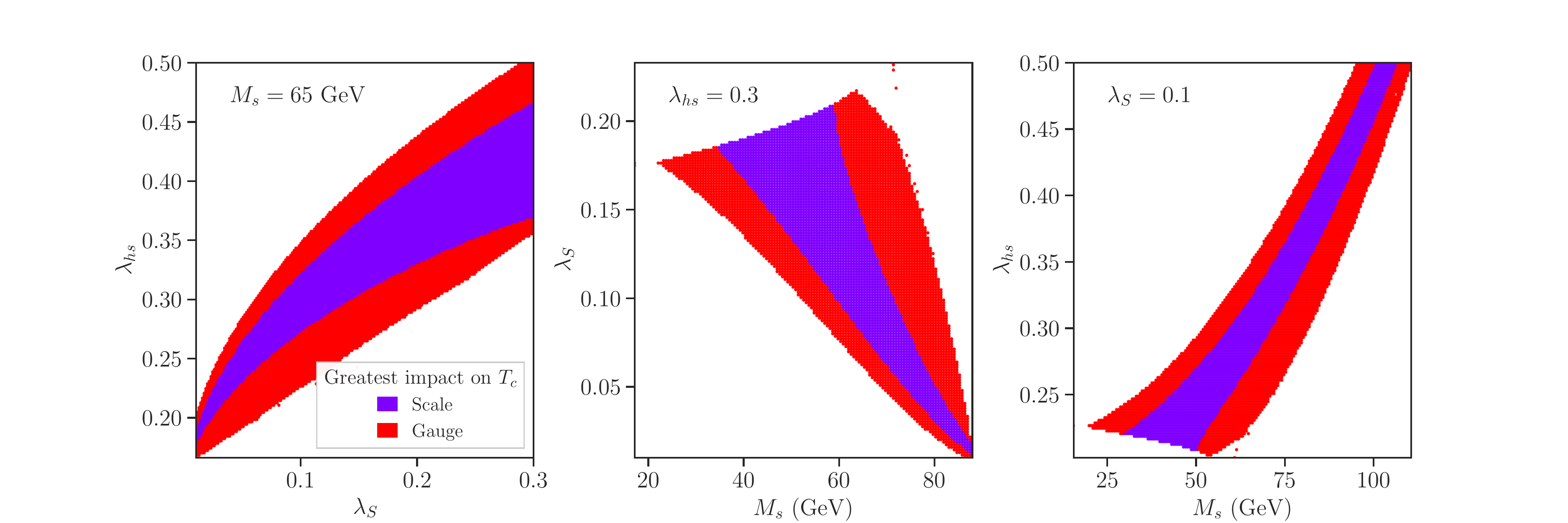}
\caption{Maximum change in the critical temperature from changing the gauge, renormalisation scale, renormalisation scheme or treatment of daisies. We show the magnitude (upper panels) and source (lower panels) of the maximum change.}
\label{fig:max_uncer}
\end{figure}

Here we compare all the possible changes investigated above in the whole parameter space. As shown in  \cref{fig:max_uncer} on 2-dimensional planes, changing the renormalisation scale between $\frac12 m_t$ and $2m_t$ or changing the gauge parameter between $\xi = 0$ and $25$ always induce the greatest changes in the properties of the FOPT. The impact of the treatment of daisies (Arnold-Espinosa versus Parwani), the GC, and the renormlisation scheme dependence (\msbar\ versus \os) are always subdominant.

\begin{figure}[t!]
\centering
\includegraphics[width=0.99\textwidth]{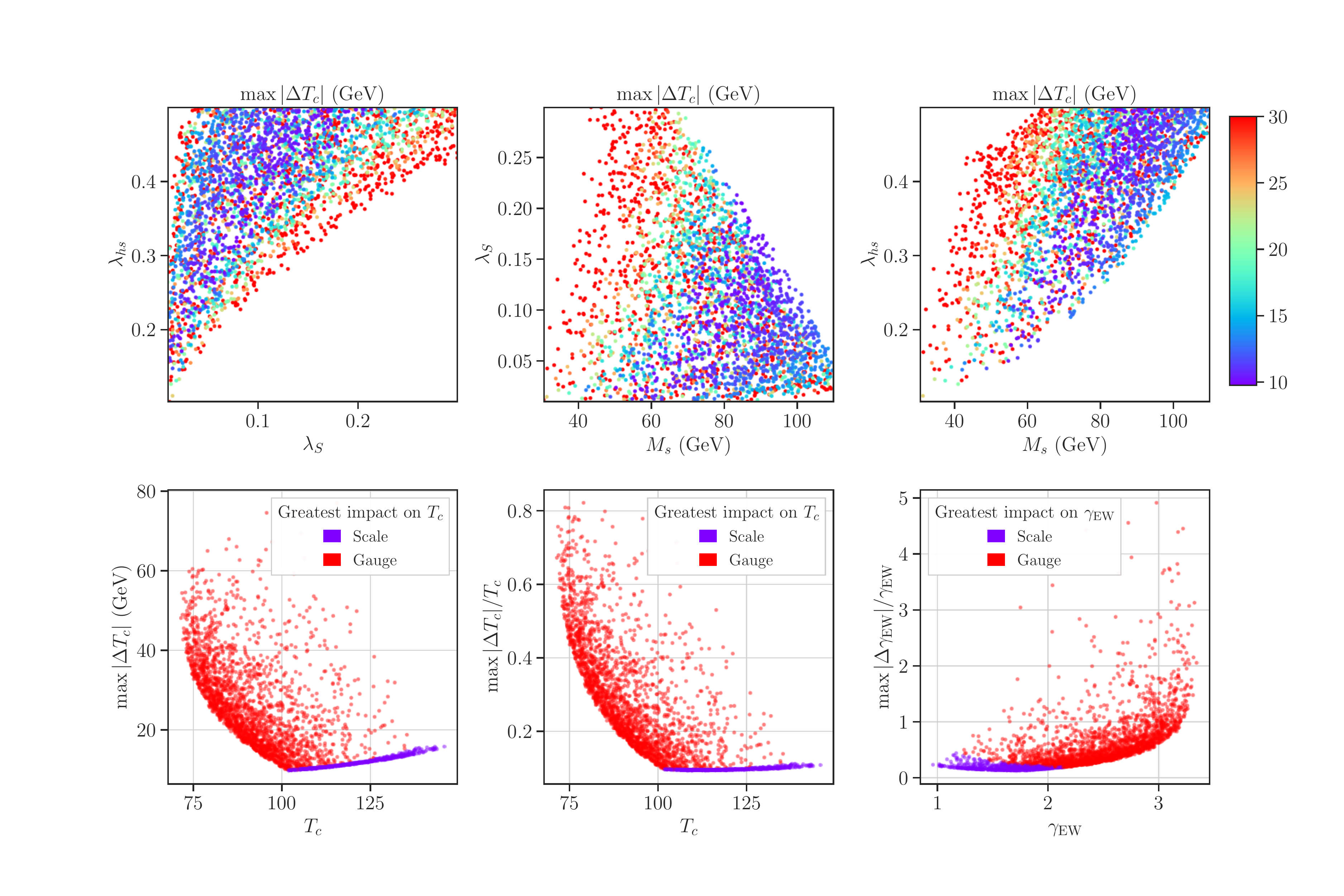}
\caption{Results from three-dimensional scans. In the top panels, we show the maximum change in the critical temperature on two-dimensional planes, where points of small uncertainty are on top of points of large uncertainty. In the bottom panels, we show the relationship between the maximum changes in $T_c$ and $\gammaew$ and their values.}
\label{fig:3d_mu}
\end{figure}

In \cref{fig:3d_mu}, we show the greatest uncertainty of each point, which was the renormalisation scale uncertainty or gauge uncertainty, in a 3-dimensional random scan.  We sampled $10^5$ points with flat priors in the parameter space of
\begin{align}
    \lambda_{hs} \in [0.1,0.5],~~~\lambda_s \in [0,0.3],~~~ m_s \in [10\gev,110\gev].
\end{align}
With these choices, about $20\%$ of points had a FOPT at $\xi = 0$ and $Q = m_t$. Of those points, about $50\%$ didn't have a FOPT at $\xi = 25$ and about $10\%$ didn't have a FOPT at either $Q = \frac12 m_t$ or $2m_t$ (or both).  In about $80\%$ of cases gauge dependence was the greatest source of uncertainty. The top panels show the samples on planes of Lagrangian parameters with points with the smallest scale uncertainty shown on top.
There are no strong trends on these planes, though samples of small uncertainty tend to lie at large $m_s$.
In the bottom panels, the uncertainties are shown versus $T_c$ and \gammaew, colored by the source of the maximal uncertainty. When $T_c<100\gev$, varying the gauge parameter between $0$ -- $25$ always results in a greater change in the critical temperature than varying the renormalisation scale between $\frac{1}{2}m_t$ -- $2m_t$. When $T_c>100\gev$, on the other hand, varying the renormalisation scale can have the greatest impact, and the lower bound of maximal uncertainty on $T_c$ is given by points for which the maximum uncertainty comes from changing the renormalisation scale. The relative maximal uncertainty on $T_c$  derived from gauge dependence decreases with $T_c$ increasing, while for scale  dependence the relative maximal uncertainty always lies around $10\%$.  Conversely,  varying the renormalisation scale can have the greatest impact on $\gammaew$ for small $\gammaew$, and $\max |\Delta \gamma_{\rm EW}|/\gamma_{\rm EW}$  derived from gauge dependence increase with $\gammaew$ increasing. The relative maximal uncertainty on $\gammaew$ can scarcely reach about 500\%, because large $\gammaew$ corresponds small $T_c$, which become sensitive to the Lagrangian parameters.

\begin{figure}[t!]
\centering
\includegraphics[width=0.66\textwidth]{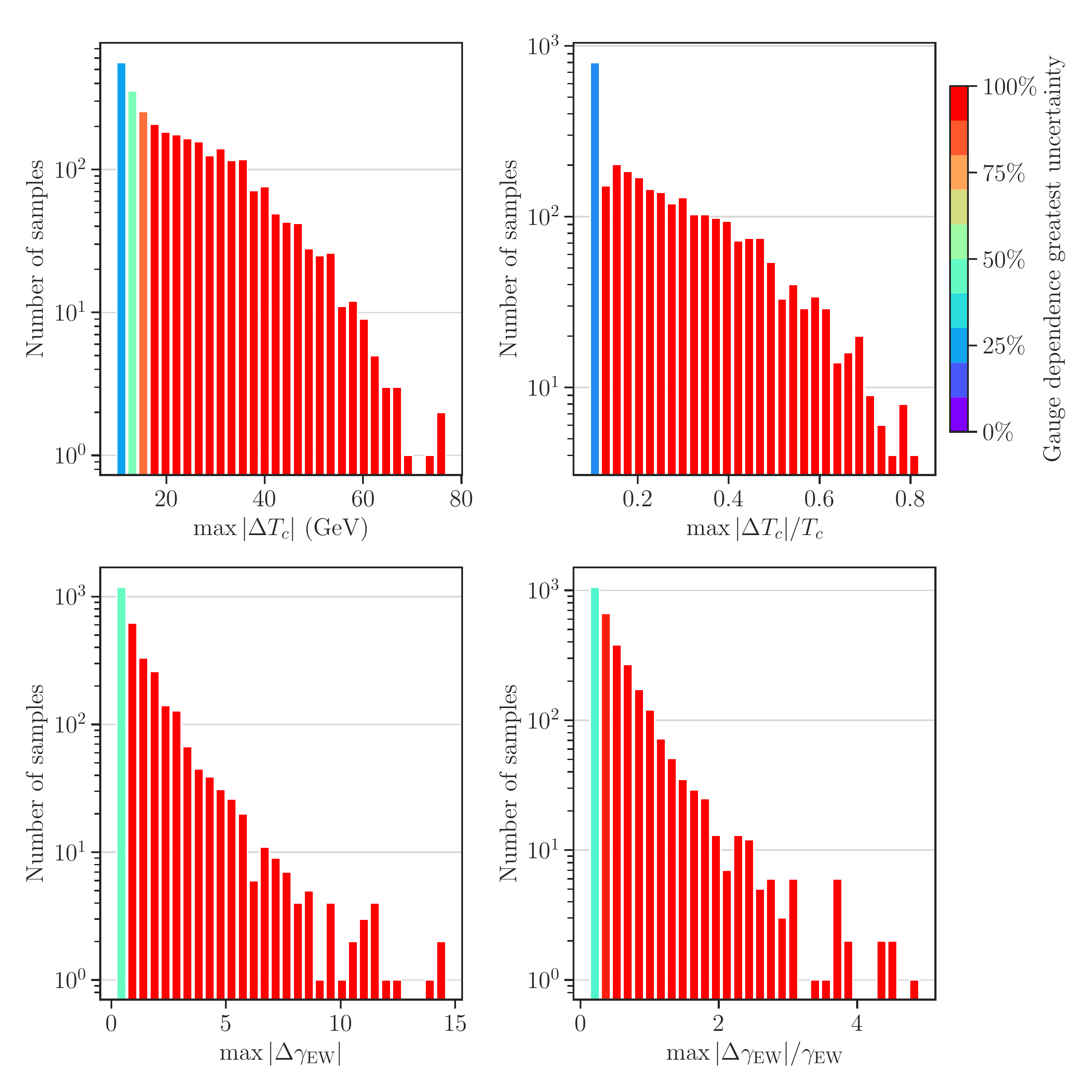}
\caption{Distributions of maximum change in the critical temperature (upper) and transition strength (lower) for samples shown in \cref{fig:3d_mu}. The bars are colored by the fraction of points for which gauge dependence was the greatest source of uncertainty.}
\label{fig:3d_mu_hist}
\end{figure}

The relative uncertainties can be as large as 100\% when one of the $T_c\simeq0\gev$, but it is rare to encounter such cases in general scans because the parameters need to be fine-tuned to achieve low $T_c$.  In \cref{fig:3d_mu_hist}, we display the distribution of relative and absolute uncertainties of $T_c$ and $\gamma$ of the 3-dimensional random scan.  
We find that absolute (relative) uncertainties of $T_c$ and $\gamma$ for most of the samples are smaller than 13\gev (10\%) and 0.4 (14\%), respectively, but that no samples have uncertainties smaller than 10\gev (9\%) and 0.19 (11\%).  
In extreme and rare cases, $\max |\Delta T_c|$ and $\max |\Delta \gamma_{\rm EW}|$ reached 77\gev and 0.49 respectively, and in an even more detailed scan the maximum uncertainty would be even larger. We see from the colors of the bars that for points with moderate or large uncertainties, the greatest source of uncertainty is almost always gauge dependence. This final remark is subject to the choice of scales and gauge parameters that we compare. With a smaller range of gauge parameters, $\xi = 0$ to $\xi = 3$, it was instead found that scale dependence was the greatest source of uncertainty for the majority of the parameter space.

\section{Conclusions}\label{sec:conc}

First-order phase transitions could have occurred in the history of our Universe, helped lead to the abundance of matter over anti-matter, and left behind observable gravitational wave signatures. In perturbative calculations in models of particle physics, however, the properties and occurrence of first-order phase transitions are sensitive to somewhat arbitrary choices, including choices of renormalisation scale and gauge. We investigated the impact of those choices in detail, using the electroweak phase transition in the SM extended by a real scalar singlet as a benchmark. We explored the three-dimensional parameter space of this model through one- and two-dimensional slices as well as three-dimensional scans. 

We refrain from making recommendations about the best choices, and instead wish to emphasise the sizes of the uncertainties and which choices influence the results the most. The scale dependence in the \msbar scheme and gauge dependence were the most significant sources of arbitrariness. Regarding the former, we found that
\begin{itemize}
    \item Changing $Q$ from $\frac12 m_t$ to $2 m_t$ induced moderate changes in the PT properties, typically less than about $10\%$.
    \item Scale dependence was particularly severe in the PRM scheme, and only somewhat alleviated by using RGE running.
    \item Counter-intuitively, applying two-loop RGE running in the \msbar scheme worsened scale dependence, possibly because this was not a strict fixed-order calculation and because we neglected the anomalous dimension of the fields.
    \item Similarly, the Parwani method worsened scale dependence, despite the possibility of cancellations in the scale dependence between the CW and finite-temperature parts of the potential.
    \item Whenever the PT properties strongly depend on the choices of Lagrangian parameters, they inevitably strongly depend on the choice of scale.
    \item Only about $10\%$ of our points with a FOPT at $Q = m_t$ didn't have a FOPT at both $Q = \frac12 m_t$ and $2m_t$.
\end{itemize}
In an OS-like scheme, on the other hand, we found quite similar results to the \msbar scheme at $Q = m_t$, though in extreme cases the differences in critical temperature reached about $10\gev$. Regarding gauge dependence, we found that
\begin{itemize}
    \item In the \msbar scheme the gauge dependence of the critical temperature and transition strength could be $\mathcal{O}(100\%)$ in extreme cases when $\xi$ was varied from $0$ to $25$ for both the $\Rxi$ and covariant gauges, though was typically milder.
    \item When varying between $\xi =0$ and $1$, with an upper bound consistent with an $\mathcal{O}(1)$  bound from perturbativity~\cite{Laine:1994bf,Garny:2012cg}, the gauge dependence was extremely mild.
    \item The gauge-independent PRM and HT predictions for PT properties were similar to each other and quite different from those from the \msbar method. 
    \item Despite being gauge independent, gauge dependence creeps into the PRM and HT methods when Lagrangian parameters are determined from tadpole conditions or mass constraints in a gauge-dependent way. In these cases the gauge dependence was similar to that in the \msbar scheme.
    \item In the $\Rxi$ gauge, we can avoid the GC at $\xi = 0$ by using the Higgs mass squared parameter obtained from one-loop tadpole conditions in the calculation of the field-dependent Goldstone masses. However, even with this approach, Goldstone catastrophes still occurred for other choices of $\xi$ for which Goldstones were massless in the EW vacuum.
    \item About $50\%$ of our points with a FOPT at $\xi = 0$ didn't have a FOPT at $\xi = 25$.
\end{itemize}
Finally, the impact of the different treatments of daisy diagrams and the GC were relatively small, always changing the critical temperature by less than about $10\gev$.

In summary, we investigated the major sources of arbitrariness in perturbative treatments of FOPTs, including renormalisation scales and schemes, and gauges and gauge fixing parameters. We found that moderate variations in these choices, especially the choice of gauge parameter, may lead to significant changes in the predictions for the FOPTs. 

\section*{Acknowledgements}

The work of P.A. in this paper has been supported by the Australian Research Council Future Fellowship grant FT160100274 and the National Natural Science Foundation of China (NNSFC) Research Fund for International Excellent Young Scientists, grant No.\ 12150610460.  The work of P.A. and C.B. was also supported by the Australian Research Council Discovery Project grants DP180102209 and DP210101636.  The work of L.M.\ was supported by an Australian Government Research Training Program (RTP) Scholarship and a Monash Graduate Excellence Scholarship (MGES). 
The work of Y.Z. was supported by the NNSFC under grant No. 12105248, by 
CAS Peng Huanwu Junior Visiting Professor Program and by Zhenghou University Young Talent Program.
This project was also undertaken with the assistance of resources and services from the National Computational Infrastructure, which is supported by the Australian Government.  We thank Astronomy Australia Limited for financial support of computing resources, and the Astronomy Supercomputer Time Allocation Committee for its generous grant of computing time. 
The work of GW is supported by World Premier International Research Center Initiative (WPI), MEXT, Japan. GW was supported by JSPS KAKENHI Grant Number JP22K14033

\appendix

\section{Phase structure}
\label{app:phasestructure}

\begin{figure}[t!]
\centering
\includegraphics[height=0.35\textwidth]{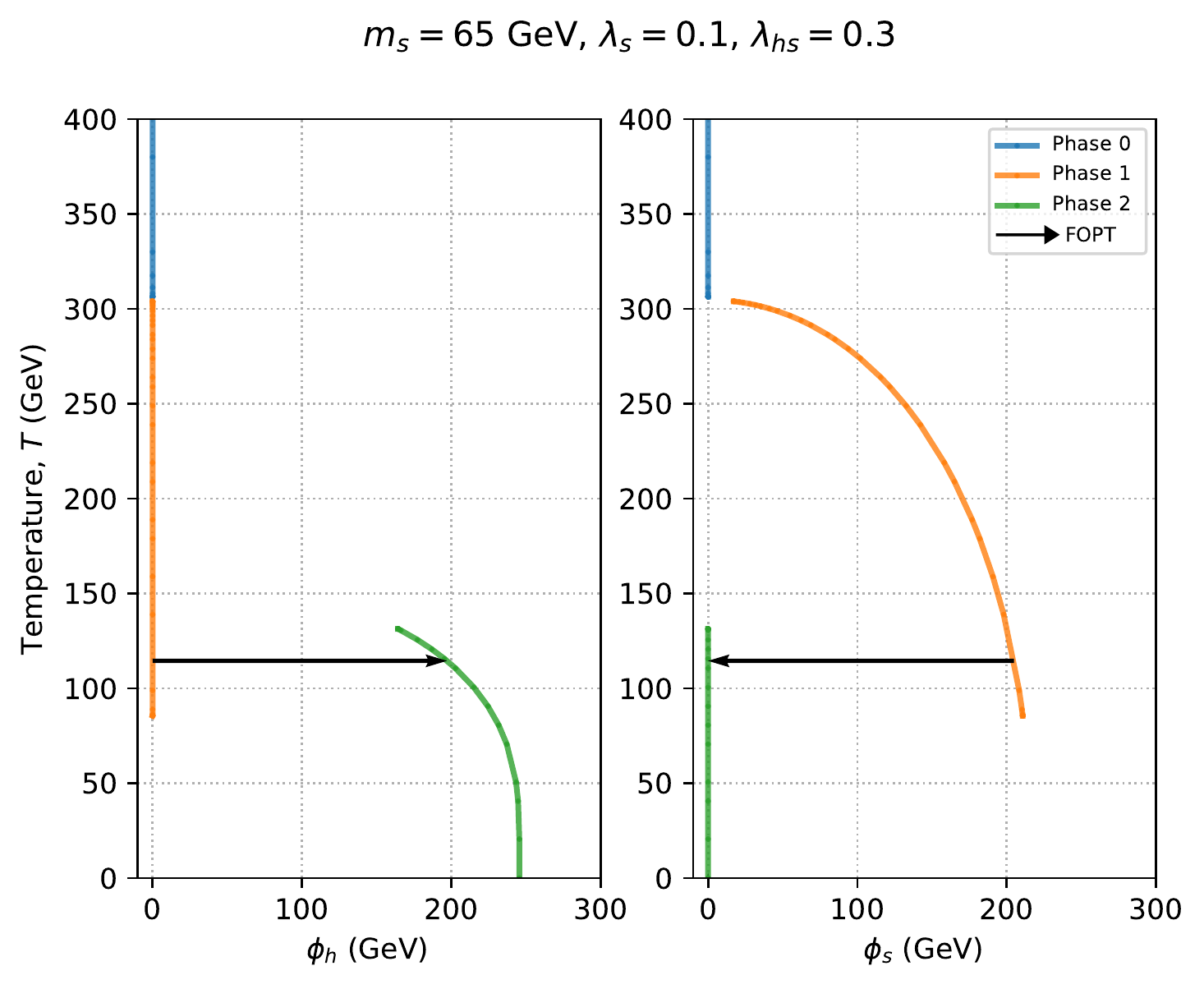}
\includegraphics[height=0.35\textwidth]{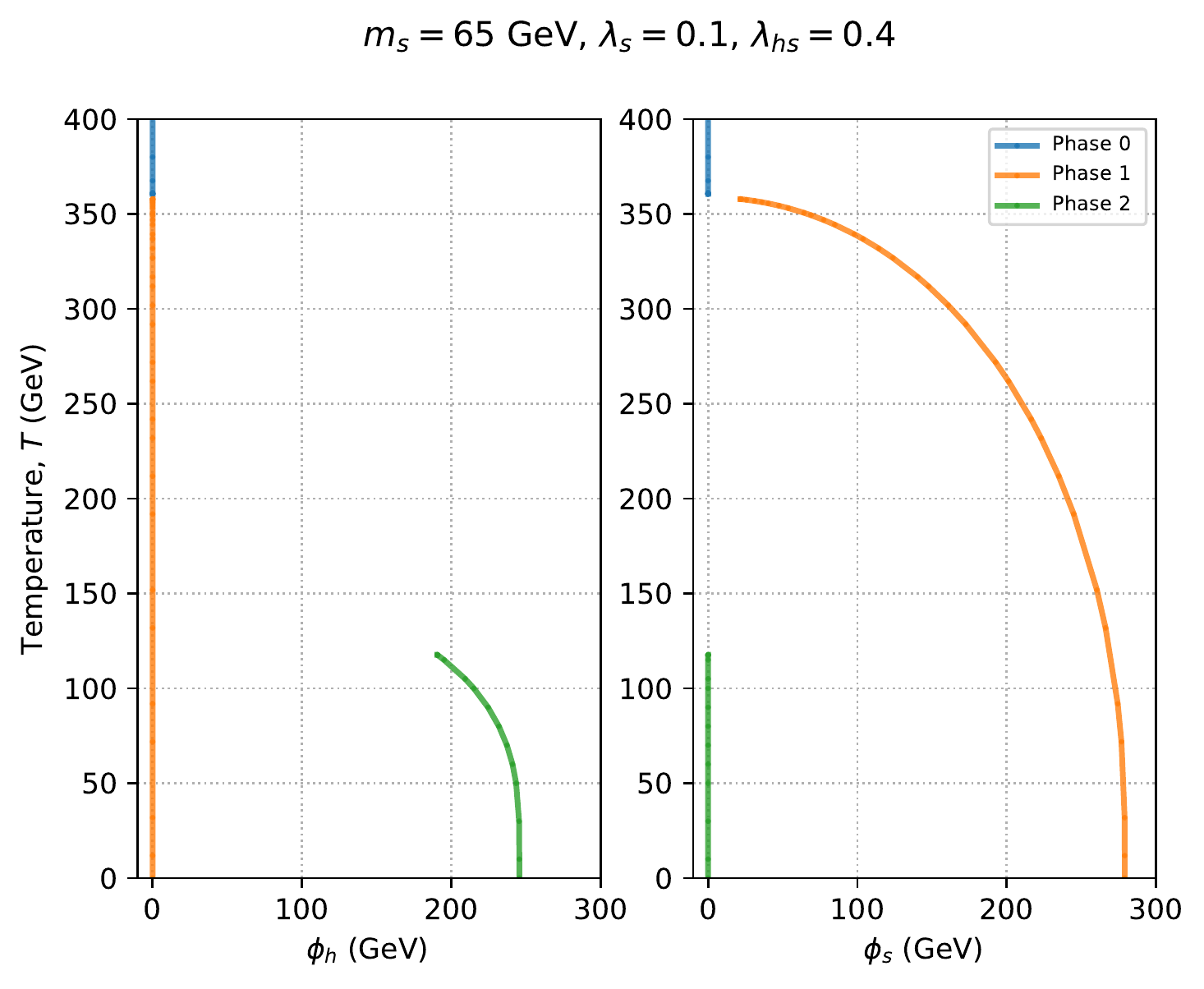}\\
\includegraphics[height=0.35\textwidth]{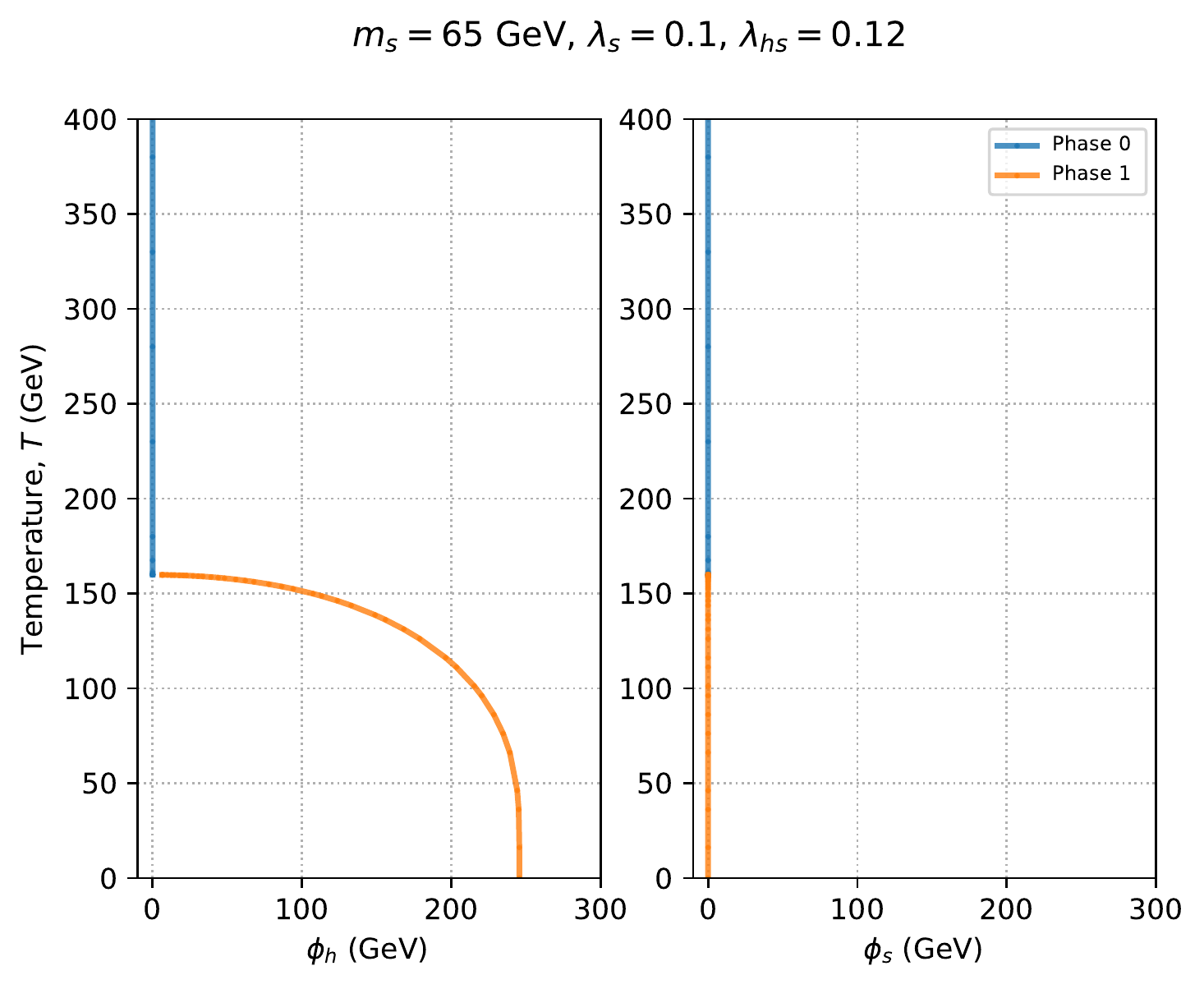}
\includegraphics[height=0.35\textwidth]{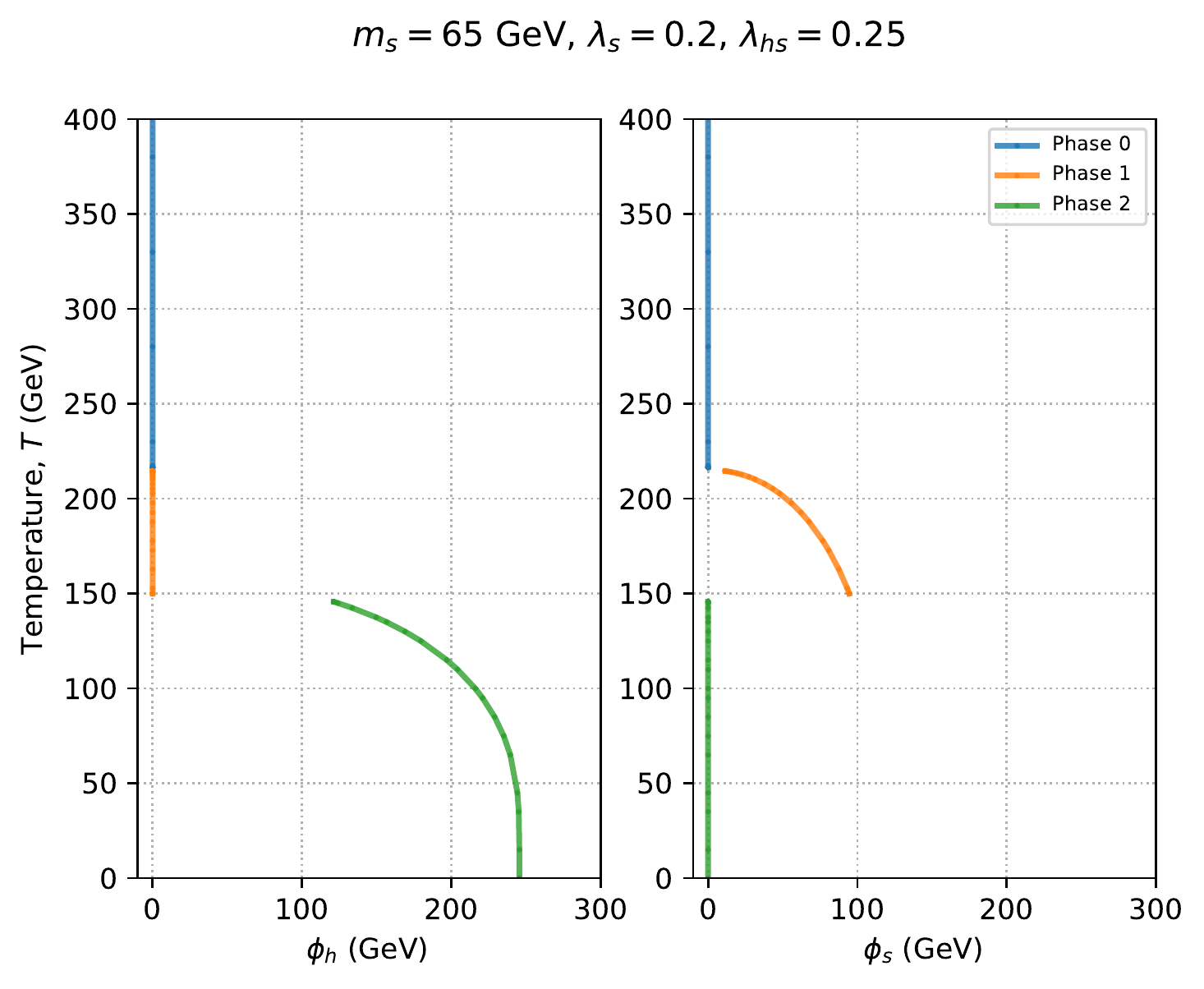}
\caption{Typical phase structures in the SSM. The lines show the field values at a particular minimum as a function of temperature. The arrows indicate that at that temperature the two phases linked by the arrows are degenerate and thus that a FOPT could occur in the direction of the arrow.}
\label{fig:phasestructure}
\end{figure}

In the SSM, the FOPT almost always happens between a minimum of the form $\field_s\equiv (v_h^{\rm high}=0, v_s^{\rm high})$ and minimum of the form $\field_h \equiv (v_h^{\rm low}, v_s^{\rm low}=0)$, as shown in the top left panel in \cref{fig:phasestructure}. \Cref{fig:phasestructure} shows the evolution of minima of the effective potential with temperature for different cases. In all the cases, there is only one minimum at high temperature, $T>400\gev$. In the top left panel, a second-order phase transition occurs at $T\simeq 300\gev$ from the origin to a minimum at $(\phi_h=0, \phi_s\neq0)$, indicated by orange lines.  With temperature decreasing, a FOPT becomes possible below $T_c= 114\gev$, after which the minimum at $(\phi_h\neq0, \phi_s=0)$ becomes the deepest minimum of the effective potential. 

There are parameter points for which no FOPT occurs during the evolution of the potential with temperature. They may be split into three cases:
\begin{itemize}
    \item The minimum at $(\phi_h\neq0, \phi_s=0)$ is never the deepest minimum, so there is no transition from the minimum at $(\phi_h=0, \phi_s\neq0)$ to it, as shown in the top right panel of \cref{fig:phasestructure}. This happens for large $\lambda_{hs}$, small $\lambda_{s}$ and $m_S$, i.e.\ the top left corners of the $\lambda_s$--$\lambda_{hs}$ and $m_s$--$\lambda_{hs}$ planes in \cref{fig:xi_msbar_2d}.
    
    \item There is never a minimum at $(\phi_h=0, \phi_s\neq0)$ during the evolution, and the phase transition from the origin to $(\phi_h\neq0, \phi_s=0)$ is not a strong first-order transition. This is similar to the situation in the SM.  This case is displayed in the bottom left panel of \cref{fig:phasestructure}, and corresponds to small $\lambda_{hs}$ and large $\lambda_{s}$ and $M_S$. This kind of transition is ignored in 2-dimensional and 3-dimensional scans.
    
    \item The phase transition from  the minimum at $(\phi_h=0, \phi_s\neq0)$ to the minimum at $(\phi_h\neq0, \phi_s=0)$  is not first-order, which is presented in the bottom right panel of \cref{fig:phasestructure}. It is caused by large $\lambda_s$ and small $m_s$. 
\end{itemize}

In our calculations, there are more complicated phase structures than above cases, caused by numerical problems, such as the fact that the thermal functions are not smooth. To maintain consistency, unless otherwise stated we only compare and show the type of FOPT shown in the top left panel of \cref{fig:phasestructure} in our results.

\section{Benchmark point}
\label{appendix:BMs}

\begin{table}[th]
\centering
\begin{tabular}{|c|c|c|c|c|c|c|c|c|c|c|}
\noalign{\hrule height 1.5pt}
$m_s$ & $\lambda_s$ & $\lambda_{hs}$ & $\xi$ & $\mu_h^2$ & $\mu_s^2$ & $\lambda_h$ & $T_c$ & \gammaew & $v_s^{\rm high}(T_c)$ & $v_h^{\rm low}(T_c)$ \\ \noalign{\hrule height 1.5pt}
 &  &  & 0 & 8720.10 & 4860.43 & 0.129 & 113.54 & 1.79 & 211.04 & 203.53\\ \cline{4-11} 
 &  &  & 25 & 7721.33 & 5267.03 & 0.106 & 114.56 & -1.90 & 220.43 & -217.13 \\ \cline{4-11} 
\multirow{-3}{*}{65} & \multirow{-3}{*}{0.1} & \multirow{-3}{*}{0.3} &  & {\color{blue}998.8} & {\color{blue}-406.6} & {\color{blue}0.023} & {\color{blue}-1.016} & {\color{blue}3.69} & {\color{blue}-9.39} & {\color{blue}-13.60} \\ \noalign{\hrule height 1.5pt}
 &  &  & 0 & 8723.64 & 5722.60 & 0.129 & 94.90 & 2.36 & 234.75 & 223.58 \\ \cline{4-11} 
 &  &  & 25 & 7725.12 & 6128.86 & 0.106 & 82.86 & 2.87 & 245.66 & 237.40 \\ \cline{4-11} 
\multirow{-3}{*}{58} &  &  &  & {\color{blue} 998.53 } & {\color{blue}-406.26 } & {\color{blue}0.023 } & {\color{blue}12.04 } & {\color{blue}-0.51 } & {\color{blue}-10.91 } & {\color{blue}-13.82 } \\ \cline{1-1} \cline{4-11} 
 &  &  & 0 & 8716.74 & 3900.27 & 0.129 & 129.87 & 1.36 & 180.70 & 176.87 \\ \cline{4-11} 
 &  &  & 25 & 7717.74 & 4307.17 & 0.106 & 142.13 & -1.31 & 185.05 & -186.62 \\ \cline{4-11} 
\multirow{-3}{*}{72} & \multirow{-6}{*}{0.1} & \multirow{-6}{*}{0.3} &  & {\color{blue} 999.00 } & {\color{blue}-406.90 } & {\color{blue}0.023 } & {\color{blue}-12.26 } & {\color{blue}2.68 } & {\color{blue}-4.36 } & {\color{blue}-9.75} \\ \noalign{\hrule height 1.5pt}
 &  &  & 0 & 8720.11 & 4862.84 & 0.129 & 108.47 & 1.94 & 224.43 & 209.92 \\ \cline{4-11} 
 &  &  & 25 & 7721.34 & 5269.44 & 0.106 & 105.38 & 2.13 & 235.65 & 224.44 \\ \cline{4-11} 
 & \multirow{-3}{*}{0.09} &  &  & {\color{blue} 998.77 } & {\color{blue}-406.60 } & {\color{blue}0.023 } & {\color{blue}3.09 } & {\color{blue}-0.19 } & {\color{blue}-11.23 } & {\color{blue}-14.52 } \\ \cline{2-2} \cline{4-11} 
 &  &  & 0 & 8720.09 & 4858.02 & 0.129 & 117.62 & 1.68 & 199.57 & 197.81 \\ \cline{4-11} 
 &  &  & 25 & 7721.32 & 5264.62 & 0.106 & 121.80 & 1.73 & 207.02 & 210.41 \\ \cline{4-11} 
\multirow{-6}{*}{65} & \multirow{-3}{*}{0.11} & \multirow{-6}{*}{0.3} &  & {\color{blue} 998.77 } & {\color{blue}-406.59 } & {\color{blue}0.023 } & {\color{blue}-4.19 } & {\color{blue}-0.05 } & {\color{blue}-7.45 } & {\color{blue}-12.61 } \\ \noalign{\hrule height 1.5pt}
 &  &  & 0 & 8718.11 & 3953.10 & 0.129 & 128.62 & 1.40 & 183.27 & 179.53 \\ \cline{4-11} 
 &  &  & 25 & 7719.28 & 4319.66 & 0.106 & 140.41 & 1.35 & 185.10 & 189.13 \\ \cline{4-11} 
 &  & \multirow{-3}{*}{0.27} &  & {\color{blue} 998.83 } & {\color{blue}-366.56 } & {\color{blue}0.023 } & {\color{blue}-11.79 } & {\color{blue}0.05 } & {\color{blue}-1.82 } & {\color{blue}-9.60 } \\ \cline{3-11} 
 &  &  & 0 & 8722.40 & 5766.98 & 0.129 & 94.00 & -2.39 & 236.00 & 224.32 \\ \cline{4-11} 
 &  &  & 25 & 7723.71 & 6213.46 & 0.106 & 78.95 & -3.03 & 248.13 & -239.03 \\ \cline{4-11} 
\multirow{-6}{*}{65} & \multirow{-6}{*}{0.1} & \multirow{-3}{*}{0.33} &  & {\color{blue} 998.69 } & {\color{blue}-446.47 } & {\color{blue}0.023 } & {\color{blue}15.05 } & {\color{blue}0.64 } & {\color{blue}-12.13 } & {\color{blue}14.71 } \\ \noalign{\hrule height 1.5pt}
\end{tabular}
\caption{Values of Lagrangian parameters, critical temperature, \gammaew, and VEVs for points around the benchmark point in the \msbar + \texttt{GC\_SelfEnergy\_Sol} scheme. All dimensionful quantities are in $\gev$ or $\gev^2$.  The blue numbers show difference between quantities evaluated at $\xi=0$ and 25. }\label{tab:gauge}
\end{table}

In \cref{tab:gauge}, we present input parameters, parameters extracted from tadpole conditions, critical temperature and transition strength of the benchmark point used in \cref{sec:results}, for different gauge parameters. Here we used the \msbar scheme in \Rxi gauges, the Arnold-Espinosa technique to resum daisy terms, \texttt{GC\_SelfEnergy\_Sol} to solve the GC and set $Q = m_t$. 

\section{Mass spectrum} \label{App:MassSpectrum}

The field-dependent Higgs and singlet masses are given by the eigenvalues of the scalar mass matrix
\begin{equation}
    M_{\field}^2 \vp = 
\begin{pmatrix}
    \displaystyle \pdv[2]{V}{\phi_h} & \displaystyle \frac{\partial^2 V}{\partial \phi_h \partial \phi_s} \\[2.5ex]
    \displaystyle \frac{\partial^2 V}{\partial \phi_h \partial \phi_s} & \displaystyle \pdv[2]{V}{\phi_s}
\end{pmatrix} . \label{eq:massmat-scalar}
\end{equation}
When evaluated at the EW VEV, they are the masses of the Higgs and singlet. The off-diagonal terms vanish in the \ztwo-symmetric model because $v_s = 0$ at $T=0$, such that the masses are the simply the diagonal entries, which are given in \cref{eq:mh,eq:ms}. 
The $Z$ boson and photon field-dependent masses are similarly obtained as eigenvalues of the neutral gauge boson mass matrix
\begin{equation}
    M_{A^0}^2 \vp =
\begin{pmatrix}
    \displaystyle \recip{4} g^2 \phi_h^2 & \displaystyle -\recip{4} g g' \phi_h^2 \\[2.5ex]
    \displaystyle -\recip{4} g g' \phi_h^2 & \displaystyle \recip{4} g'^2 \phi_h^2
\end{pmatrix} . \label{eq:massmat-neutral}
\end{equation}
We use the third family approximation for the fermions, and so include the masses
\begin{equation}
m_{b,t,\tau}^2 \vp = \frac{y_{b,t,\tau}^2}{2} \phi_h^2 .
\end{equation}
This ignores fermions lighter than the bottom quark.

The scalars and longitudinal components of the gauge bosons receive Debye corrections at finite temperature. The Debye corrections are diagonal matrices that are added to the corresponding mass matrices before diagonalisation, with \cref{eq:massmat-scalar} becoming
\begin{equation}
    M_{\field}^2 (\field, T) =
\begin{pmatrix}
    \displaystyle \pdv[2]{V}{\phi_h} + c_h T^2 & \displaystyle \frac{\partial^2 V}{\partial \phi_h \partial \phi_s} \\[2.5ex]
    \displaystyle \frac{\partial^2 V}{\partial \phi_h \partial \phi_s} & \displaystyle \pdv[2]{V}{\phi_s} + c_s T^2
\end{pmatrix} , \label{eq:massmat-scalar-ft}
\end{equation}
where the scalar Debye coefficients are
\begin{align}\label{eq:debye_coefficient_h}
    c_h & = \recip{48} \left[9g^2 + 3g'^2 + 2 \left(6y_t^2 + 6y_b^2 + 2y_{\tau}^2 + 12\lambda_h + \lambda_{hs} \right) \right] , \\
    c_s & = \recip{12} \left[2\lambda_{hs} + 3\lambda_s \right] . \label{eq:debye_coefficient_s}
\end{align}
For the longitudinal components of the gauge bosons, \cref{eq:massmat-neutral} becomes
\begin{equation}
    M_{A^0_L}^2 \vp =
\begin{pmatrix}
    \displaystyle \recip{4} g^2 \phi_h^2 + \frac{11}{6} g^2 T^2 & \displaystyle -\recip{4} g g' \phi_h^2 \\[2.5ex]
    \displaystyle -\recip{4} g g' \phi_h^2 & \displaystyle \recip{4} g'^2 \phi_h^2 + \frac{11}{6} g'^2 T^2
\end{pmatrix} .
\end{equation}
Note that the ghosts do not get a thermal mass \cite{Laine:2016hma}.
Lastly, the Goldstone boson masses receive the Debye correction $c_h T^2$ in both the $\Rxi$ and covariant gauges, and the longitudinal component of the $W$ boson receives the Debye correction $\frac{11}{6} g^2 T^2$. Performing the diagonalisation leaves the following field-dependent thermal boson masses, 
\begin{align}
	\begin{split}
	    \overline{m}_h^2 (\field, T) & = -\left(\mu_h^2 + \mu_s^2 \right) + 3 \left(\lambda_h \phi_h^2 + \lambda_s \phi_s^2 \right) + \half \lambda_{hs} \left(\phi_h^2 + \phi_s^2 \right) + \left(c_h + c_s \right) T^2 \\
	    & \qquad {}+{} \left[\left( -\left(\mu_h^2 - \mu_s^2 \right) + 3\left(\lambda_h \phi_h^2 - \lambda_s \phi_s^2 \right) - \half \lambda_{hs} \left(\phi_h^2 - \phi_s^2 \right) + \left(c_h - c_s \right) T^2 \right)^2 \right. \\
	    & \left. \qquad\quad\quad {}+{} 4 \lambda_{hs}^2 \phi_h^2 \phi_s^2 \vphantom{\left(\half \right)^2} \right]^{\half} ,
	\end{split} \\
	\begin{split}
	    \overline{m}_s^2 (\field, T) & = -\left(\mu_h^2 + \mu_s^2 \right) + 3 \left(\lambda_h \phi_h^2 + \lambda_s \phi_s^2 \right) + \half \lambda_{hs} \left(\phi_h^2 + \phi_s^2 \right) + \left(c_h + c_s \right) T^2 \\
	    & \qquad {}-{} \left[\left( -\left(\mu_h^2 - \mu_s^2 \right) + 3\left(\lambda_h \phi_h^2 - \lambda_s \phi_s^2 \right) - \half \lambda_{hs} \left(\phi_h^2 - \phi_s^2 \right) + \left(c_h - c_s \right) T^2 \right)^2 \right. \\
	    & \left. \qquad\quad\quad {}+{} 4 \lambda_{hs}^2 \phi_h^2 \phi_s^2 \vphantom{\left(\half \right)^2} \right]^{\half} ,
	\end{split} \\
	m_G^2 (\field, T) & = \mu_h^2 + \lambda_h \phi_h^2 + \half \lambda_{hs} \phi_s^2 + c_h T^2, \\
	m_{W_T}^2 (\field) & = \frac{g^2}{4} \phi_h^2, \\
	m_{W_L}^2 (\field, T) & = \frac{g^2}{4} \phi_h^2 + \frac{11}{6} g^2 T^2 , \\
	m_{\gamma_T}^2 (\field) & = 0, \\
    \begin{split}
        m_{\gamma_L}^2 (\field, T) & = \recip{24} \left[\vphantom{\left(g^2 + g'^2 \right)^2} \left(g^2 + g'^2 \right) \left(3 \phi_h^2 + 22 T^2 \right) \right. \\
        & \left. \qquad\quad {}-{} \sqrt{9 \left(g^2 + g'^2 \right)^2 \phi_h^4 + 44 T^2 \left(g^2 - g'^2 \right)^2 \left(3 \phi_h^2 + 11 T^2 \right)} \right] ,
    \end{split}	\\
	m_{Z_T}^2 (\field) & = \frac{g^2 + g'^2}{4} \phi_h^2, \\	
    \begin{split}
        m_{Z_L}^2 (\field, T) & = \recip{24} \left[\vphantom{\left(g^2 + g'^2 \right)^2} \left(g^2 + g'^2 \right) \left(3 \phi_h^2 + 22 T^2 \right) \right. \\
        & \left. \qquad\quad {}+{} \sqrt{9 \left(g^2 + g'^2 \right)^2 \phi_h^4 + 44 T^2 \left(g^2 - g'^2 \right)^2 \left(3 \phi_h^2 + 11 T^2 \right)} \right] .
    \end{split}
\end{align}
At zero temperature, the photon mass vanishes as expected. The degrees of freedom for these masses are $n_h = n_s = 1$, $n_G = 3$, $n_{\gamma_L} = n_{Z_L} = 2$, $n_{\gamma_T} = n_{Z_T} = 1$, $n_{W_L} = 4$, $n_{W_T} = 2, $ $n_{\tau} = 4$, and $n_b = n_t = 12$. The degrees of freedom for the scalar particles differs in the covariant gauge, as discussed in \cref{sec:covariant}. 

\section{Numerical uncertainties}\label{app:numerical}

All the numerical results presented above are obtained using \texttt{PhaseTracer}~\cite{Athron:2020sbe}.
In the manual~\cite{Athron:2020sbe}, we have shown the numerical uncertainty in the calculation of a phase transition using the HT method and the SSM as an example. The differences between the analytic formulae and the numerical results from \pt are less than about 0.01\%. In this section, we further discuss numerical uncertainties and issues we encountered in this work.

The SM parameters involved in the effective potential models include mass of the SM-like Higgs $m_h$, SM EW VEV at zero-temperature $v$, the gauge coupling constants $g$ and $g'$, and the Yukawa coupling $y_t$, $y_b$ and $y_{\tau}$. Since we use \fs for our RGE running, we use the default values in \fs in all the models to be consistent, 
\begin{equation}
\begin{split}
    & m_h = 125.25\gev, ~ v = 247.4554\gev, ~ g = 0.6477, ~ g' = 0.3586, \\
    & y_t = 0.9341, ~ y_b = 0.01547, ~ y_\tau = 0.010014.
\end{split}
\end{equation}
Modifying these values will of course affect the values of $T_c$ ad \gammaew. However, by regenerating all the results with $m_h = 126.0\gev$ and $v = 246\gev$ we have checked that it will not qualitatively change our above discussions and conclusions. 

There is no closed-form solution for calculation of the thermal functions, $J_B(y^2)$ and $J_F(y^2)$, in \cref{eq:JB}. The precision of several numerical methods to do this have been fully discussed in \cite{Fowlie:2018eiu}. \pt adopts the same method as \texttt{CosmoTransitions}~\cite{Wainwright:2011kj}, calculating a series of points by quadrature integration in \texttt{SciPy} library for \texttt{Python} and then using B-spline interpolation to get thermal functions of any point. This method is very fast and pretty accurate except around $y^2=0$. We find that the second-order derivative of the thermal function obtained in this way is not continuous at $y^2=0$. Fortunately, it is very rare that the minima at $T_c$ encounter $y^2=0$, so the $T_c$ of the FOPT is not ruined.

Nevertheless, the discontinuity of the second-order derivatives of the thermal function around zero mass may introduce an additional artificial minimum in the effective potential, which will form an artificial phase. In some cases, there is a transition between this artificial phase and an ordinary phase, especially if the Arnold-Espinosa method is used for daisy resummation. The Parwani method can partially avoid this situation, because the Debye corrections to the field-dependent masses are positive such that we do not encounter zero masses at the minimum. In any case, the artificial phase cannot affect our results because we focus on the FOPT described in \cref{app:phasestructure}, and ignore other transitions. 

As already discussed, the number density of parameter points sampled in the scan will impact the maximum change in the critical temperature and transition strength found. Since the maximum change occurs when $T_c$ is very sensitive to the input parameters, the more detailed scan we have, the larger the maximum uncertainty we are likely to obtain. Besides, the local minimum finding algorithm adopted in \pt has large numerical uncertainty when the potential around the minimum tends to be flat, and has a chance of misidentifying a saddle point as a minimum. This causes noise in our plots. The more detailed scan we have, the more noisy points will appear. 

These numerical problems do not only exist in the SSM or only when using \pt. As the SSM is quite a simple model, we anticipate that further issues and larger uncertainties could be found in more complicated models. Therefore, they should be carefully considered in any perturbative calculations of phase transitions.

\addcontentsline{toc}{section}{References}
\bibliographystyle{JHEP}
\bibliography{bibliography}

\providecommand{\href}[2]{#2}\begingroup\raggedright\begin{thebibliography}{100}

\bibitem{Caprini:2019egz}
C.~Caprini et~al., \emph{{Detecting gravitational waves from cosmological phase
  transitions with LISA: an update}},
  \href{https://doi.org/10.1088/1475-7516/2020/03/024}{\emph{JCAP} {\bfseries
  03} (2020) 024}, [\href{https://arxiv.org/abs/1910.13125}{{\ttfamily
  1910.13125}}].

\bibitem{Ramsey-Musolf:2019lsf}
M.~J. Ramsey-Musolf, \emph{{The electroweak phase transition: a collider
  target}}, \href{https://doi.org/10.1007/JHEP09(2020)179}{\emph{JHEP}
  {\bfseries 09} (2020) 179},
  [\href{https://arxiv.org/abs/1912.07189}{{\ttfamily 1912.07189}}].

\bibitem{Barrow:2022gsu}
J.~L. Barrow et~al., \emph{{Theories and Experiments for Testable Baryogenesis
  Mechanisms: A Snowmass White Paper}},
  \href{https://arxiv.org/abs/2203.07059}{{\ttfamily 2203.07059}}.

\bibitem{Sakharov:1967dj}
A.~D. Sakharov, \emph{{Violation of CP Invariance, C asymmetry, and baryon
  asymmetry of the universe}},
  \href{https://doi.org/10.1070/PU1991v034n05ABEH002497}{\emph{Pisma Zh. Eksp.
  Teor. Fiz.} {\bfseries 5} (1967) 32--35}.

\bibitem{Kajantie:1995kf}
K.~Kajantie, M.~Laine, K.~Rummukainen and M.~E. Shaposhnikov, \emph{{The
  Electroweak phase transition: A Nonperturbative analysis}},
  \href{https://doi.org/10.1016/0550-3213(96)00052-1}{\emph{Nucl. Phys. B}
  {\bfseries 466} (1996) 189--258},
  [\href{https://arxiv.org/abs/hep-lat/9510020}{{\ttfamily hep-lat/9510020}}].

\bibitem{Kajantie:1996mn}
K.~Kajantie, M.~Laine, K.~Rummukainen and M.~E. Shaposhnikov, \emph{{Is there a
  hot electroweak phase transition at $m_H \gtrsim m_W$?}},
  \href{https://doi.org/10.1103/PhysRevLett.77.2887}{\emph{Phys. Rev. Lett.}
  {\bfseries 77} (1996) 2887--2890},
  [\href{https://arxiv.org/abs/hep-ph/9605288}{{\ttfamily hep-ph/9605288}}].

\bibitem{Kajantie:1996qd}
K.~Kajantie, M.~Laine, K.~Rummukainen and M.~E. Shaposhnikov, \emph{{A
  Nonperturbative analysis of the finite-$T$ phase transition in $\text{SU}(2)
  \times \text{U}(1)$ electroweak theory}},
  \href{https://doi.org/10.1016/S0550-3213(97)00164-8}{\emph{Nucl. Phys. B}
  {\bfseries 493} (1997) 413--438},
  [\href{https://arxiv.org/abs/hep-lat/9612006}{{\ttfamily hep-lat/9612006}}].

\bibitem{Csikor:1998eu}
F.~Csikor, Z.~Fodor and J.~Heitger, \emph{{Endpoint of the hot electroweak
  phase transition}},
  \href{https://doi.org/10.1103/PhysRevLett.82.21}{\emph{Phys. Rev. Lett.}
  {\bfseries 82} (1999) 21--24},
  [\href{https://arxiv.org/abs/hep-ph/9809291}{{\ttfamily hep-ph/9809291}}].

\bibitem{DOnofrio:2015gop}
M.~D'Onofrio and K.~Rummukainen, \emph{{Standard model cross-over on the
  lattice}}, \href{https://doi.org/10.1103/PhysRevD.93.025003}{\emph{Phys. Rev.
  D} {\bfseries 93} (2016) 025003},
  [\href{https://arxiv.org/abs/1508.07161}{{\ttfamily 1508.07161}}].

\bibitem{Pietroni:1992in}
M.~Pietroni, \emph{{The Electroweak phase transition in a nonminimal
  supersymmetric model}},
  \href{https://doi.org/10.1016/0550-3213(93)90635-3}{\emph{Nucl. Phys. B}
  {\bfseries 402} (1993) 27--45},
  [\href{https://arxiv.org/abs/hep-ph/9207227}{{\ttfamily hep-ph/9207227}}].

\bibitem{Cline:1996mga}
J.~M. Cline and P.-A. Lemieux, \emph{{Electroweak phase transition in two Higgs
  doublet models}}, \href{https://doi.org/10.1103/PhysRevD.55.3873}{\emph{Phys.
  Rev. D} {\bfseries 55} (1997) 3873--3881},
  [\href{https://arxiv.org/abs/hep-ph/9609240}{{\ttfamily hep-ph/9609240}}].

\bibitem{Ham:2004nv}
S.~W. Ham, S.~K. OH, C.~M. Kim, E.~J. Yoo and D.~Son, \emph{{Electroweak phase
  transition in a nonminimal supersymmetric model}},
  \href{https://doi.org/10.1103/PhysRevD.70.075001}{\emph{Phys. Rev. D}
  {\bfseries 70} (2004) 075001},
  [\href{https://arxiv.org/abs/hep-ph/0406062}{{\ttfamily hep-ph/0406062}}].

\bibitem{Funakubo:2005pu}
K.~Funakubo, S.~Tao and F.~Toyoda, \emph{{Phase transitions in the NMSSM}},
  \href{https://doi.org/10.1143/PTP.114.369}{\emph{Prog. Theor. Phys.}
  {\bfseries 114} (2005) 369--389},
  [\href{https://arxiv.org/abs/hep-ph/0501052}{{\ttfamily hep-ph/0501052}}].

\bibitem{Barger:2008jx}
V.~Barger, P.~Langacker, M.~McCaskey, M.~Ramsey-Musolf and G.~Shaughnessy,
  \emph{{Complex Singlet Extension of the Standard Model}},
  \href{https://doi.org/10.1103/PhysRevD.79.015018}{\emph{Phys. Rev. D}
  {\bfseries 79} (2009) 015018},
  [\href{https://arxiv.org/abs/0811.0393}{{\ttfamily 0811.0393}}].

\bibitem{Chung:2010cd}
D.~J.~H. Chung and A.~J. Long, \emph{{Electroweak Phase Transition in the
  $\mu\nu$SSM}}, \href{https://doi.org/10.1103/PhysRevD.81.123531}{\emph{Phys.
  Rev. D} {\bfseries 81} (2010) 123531},
  [\href{https://arxiv.org/abs/1004.0942}{{\ttfamily 1004.0942}}].

\bibitem{Espinosa:2011ax}
J.~R. Espinosa, T.~Konstandin and F.~Riva, \emph{{Strong Electroweak Phase
  Transitions in the Standard Model with a Singlet}},
  \href{https://doi.org/10.1016/j.nuclphysb.2011.09.010}{\emph{Nucl. Phys. B}
  {\bfseries 854} (2012) 592--630},
  [\href{https://arxiv.org/abs/1107.5441}{{\ttfamily 1107.5441}}].

\bibitem{Chowdhury:2011ga}
T.~A. Chowdhury, M.~Nemevsek, G.~Senjanovic and Y.~Zhang, \emph{{Dark Matter as
  the Trigger of Strong Electroweak Phase Transition}},
  \href{https://doi.org/10.1088/1475-7516/2012/02/029}{\emph{JCAP} {\bfseries
  02} (2012) 029}, [\href{https://arxiv.org/abs/1110.5334}{{\ttfamily
  1110.5334}}].

\bibitem{Gil:2012ya}
G.~Gil, P.~Chankowski and M.~Krawczyk, \emph{{Inert Dark Matter and Strong
  Electroweak Phase Transition}},
  \href{https://doi.org/10.1016/j.physletb.2012.09.052}{\emph{Phys. Lett. B}
  {\bfseries 717} (2012) 396--402},
  [\href{https://arxiv.org/abs/1207.0084}{{\ttfamily 1207.0084}}].

\bibitem{Carena:2012np}
M.~Carena, G.~Nardini, M.~Quiros and C.~E. Wagner, \emph{{MSSM Electroweak
  Baryogenesis and LHC Data}},
  \href{https://doi.org/10.1007/JHEP02(2013)001}{\emph{JHEP} {\bfseries 02}
  (2013) 001}, [\href{https://arxiv.org/abs/1207.6330}{{\ttfamily 1207.6330}}].

\bibitem{No:2013wsa}
J.~M. No and M.~Ramsey-Musolf, \emph{{Probing the Higgs Portal at the LHC
  Through Resonant di-Higgs Production}},
  \href{https://doi.org/10.1103/PhysRevD.89.095031}{\emph{Phys. Rev. D}
  {\bfseries 89} (2014) 095031},
  [\href{https://arxiv.org/abs/1310.6035}{{\ttfamily 1310.6035}}].

\bibitem{Dorsch:2013wja}
G.~C. Dorsch, S.~J. Huber and J.~M. No, \emph{{A strong electroweak phase
  transition in the 2HDM after LHC8}},
  \href{https://doi.org/10.1007/JHEP10(2013)029}{\emph{JHEP} {\bfseries 10}
  (2013) 029}, [\href{https://arxiv.org/abs/1305.6610}{{\ttfamily 1305.6610}}].

\bibitem{Curtin:2014jma}
D.~Curtin, P.~Meade and C.-T. Yu, \emph{{Testing Electroweak Baryogenesis with
  Future Colliders}},
  \href{https://doi.org/10.1007/JHEP11(2014)127}{\emph{JHEP} {\bfseries 11}
  (2014) 127}, [\href{https://arxiv.org/abs/1409.0005}{{\ttfamily 1409.0005}}].

\bibitem{Huang:2014ifa}
W.~Huang, Z.~Kang, J.~Shu, P.~Wu and J.~M. Yang, \emph{{New insights in the
  electroweak phase transition in the NMSSM}},
  \href{https://doi.org/10.1103/PhysRevD.91.025006}{\emph{Phys. Rev. D}
  {\bfseries 91} (2015) 025006},
  [\href{https://arxiv.org/abs/1405.1152}{{\ttfamily 1405.1152}}].

\bibitem{Profumo:2014opa}
S.~Profumo, M.~J. Ramsey-Musolf, C.~L. Wainwright and P.~Winslow,
  \emph{{Singlet-catalyzed electroweak phase transitions and precision Higgs
  boson studies}},
  \href{https://doi.org/10.1103/PhysRevD.91.035018}{\emph{Phys. Rev. D}
  {\bfseries 91} (2015) 035018},
  [\href{https://arxiv.org/abs/1407.5342}{{\ttfamily 1407.5342}}].

\bibitem{Kozaczuk:2014kva}
J.~Kozaczuk, S.~Profumo, L.~S. Haskins and C.~L. Wainwright,
  \emph{{Cosmological Phase Transitions and their Properties in the NMSSM}},
  \href{https://doi.org/10.1007/JHEP01(2015)144}{\emph{JHEP} {\bfseries 01}
  (2015) 144}, [\href{https://arxiv.org/abs/1407.4134}{{\ttfamily 1407.4134}}].

\bibitem{Jiang:2015cwa}
M.~Jiang, L.~Bian, W.~Huang and J.~Shu, \emph{{Impact of a complex singlet:
  Electroweak baryogenesis and dark matter}},
  \href{https://doi.org/10.1103/PhysRevD.93.065032}{\emph{Phys. Rev. D}
  {\bfseries 93} (2016) 065032},
  [\href{https://arxiv.org/abs/1502.07574}{{\ttfamily 1502.07574}}].

\bibitem{Curtin:2016urg}
D.~Curtin, P.~Meade and H.~Ramani, \emph{{Thermal Resummation and Phase
  Transitions}},
  \href{https://doi.org/10.1140/epjc/s10052-018-6268-0}{\emph{Eur. Phys. J. C}
  {\bfseries 78} (2018) 787},
  [\href{https://arxiv.org/abs/1612.00466}{{\ttfamily 1612.00466}}].

\bibitem{Vaskonen:2016yiu}
V.~Vaskonen, \emph{{Electroweak baryogenesis and gravitational waves from a
  real scalar singlet}},
  \href{https://doi.org/10.1103/PhysRevD.95.123515}{\emph{Phys. Rev. D}
  {\bfseries 95} (2017) 123515},
  [\href{https://arxiv.org/abs/1611.02073}{{\ttfamily 1611.02073}}].

\bibitem{Dorsch:2016nrg}
G.~Dorsch, S.~Huber, T.~Konstandin and J.~No, \emph{{A Second Higgs Doublet in
  the Early Universe: Baryogenesis and Gravitational Waves}},
  \href{https://doi.org/10.1088/1475-7516/2017/05/052}{\emph{JCAP} {\bfseries
  05} (2017) 052}, [\href{https://arxiv.org/abs/1611.05874}{{\ttfamily
  1611.05874}}].

\bibitem{Huang:2016cjm}
P.~Huang, A.~J. Long and L.-T. Wang, \emph{{Probing the Electroweak Phase
  Transition with Higgs Factories and Gravitational Waves}},
  \href{https://doi.org/10.1103/PhysRevD.94.075008}{\emph{Phys. Rev. D}
  {\bfseries 94} (2016) 075008},
  [\href{https://arxiv.org/abs/1608.06619}{{\ttfamily 1608.06619}}].

\bibitem{Chala:2016ykx}
M.~Chala, G.~Nardini and I.~Sobolev, \emph{{Unified explanation for dark matter
  and electroweak baryogenesis with direct detection and gravitational wave
  signatures}}, \href{https://doi.org/10.1103/PhysRevD.94.055006}{\emph{Phys.
  Rev. D} {\bfseries 94} (2016) 055006},
  [\href{https://arxiv.org/abs/1605.08663}{{\ttfamily 1605.08663}}].

\bibitem{Basler:2016obg}
P.~Basler, M.~Krause, M.~Muhlleitner, J.~Wittbrodt and A.~Wlotzka,
  \emph{{Strong First Order Electroweak Phase Transition in the CP-Conserving
  2HDM Revisited}}, \href{https://doi.org/10.1007/JHEP02(2017)121}{\emph{JHEP}
  {\bfseries 02} (2017) 121},
  [\href{https://arxiv.org/abs/1612.04086}{{\ttfamily 1612.04086}}].

\bibitem{Beniwal:2017eik}
A.~Beniwal, M.~Lewicki, J.~D. Wells, M.~White and A.~G. Williams,
  \emph{{Gravitational wave, collider and dark matter signals from a scalar
  singlet electroweak baryogenesis}},
  \href{https://doi.org/10.1007/JHEP08(2017)108}{\emph{JHEP} {\bfseries 08}
  (2017) 108}, [\href{https://arxiv.org/abs/1702.06124}{{\ttfamily
  1702.06124}}].

\bibitem{Bernon:2017jgv}
J.~Bernon, L.~Bian and Y.~Jiang, \emph{{A new insight into the phase transition
  in the early Universe with two Higgs doublets}},
  \href{https://doi.org/10.1007/JHEP05(2018)151}{\emph{JHEP} {\bfseries 05}
  (2018) 151}, [\href{https://arxiv.org/abs/1712.08430}{{\ttfamily
  1712.08430}}].

\bibitem{Kurup:2017dzf}
G.~Kurup and M.~Perelstein, \emph{{Dynamics of Electroweak Phase Transition In
  Singlet-Scalar Extension of the Standard Model}},
  \href{https://doi.org/10.1103/PhysRevD.96.015036}{\emph{Phys. Rev. D}
  {\bfseries 96} (2017) 015036},
  [\href{https://arxiv.org/abs/1704.03381}{{\ttfamily 1704.03381}}].

\bibitem{Andersen:2017ika}
J.~O. Andersen, T.~Gorda, A.~Helset, L.~Niemi, T.~V.~I. Tenkanen, A.~Tranberg
  et~al., \emph{{Nonperturbative Analysis of the Electroweak Phase Transition
  in the Two Higgs Doublet Model}},
  \href{https://doi.org/10.1103/PhysRevLett.121.191802}{\emph{Phys. Rev. Lett.}
  {\bfseries 121} (2018) 191802},
  [\href{https://arxiv.org/abs/1711.09849}{{\ttfamily 1711.09849}}].

\bibitem{Chiang:2017nmu}
C.-W. Chiang, M.~J. Ramsey-Musolf and E.~Senaha, \emph{{Standard Model with a
  Complex Scalar Singlet: Cosmological Implications and Theoretical
  Considerations}},
  \href{https://doi.org/10.1103/PhysRevD.97.015005}{\emph{Phys. Rev. D}
  {\bfseries 97} (2018) 015005},
  [\href{https://arxiv.org/abs/1707.09960}{{\ttfamily 1707.09960}}].

\bibitem{Dorsch:2017nza}
G.~C. Dorsch, S.~J. Huber, K.~Mimasu and J.~M. No, \emph{{The Higgs Vacuum
  Uplifted: Revisiting the Electroweak Phase Transition with a Second Higgs
  Doublet}}, \href{https://doi.org/10.1007/JHEP12(2017)086}{\emph{JHEP}
  {\bfseries 12} (2017) 086},
  [\href{https://arxiv.org/abs/1705.09186}{{\ttfamily 1705.09186}}].

\bibitem{Beniwal:2018hyi}
A.~Beniwal, M.~Lewicki, M.~White and A.~G. Williams, \emph{{Gravitational waves
  and electroweak baryogenesis in a global study of the extended scalar singlet
  model}}, \href{https://doi.org/10.1007/JHEP02(2019)183}{\emph{JHEP}
  {\bfseries 02} (2019) 183},
  [\href{https://arxiv.org/abs/1810.02380}{{\ttfamily 1810.02380}}].

\bibitem{Bruggisser:2018mrt}
S.~Bruggisser, B.~Von~Harling, O.~Matsedonskyi and G.~Servant,
  \emph{{Electroweak Phase Transition and Baryogenesis in Composite Higgs
  Models}}, \href{https://doi.org/10.1007/JHEP12(2018)099}{\emph{JHEP}
  {\bfseries 12} (2018) 099},
  [\href{https://arxiv.org/abs/1804.07314}{{\ttfamily 1804.07314}}].

\bibitem{Athron:2019teq}
P.~Athron, C.~Balazs, A.~Fowlie, G.~Pozzo, G.~White and Y.~Zhang, \emph{{Strong
  first-order phase transitions in the NMSSM --- a comprehensive survey}},
  \href{https://doi.org/10.1007/JHEP11(2019)151}{\emph{JHEP} {\bfseries 11}
  (2019) 151}, [\href{https://arxiv.org/abs/1908.11847}{{\ttfamily
  1908.11847}}].

\bibitem{Kainulainen:2019kyp}
K.~Kainulainen, V.~Keus, L.~Niemi, K.~Rummukainen, T.~V.~I. Tenkanen and
  V.~Vaskonen, \emph{{On the validity of perturbative studies of the
  electroweak phase transition in the Two Higgs Doublet model}},
  \href{https://doi.org/10.1007/JHEP06(2019)075}{\emph{JHEP} {\bfseries 06}
  (2019) 075}, [\href{https://arxiv.org/abs/1904.01329}{{\ttfamily
  1904.01329}}].

\bibitem{Bian:2019kmg}
L.~Bian, Y.~Wu and K.-P. Xie, \emph{{Electroweak phase transition with
  composite Higgs models: calculability, gravitational waves and collider
  searches}}, \href{https://doi.org/10.1007/JHEP12(2019)028}{\emph{JHEP}
  {\bfseries 12} (2019) 028},
  [\href{https://arxiv.org/abs/1909.02014}{{\ttfamily 1909.02014}}].

\bibitem{Li:2019tfd}
H.-L. Li, M.~Ramsey-Musolf and S.~Willocq, \emph{{Probing a scalar
  singlet-catalyzed electroweak phase transition with resonant di-Higgs boson
  production in the $4b$ channel}},
  \href{https://doi.org/10.1103/PhysRevD.100.075035}{\emph{Phys. Rev. D}
  {\bfseries 100} (2019) 075035},
  [\href{https://arxiv.org/abs/1906.05289}{{\ttfamily 1906.05289}}].

\bibitem{Chiang:2019oms}
C.-W. Chiang and B.-Q. Lu, \emph{{First-order electroweak phase transition in a
  complex singlet model with $\mathbb{Z}_3$ symmetry}},
  \href{https://doi.org/10.1007/JHEP07(2020)082}{\emph{JHEP} {\bfseries 07}
  (2020) 082}, [\href{https://arxiv.org/abs/1912.12634}{{\ttfamily
  1912.12634}}].

\bibitem{Xie:2020bkl}
K.-P. Xie, L.~Bian and Y.~Wu, \emph{{Electroweak baryogenesis and gravitational
  waves in a composite Higgs model with high dimensional fermion
  representations}}, \href{https://doi.org/10.1007/JHEP12(2020)047}{\emph{JHEP}
  {\bfseries 12} (2020) 047},
  [\href{https://arxiv.org/abs/2005.13552}{{\ttfamily 2005.13552}}].

\bibitem{Bell:2020gug}
N.~F. Bell, M.~J. Dolan, L.~S. Friedrich, M.~J. Ramsey-Musolf and R.~R. Volkas,
  \emph{{Two-Step Electroweak Symmetry-Breaking: Theory Meets Experiment}},
  \href{https://doi.org/10.1007/JHEP05(2020)050}{\emph{JHEP} {\bfseries 05}
  (2020) 050}, [\href{https://arxiv.org/abs/2001.05335}{{\ttfamily
  2001.05335}}].

\bibitem{Papaefstathiou:2020iag}
A.~Papaefstathiou and G.~White, \emph{{The electro-weak phase transition at
  colliders: confronting theoretical uncertainties and complementary
  channels}}, \href{https://doi.org/10.1007/JHEP05(2021)099}{\emph{JHEP}
  {\bfseries 05} (2021) 099},
  [\href{https://arxiv.org/abs/2010.00597}{{\ttfamily 2010.00597}}].

\bibitem{Kotwal:2016tex}
A.~V. Kotwal, M.~J. Ramsey-Musolf, J.~M. No and P.~Winslow,
  \emph{{Singlet-catalyzed electroweak phase transitions in the 100 TeV
  frontier}}, \href{https://doi.org/10.1103/PhysRevD.94.035022}{\emph{Phys.
  Rev. D} {\bfseries 94} (2016) 035022},
  [\href{https://arxiv.org/abs/1605.06123}{{\ttfamily 1605.06123}}].

\bibitem{Patel:2011th}
H.~H. Patel and M.~J. Ramsey-Musolf, \emph{{Baryon Washout, Electroweak Phase
  Transition, and Perturbation Theory}},
  \href{https://doi.org/10.1007/JHEP07(2011)029}{\emph{JHEP} {\bfseries 07}
  (2011) 029}, [\href{https://arxiv.org/abs/1101.4665}{{\ttfamily 1101.4665}}].

\bibitem{Martin:2001vx}
S.~P. Martin, \emph{{Two Loop Effective Potential for a General Renormalizable
  Theory and Softly Broken Supersymmetry}},
  \href{https://doi.org/10.1103/PhysRevD.65.116003}{\emph{Phys. Rev. D}
  {\bfseries 65} (2002) 116003},
  [\href{https://arxiv.org/abs/hep-ph/0111209}{{\ttfamily hep-ph/0111209}}].

\bibitem{McKeon:2015zxa}
D.~G.~C. McKeon, \emph{{Renormalization Scheme Dependence with Renormalization
  Group Summation}},
  \href{https://doi.org/10.1103/PhysRevD.92.045031}{\emph{Phys. Rev. D}
  {\bfseries 92} (2015) 045031},
  [\href{https://arxiv.org/abs/1503.03823}{{\ttfamily 1503.03823}}].

\bibitem{Coleman:1973jx}
S.~R. Coleman and E.~J. Weinberg, \emph{{Radiative Corrections as the Origin of
  Spontaneous Symmetry Breaking}},
  \href{https://doi.org/10.1103/PhysRevD.7.1888}{\emph{Phys. Rev. D} {\bfseries
  7} (1973) 1888--1910}.

\bibitem{Arnold:1992fb}
P.~B. Arnold, \emph{{Phase transition temperatures at next-to-leading order}},
  \href{https://doi.org/10.1103/PhysRevD.46.2628}{\emph{Phys. Rev. D}
  {\bfseries 46} (1992) 2628--2635},
  [\href{https://arxiv.org/abs/hep-ph/9204228}{{\ttfamily hep-ph/9204228}}].

\bibitem{Gould:2022ran}
O.~Gould, S.~G\"uyer and K.~Rummukainen, \emph{{First-order electroweak phase
  transitions: a nonperturbative update}},
  \href{https://arxiv.org/abs/2205.07238}{{\ttfamily 2205.07238}}.

\bibitem{Schicho:2022wty}
P.~Schicho, T.~V.~I. Tenkanen and G.~White, \emph{{Combining thermal
  resummation and gauge invariance for electroweak phase transition}},
  \href{https://arxiv.org/abs/2203.04284}{{\ttfamily 2203.04284}}.

\bibitem{Ekstedt:2022ceo}
A.~Ekstedt, \emph{{Convergence of the nucleation rate for first-order phase
  transitions}},  \href{https://arxiv.org/abs/2205.05145}{{\ttfamily
  2205.05145}}.

\bibitem{Ekstedt:2022zro}
A.~Ekstedt, O.~Gould and J.~L\"ofgren, \emph{{Radiative first-order phase
  transitions to next-to-next-to-leading-order}},
  \href{https://arxiv.org/abs/2205.07241}{{\ttfamily 2205.07241}}.

\bibitem{Ekstedt:2022bff}
A.~Ekstedt, P.~Schicho and T.~V.~I. Tenkanen, \emph{{DRalgo: a package for
  effective field theory approach for thermal phase transitions}},
  \href{https://arxiv.org/abs/2205.08815}{{\ttfamily 2205.08815}}.

\bibitem{Garny:2012cg}
M.~Garny and T.~Konstandin, \emph{{On the gauge dependence of vacuum
  transitions at finite temperature}},
  \href{https://doi.org/10.1007/JHEP07(2012)189}{\emph{JHEP} {\bfseries 07}
  (2012) 189}, [\href{https://arxiv.org/abs/1205.3392}{{\ttfamily 1205.3392}}].

\bibitem{Fuyuto:2014yia}
K.~Fuyuto and E.~Senaha, \emph{{Improved sphaleron decoupling condition and the
  Higgs coupling constants in the real singlet-extended standard model}},
  \href{https://doi.org/10.1103/PhysRevD.90.015015}{\emph{Phys. Rev. D}
  {\bfseries 90} (2014) 015015},
  [\href{https://arxiv.org/abs/1406.0433}{{\ttfamily 1406.0433}}].

\bibitem{Fuyuto:2015jha}
K.~Fuyuto and E.~Senaha, \emph{{Sphaleron and critical bubble in the scale
  invariant two Higgs doublet model}},
  \href{https://doi.org/10.1016/j.physletb.2015.05.061}{\emph{Phys. Lett. B}
  {\bfseries 747} (2015) 152--157},
  [\href{https://arxiv.org/abs/1504.04291}{{\ttfamily 1504.04291}}].

\bibitem{Athron:2020sbe}
P.~Athron, C.~Bal\'azs, A.~Fowlie and Y.~Zhang, \emph{{PhaseTracer: tracing
  cosmological phases and calculating transition properties}},
  \href{https://doi.org/10.1140/epjc/s10052-020-8035-2}{\emph{Eur. Phys. J. C}
  {\bfseries 80} (2020) 567},
  [\href{https://arxiv.org/abs/2003.02859}{{\ttfamily 2003.02859}}].

\bibitem{Chiang:2017zbz}
C.-W. Chiang and E.~Senaha, \emph{{On gauge dependence of gravitational waves
  from a first-order phase transition in classical scale-invariant $U(1)'$
  models}}, \href{https://doi.org/10.1016/j.physletb.2017.09.064}{\emph{Phys.
  Lett. B} {\bfseries 774} (2017) 489--493},
  [\href{https://arxiv.org/abs/1707.06765}{{\ttfamily 1707.06765}}].

\bibitem{Arunasalam:2021zrs}
S.~Arunasalam and M.~J. Ramsey-Musolf, \emph{{Tunneling Potentials for the
  Tunneling Action: Gauge Invariance}},
  \href{https://arxiv.org/abs/2105.07588}{{\ttfamily 2105.07588}}.

\bibitem{Chiang:2018gsn}
C.-W. Chiang, Y.-T. Li and E.~Senaha, \emph{{Revisiting electroweak phase
  transition in the standard model with a real singlet scalar}},
  \href{https://doi.org/10.1016/j.physletb.2018.12.017}{\emph{Phys. Lett. B}
  {\bfseries 789} (2019) 154--159},
  [\href{https://arxiv.org/abs/1808.01098}{{\ttfamily 1808.01098}}].

\bibitem{Croon:2020cgk}
D.~Croon, O.~Gould, P.~Schicho, T.~V.~I. Tenkanen and G.~White,
  \emph{{Theoretical uncertainties for cosmological first-order phase
  transitions}}, \href{https://doi.org/10.1007/JHEP04(2021)055}{\emph{JHEP}
  {\bfseries 04} (2021) 055},
  [\href{https://arxiv.org/abs/2009.10080}{{\ttfamily 2009.10080}}].

\bibitem{Gould:2021oba}
O.~Gould and T.~V.~I. Tenkanen, \emph{{On the perturbative expansion at high
  temperature and implications for cosmological phase transitions}},
  \href{https://doi.org/10.1007/JHEP06(2021)069}{\emph{JHEP} {\bfseries 06}
  (2021) 069}, [\href{https://arxiv.org/abs/2104.04399}{{\ttfamily
  2104.04399}}].

\bibitem{Jackiw:1974cv}
R.~Jackiw, \emph{{Functional evaluation of the effective potential}},
  \href{https://doi.org/10.1103/PhysRevD.9.1686}{\emph{Phys. Rev. D} {\bfseries
  9} (1974) 1686}.

\bibitem{PhysRevD.9.3320}
L.~Dolan and R.~Jackiw, \emph{Symmetry behavior at finite temperature},
  \href{https://doi.org/10.1103/PhysRevD.9.3320}{\emph{Phys. Rev. D} {\bfseries
  9} (Jun, 1974) 3320--3341}.

\bibitem{Weinberg:1974hy}
S.~Weinberg, \emph{{Gauge and Global Symmetries at High Temperature}},
  \href{https://doi.org/10.1103/PhysRevD.9.3357}{\emph{Phys. Rev. D} {\bfseries
  9} (1974) 3357--3378}.

\bibitem{Kirzhnits:1974as}
D.~A. Kirzhnits and A.~D. Linde, \emph{{A Relativistic phase transition}},
  {\emph{Zh. Eksp. Teor. Fiz.} {\bfseries 67} (1974) 1263--1275}.

\bibitem{Ekstedt:2021kyx}
A.~Ekstedt, \emph{{Higher-order corrections to the bubble-nucleation rate at
  finite temperature}},
  \href{https://doi.org/10.1140/epjc/s10052-022-10130-5}{\emph{Eur. Phys. J. C}
  {\bfseries 82} (2022) 173},
  [\href{https://arxiv.org/abs/2104.11804}{{\ttfamily 2104.11804}}].

\bibitem{Elias-Miro:2014pca}
J.~Elias-Miro, J.~R. Espinosa and T.~Konstandin, \emph{{Taming Infrared
  Divergences in the Effective Potential}},
  \href{https://doi.org/10.1007/JHEP08(2014)034}{\emph{JHEP} {\bfseries 08}
  (2014) 034}, [\href{https://arxiv.org/abs/1406.2652}{{\ttfamily 1406.2652}}].

\bibitem{Martin:2014bca}
S.~P. Martin, \emph{{Taming the Goldstone contributions to the effective
  potential}}, \href{https://doi.org/10.1103/PhysRevD.90.016013}{\emph{Phys.
  Rev. D} {\bfseries 90} (2014) 016013},
  [\href{https://arxiv.org/abs/1406.2355}{{\ttfamily 1406.2355}}].

\bibitem{PhysRevD.97.056020}
J.~R. Espinosa and T.~Konstandin, \emph{{Resummation of Goldstone Infrared
  Divergences: A Proof to All Orders}},
  \href{https://doi.org/10.1103/PhysRevD.97.056020}{\emph{Phys. Rev. D}
  {\bfseries 97} (2018) 056020},
  [\href{https://arxiv.org/abs/1712.08068}{{\ttfamily 1712.08068}}].

\bibitem{Ghorbani:2020xqv}
P.~Ghorbani, \emph{{Vacuum structure and electroweak phase transition in
  singlet scalar dark matter}},
  \href{https://doi.org/10.1016/j.dark.2021.100861}{\emph{Phys. Dark Univ.}
  {\bfseries 33} (2021) 100861},
  [\href{https://arxiv.org/abs/2010.15708}{{\ttfamily 2010.15708}}].

\bibitem{ParticleDataGroup:2020ssz}
{\scshape Particle Data Group} collaboration, P.~A. Zyla et~al., \emph{{Review
  of Particle Physics}},
  \href{https://doi.org/10.1093/ptep/ptaa104}{\emph{PTEP} {\bfseries 2020}
  (2020) 083C01}.

\bibitem{Fujikawa:1972fe}
K.~Fujikawa, B.~W. Lee and A.~I. Sanda, \emph{{Generalized Renormalizable Gauge
  Formulation of Spontaneously Broken Gauge Theories}},
  \href{https://doi.org/10.1103/PhysRevD.6.2923}{\emph{Phys. Rev. D} {\bfseries
  6} (1972) 2923--2943}.

\bibitem{Martin:2018emo}
S.~P. Martin and H.~H. Patel, \emph{{Two-loop effective potential for
  generalized gauge fixing}},
  \href{https://doi.org/10.1103/PhysRevD.98.076008}{\emph{Phys. Rev. D}
  {\bfseries 98} (2018) 076008},
  [\href{https://arxiv.org/abs/1808.07615}{{\ttfamily 1808.07615}}].

\bibitem{Laine:1994bf}
M.~Laine, \emph{{The Two loop effective potential of the 3-d SU(2) Higgs model
  in a general covariant gauge}},
  \href{https://doi.org/10.1016/0370-2693(94)91409-5}{\emph{Phys. Lett. B}
  {\bfseries 335} (1994) 173--178},
  [\href{https://arxiv.org/abs/hep-ph/9406268}{{\ttfamily hep-ph/9406268}}].

\bibitem{Andreassen:2013hpa}
A.~J. Andreassen, \emph{{Gauge Dependence of the Quantum Field Theory Effective
  Potential}},  Master's thesis, Norwegian U. Sci. Tech., 2013.

\bibitem{Andreassen:2014gha}
A.~Andreassen, W.~Frost and M.~D. Schwartz, \emph{{Consistent Use of the
  Standard Model Effective Potential}},
  \href{https://doi.org/10.1103/PhysRevLett.113.241801}{\emph{Phys. Rev. Lett.}
  {\bfseries 113} (2014) 241801},
  [\href{https://arxiv.org/abs/1408.0292}{{\ttfamily 1408.0292}}].

\bibitem{Andreassen:2014eha}
A.~Andreassen, W.~Frost and M.~D. Schwartz, \emph{{Consistent Use of Effective
  Potentials}}, \href{https://doi.org/10.1103/PhysRevD.91.016009}{\emph{Phys.
  Rev. D} {\bfseries 91} (2015) 016009},
  [\href{https://arxiv.org/abs/1408.0287}{{\ttfamily 1408.0287}}].

\bibitem{Nielsen:1975fs}
N.~Nielsen, \emph{{On the Gauge Dependence of Spontaneous Symmetry Breaking in
  Gauge Theories}},
  \href{https://doi.org/10.1016/0550-3213(75)90301-6}{\emph{Nucl. Phys. B}
  {\bfseries 101} (1975) 173--188}.

\bibitem{Fleischer:1980ub}
J.~Fleischer and F.~Jegerlehner, \emph{{Radiative Corrections to Higgs Decays
  in the Extended Weinberg-Salam Model}},
  \href{https://doi.org/10.1103/PhysRevD.23.2001}{\emph{Phys. Rev. D}
  {\bfseries 23} (1981) 2001--2026}.

\bibitem{Braathen:2021fyq}
J.~Braathen, M.~D. Goodsell, S.~Pa\ss{}ehr and E.~Pinsard, \emph{{Expectation
  management}},
  \href{https://doi.org/10.1140/epjc/s10052-021-09285-4}{\emph{Eur. Phys. J. C}
  {\bfseries 81} (2021) 498},
  [\href{https://arxiv.org/abs/2103.06773}{{\ttfamily 2103.06773}}].

\bibitem{Ekstedt:2018ftj}
A.~Ekstedt and J.~L\"ofgren, \emph{{On the relationship between gauge
  dependence and IR divergences in the $\hbar$-expansion of the effective
  potential}}, \href{https://doi.org/10.1007/JHEP01(2019)226}{\emph{JHEP}
  {\bfseries 01} (2019) 226},
  [\href{https://arxiv.org/abs/1810.01416}{{\ttfamily 1810.01416}}].

\bibitem{Anderson:1991zb}
G.~W. Anderson and L.~J. Hall, \emph{{The Electroweak phase transition and
  baryogenesis}}, \href{https://doi.org/10.1103/PhysRevD.45.2685}{\emph{Phys.
  Rev. D} {\bfseries 45} (1992) 2685--2698}.

\bibitem{Megevand:2007sv}
A.~Megevand and A.~D. Sanchez, \emph{{Supercooling and phase coexistence in
  cosmological phase transitions}},
  \href{https://doi.org/10.1103/PhysRevD.77.063519}{\emph{Phys. Rev. D}
  {\bfseries 77} (2008) 063519},
  [\href{https://arxiv.org/abs/0712.1031}{{\ttfamily 0712.1031}}].

\bibitem{Ashoorioon:2009nf}
A.~Ashoorioon and T.~Konstandin, \emph{{Strong electroweak phase transitions
  without collider traces}},
  \href{https://doi.org/10.1088/1126-6708/2009/07/086}{\emph{JHEP} {\bfseries
  07} (2009) 086}, [\href{https://arxiv.org/abs/0904.0353}{{\ttfamily
  0904.0353}}].

\bibitem{Megevand:2014dua}
A.~Megevand, F.~A. Membiela and A.~D. Sanchez, \emph{{Lower bound on the
  electroweak wall velocity from hydrodynamic instability}},
  \href{https://doi.org/10.1088/1475-7516/2015/03/051}{\emph{JCAP} {\bfseries
  03} (2015) 051}, [\href{https://arxiv.org/abs/1412.8064}{{\ttfamily
  1412.8064}}].

\bibitem{Alanne:2019bsm}
T.~Alanne, T.~Hugle, M.~Platscher and K.~Schmitz, \emph{{A fresh look at the
  gravitational-wave signal from cosmological phase transitions}},
  \href{https://doi.org/10.1007/JHEP03(2020)004}{\emph{JHEP} {\bfseries 03}
  (2020) 004}, [\href{https://arxiv.org/abs/1909.11356}{{\ttfamily
  1909.11356}}].

\bibitem{Abdussalam:2020ssl}
S.~Abdussalam, M.~J. Kazemi and L.~Kalhor, \emph{{Upper limit on first-order
  electroweak phase transition strength}},
  \href{https://doi.org/10.1142/S0217751X21920032}{\emph{Int. J. Mod. Phys. A}
  {\bfseries 36} (2021) 2150024},
  [\href{https://arxiv.org/abs/2001.05973}{{\ttfamily 2001.05973}}].

\bibitem{Xie:2020wzn}
K.-P. Xie, \emph{{Lepton-mediated electroweak baryogenesis, gravitational waves
  and the $4\tau$ final state at the collider}},
  \href{https://doi.org/10.1007/JHEP02(2021)090}{\emph{JHEP} {\bfseries 02}
  (2021) 090}, [\href{https://arxiv.org/abs/2011.04821}{{\ttfamily
  2011.04821}}].

\bibitem{Azatov:2022tii}
A.~Azatov, G.~Barni, S.~Chakraborty, M.~Vanvlasselaer and W.~Yin,
  \emph{{Ultra-relativistic bubbles from the simplest Higgs portal and their
  cosmological consequences}},
  \href{https://doi.org/10.1007/JHEP10(2022)017}{\emph{JHEP} {\bfseries 10}
  (2022) 017}, [\href{https://arxiv.org/abs/2207.02230}{{\ttfamily
  2207.02230}}].

\bibitem{Ellis:2022lft}
J.~Ellis, M.~Lewicki, M.~Merchand, J.~M. No and M.~Zych, \emph{{The Scalar
  Singlet Extension of the Standard Model: Gravitational Waves versus
  Baryogenesis}},  \href{https://arxiv.org/abs/2210.16305}{{\ttfamily
  2210.16305}}.

\bibitem{Parwani:1991gq}
R.~R. Parwani, \emph{{Resummation in a hot scalar field theory}},
  \href{https://doi.org/10.1103/PhysRevD.45.4695}{\emph{Phys. Rev. D}
  {\bfseries 45} (1992) 4695},
  [\href{https://arxiv.org/abs/hep-ph/9204216}{{\ttfamily hep-ph/9204216}}].

\bibitem{Dine:1992vs}
M.~Dine, R.~G. Leigh, P.~Huet, A.~D. Linde and D.~A. Linde, \emph{{Comments on
  the electroweak phase transition}},
  \href{https://doi.org/10.1016/0370-2693(92)90026-Z}{\emph{Phys. Lett. B}
  {\bfseries 283} (1992) 319--325},
  [\href{https://arxiv.org/abs/hep-ph/9203201}{{\ttfamily hep-ph/9203201}}].

\bibitem{Dine:1992wr}
M.~Dine, R.~G. Leigh, P.~Y. Huet, A.~D. Linde and D.~A. Linde, \emph{{Towards
  the theory of the electroweak phase transition}},
  \href{https://doi.org/10.1103/PhysRevD.46.550}{\emph{Phys. Rev. D} {\bfseries
  46} (1992) 550--571}, [\href{https://arxiv.org/abs/hep-ph/9203203}{{\ttfamily
  hep-ph/9203203}}].

\bibitem{Arnold:1992rz}
P.~B. Arnold and O.~Espinosa, \emph{{The Effective potential and first order
  phase transitions: Beyond leading-order}},
  \href{https://doi.org/10.1103/PhysRevD.47.3546}{\emph{Phys. Rev. D}
  {\bfseries 47} (1993) 3546},
  [\href{https://arxiv.org/abs/hep-ph/9212235}{{\ttfamily hep-ph/9212235}}].

\bibitem{Schicho:2021gca}
P.~M. Schicho, T.~V.~I. Tenkanen and J.~\"Osterman, \emph{{Robust approach to
  thermal resummation: Standard Model meets a singlet}},
  \href{https://doi.org/10.1007/JHEP06(2021)130}{\emph{JHEP} {\bfseries 06}
  (2021) 130}, [\href{https://arxiv.org/abs/2102.11145}{{\ttfamily
  2102.11145}}].

\bibitem{Niemi:2021qvp}
L.~Niemi, P.~Schicho and T.~V.~I. Tenkanen, \emph{{Singlet-assisted electroweak
  phase transition at two loops}},
  \href{https://doi.org/10.1103/PhysRevD.103.115035}{\emph{Phys. Rev. D}
  {\bfseries 103} (2021) 115035},
  [\href{https://arxiv.org/abs/2103.07467}{{\ttfamily 2103.07467}}].

\bibitem{Athron:2014yba}
P.~Athron, J.-h. Park, D.~St\"ockinger and A.~Voigt,
  \emph{{FlexibleSUSY\textemdash{}A spectrum generator generator for
  supersymmetric models}},
  \href{https://doi.org/10.1016/j.cpc.2014.12.020}{\emph{Comput. Phys. Commun.}
  {\bfseries 190} (2015) 139--172},
  [\href{https://arxiv.org/abs/1406.2319}{{\ttfamily 1406.2319}}].

\bibitem{Athron:2017fvs}
P.~Athron, M.~Bach, D.~Harries, T.~Kwasnitza, J.-h. Park, D.~St\"ockinger
  et~al., \emph{{FlexibleSUSY 2.0: Extensions to investigate the phenomenology
  of SUSY and non-SUSY models}},
  \href{https://doi.org/10.1016/j.cpc.2018.04.016}{\emph{Comput. Phys. Commun.}
  {\bfseries 230} (2018) 145--217},
  [\href{https://arxiv.org/abs/1710.03760}{{\ttfamily 1710.03760}}].

\bibitem{Staub:2009bi}
F.~Staub, \emph{{From Superpotential to Model Files for FeynArts and
  CalcHep/CompHep}},
  \href{https://doi.org/10.1016/j.cpc.2010.01.011}{\emph{Comput.Phys.Commun.}
  {\bfseries 181} (2010) 1077--1086},
  [\href{https://arxiv.org/abs/0909.2863}{{\ttfamily 0909.2863}}].

\bibitem{Staub:2010jh}
F.~Staub, \emph{{Automatic Calculation of supersymmetric Renormalization Group
  Equations and Self Energies}},
  \href{https://doi.org/10.1016/j.cpc.2010.11.030}{\emph{Comput.Phys.Commun.}
  {\bfseries 182} (2011) 808--833},
  [\href{https://arxiv.org/abs/1002.0840}{{\ttfamily 1002.0840}}].

\bibitem{Staub:2012pb}
F.~Staub, \emph{{SARAH 3.2: Dirac Gauginos, UFO output, and more}},
  \href{https://doi.org/10.1016/j.cpc.2013.02.019}{\emph{Computer Physics
  Communications} {\bfseries 184} (2013) pp. 1792--1809},
  [\href{https://arxiv.org/abs/1207.0906}{{\ttfamily 1207.0906}}].

\bibitem{Staub:2013tta}
F.~Staub, \emph{{SARAH 4 : A tool for (not only SUSY) model builders}},
  \href{https://doi.org/10.1016/j.cpc.2014.02.018}{\emph{Comput. Phys. Commun.}
  {\bfseries 185} (2014) 1773--1790},
  [\href{https://arxiv.org/abs/1309.7223}{{\ttfamily 1309.7223}}].

\bibitem{Allanach:2001kg}
B.~C. Allanach, \emph{{SOFTSUSY: a program for calculating supersymmetric
  spectra}}, \href{https://doi.org/10.1016/S0010-4655(01)00460-X}{\emph{Comput.
  Phys. Commun.} {\bfseries 143} (2002) 305--331},
  [\href{https://arxiv.org/abs/hep-ph/0104145}{{\ttfamily hep-ph/0104145}}].

\bibitem{Allanach:2013kza}
B.~Allanach, P.~Athron, L.~C. Tunstall, A.~Voigt and A.~Williams,
  \emph{{Next-to-Minimal SOFTSUSY}},
  \href{https://doi.org/10.1016/j.cpc.2014.04.015}{\emph{Comput.Phys.Commun.}
  {\bfseries 185} (2014) 2322--2339},
  [\href{https://arxiv.org/abs/1311.7659}{{\ttfamily 1311.7659}}].

\bibitem{Harman:2015gif}
C.~P.~D. Harman and S.~J. Huber, \emph{{Does zero temperature decide on the
  nature of the electroweak phase transition?}},
  \href{https://doi.org/10.1007/JHEP06(2016)005}{\emph{JHEP} {\bfseries 06}
  (2016) 005}, [\href{https://arxiv.org/abs/1512.05611}{{\ttfamily
  1512.05611}}].

\bibitem{Senaha:2018xek}
E.~Senaha, \emph{{Radiative Corrections to Triple Higgs Coupling and
  Electroweak Phase Transition: Beyond One-loop Analysis}},
  \href{https://doi.org/10.1103/PhysRevD.100.055034}{\emph{Phys. Rev. D}
  {\bfseries 100} (2019) 055034},
  [\href{https://arxiv.org/abs/1811.00336}{{\ttfamily 1811.00336}}].

\bibitem{Biekotter:2021ysx}
T.~Biek\"otter, S.~Heinemeyer, J.~M. No, M.~O. Olea and G.~Weiglein,
  \emph{{Fate of electroweak symmetry in the early Universe: Non-restoration
  and trapped vacua in the N2HDM}},
  \href{https://doi.org/10.1088/1475-7516/2021/06/018}{\emph{JCAP} {\bfseries
  06} (2021) 018}, [\href{https://arxiv.org/abs/2103.12707}{{\ttfamily
  2103.12707}}].

\bibitem{Laine:2016hma}
M.~Laine and A.~Vuorinen, \emph{{Basics of Thermal Field Theory}}, vol.~925.
\newblock Springer, 2016,
  \href{https://doi.org/10.1007/978-3-319-31933-9}{10.1007/978-3-319-31933-9}.

\bibitem{Fowlie:2018eiu}
A.~Fowlie, \emph{{A fast C++ implementation of thermal functions}},
  \href{https://doi.org/10.1016/j.cpc.2018.02.015}{\emph{Comput. Phys. Commun.}
  {\bfseries 228} (2018) 264--272},
  [\href{https://arxiv.org/abs/1802.02720}{{\ttfamily 1802.02720}}].

\bibitem{Wainwright:2011kj}
C.~L. Wainwright, \emph{{CosmoTransitions: Computing Cosmological Phase
  Transition Temperatures and Bubble Profiles with Multiple Fields}},
  \href{https://doi.org/10.1016/j.cpc.2012.04.004}{\emph{Comput. Phys. Commun.}
  {\bfseries 183} (2012) 2006--2013},
  [\href{https://arxiv.org/abs/1109.4189}{{\ttfamily 1109.4189}}].

\end{thebibliography}\endgroup

\end{document}